%% file: main.tex
\documentclass[%
 prx,
 twocolumn,
 superscriptaddress,
 amsmath,amssymb,
 longbibliography
]{revtex4-2}
\usepackage{graphicx}
\usepackage{dcolumn}
\usepackage{bm}
\usepackage{siunitx}
\usepackage{color}
\usepackage{mathtools}
\usepackage{physics}
\usepackage{array}
\usepackage[colorlinks]{hyperref}
\hypersetup{%
    plainpages=true,
    breaklinks=true,
    hypertexnames=false,
    pageanchor=true,
    colorlinks=true,
    linkcolor={blue},
    citecolor={blue},
    urlcolor={blue},
    anchorcolor={black}
}
\usepackage{enumitem}   
\usepackage{titletoc}

\usepackage[utf8]{inputenc}
\usepackage{dsfont}
\usepackage{cleveref}
\usepackage{xcolor}
\usepackage{pgf}
\usepackage{placeins}
\usepackage{newfloat}

\usepackage{mleftright} 

\begin{document}

\preprint{APS/123-QED}

\title{Driven-dissipative entanglement of distant giant atoms}

\def\RLEaffil{Research Laboratory of Electronics, Massachusetts Institute of Technology, Cambridge, MA 02139, USA}
\def\NYUaffil{Department of Electrical and Computer Engineering, New York University, Brooklyn, NY, 11201, USA}
\def\LLaffil{MIT Lincoln Laboratory, Lexington, MA 02421, USA}
\def\Physaffil{Department of Physics, Massachusetts Institute of Technology, Cambridge, MA 02139, USA}
\def\EECSaffil{Department of Electrical Engineering and Computer Science, Massachusetts Institute of Technology, Cambridge, MA 02139, USA}
\def\chalmersaffil{Department of Microtechnology and Nanoscience, Chalmers University of Technology, 412 96 Gothenburg, Sweden}

\author{Aziza Almanakly}
\email{azizaalm@mit.edu} 
\affiliation{\RLEaffil}
\affiliation{\EECSaffil}

\author{Ariadna Soro}
\author{Alejandro~Vivas-Via\~na}
\affiliation{\chalmersaffil}

\author{Beatriz Yankelevich}
\affiliation{\RLEaffil}
\affiliation{\EECSaffil}

\author{Caspar Groiseau}
\affiliation{\chalmersaffil}

\author{David Pahl}
\author{Junyoung An}
\author{Gabriel Cutter}
\affiliation{\RLEaffil}
\affiliation{\EECSaffil}

\author{Michael E. Gingras}
\author{Bethany~M.~Niedzielski}
\author{Hannah Stickler}
\author{Ren\'ee DeP\'encier Pi\~nero}
\author{Mollie E.~Schwartz}
\affiliation{\LLaffil}

\author{Kyle~Serniak}
\affiliation{\RLEaffil}
\affiliation{\LLaffil}

\author{Max Hays}
\affiliation{\RLEaffil}

\author{Jeffrey A.~Grover}
\affiliation{\RLEaffil}

\author{Anton Frisk Kockum}
\affiliation{\chalmersaffil}
 
\author{William D.~Oliver}
\email{william.oliver@mit.edu}
\affiliation{\RLEaffil}
\affiliation{\EECSaffil}
\affiliation{\Physaffil}

\begin{abstract}
%
Quantum interconnects distribute entanglement via controlled light-matter interactions for quantum computing and sensing applications~\cite{Kimble2008, Cirac1997, Cirac1999}.
%
Many entanglement generation schemes use coherent, reversible interactions that require precisely calibrated pulses to execute.
In contrast, driven-dissipative protocols use a 
continuous-wave drive in the presence of correlated dissipation to stabilize entanglement in protected (dark) states~\cite{Pichler2015, Lalumiere2013}. 
However, the same dissipation that generates the entanglement 
also limits its utility once the stabilization protocol ends.
Here, we engineer a superconducting system of two giant artificial atoms coupled sequentially to a waveguide, with tunable individual and correlated dissipation enabled by interference between coupling points~\cite{Kannan2020, Kockum2018, Soro2022}.
Continuously driving the atoms through the waveguide exploits correlated dissipation to generate remote entanglement.
We then tune the qubit frequencies~\textit{in situ} to suppress individual dissipation and thereby preserve the entanglement, achieving a Bell-state fidelity $F = 0.89 \pm 0.02$.
%
%
This demonstration indicates that the driven dissipation of giant atoms is a viable approach for distributing entanglement across quantum networks.
\end{abstract}

\maketitle


Entanglement is a fundamental resource in quantum computing~\cite{NielsenQuantumComputation2012,Chitambar2019}. In one approach to achieving quantum advantage, processors use quantum interconnects to distribute entanglement across a large-scale, modular architecture~\cite{Dalzell2023, Eisert2025}. Distributed entanglement also functions as a resource in quantum metrology, where entangled sensor networks leverage quantum correlations to achieve precision beyond classical limits~\cite{GiovannettiQuantumMetrology2006,GiovannettiAdvancesQuantum2011,Zhang2026, PezzeQuantumMetrology2018,BraskImprovedQuantum2015}.
 Because these interconnects rely on bosonic modes to mediate coherent quantum information transfer, remote entanglement protocols exploit light-matter interactions.

\begin{figure*}[t!]
    \centering
    \includegraphics[width=\textwidth]{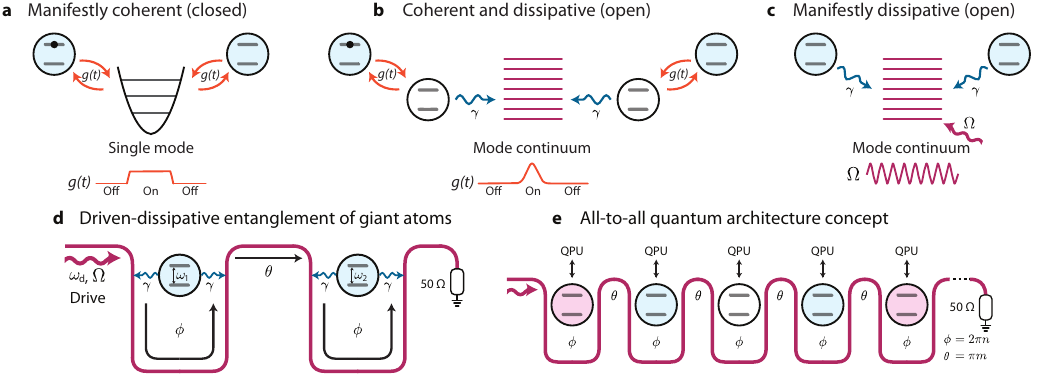}
    \caption{
    \textbf{The interactions underpinning remote entanglement generation.}
    \textbf{a)} Coherent interactions in closed quantum systems. Remote entanglement generation is often mediated by another mode. The qubit on the left begins in the excited state. The distant qubits interact with the shared mode using a time-dependent coherent coupling $g(t)$, resulting in a two-qubit entangled state indicated by the color blue. 
    \textbf{b)} Combined coherent and dissipative interactions in open systems. Alternatively, entanglement generation schemes emit and absorb itinerant photons into a mode continuum, such as in waveguide. These approaches combine dissipation ($\gamma$) and time-dependent coherent interactions ($g(t)$) to precisely shape the photon for maximum absorption efficiency.
    \textbf{c)} Driven-dissipative interactions in open systems. Two qubits coupled to a shared mode environment exhibit correlated dissipation due to interference in their spontaneous emission. Continuously driving (amplitude $\Omega$) two qubits through the environment generates steady-state entanglement. 
    \textbf{d)} Driven-dissipative entanglement scheme using sequential superconducting giant atoms coupled twice to a waveguide with strength $\gamma$. A coherent drive of amplitude $\Omega$ at frequency equivalent to the superradiant frequency ($\omega_\mathrm{d} =  \omega_\mathrm{super}$) entangles the qubits at frequencies $\omega_1 = \omega_\mathrm{super} + \delta$ and $\omega_2 = \omega_\mathrm{super} - \delta$, with opposite detuning from the superradiant frequency ($\delta$). The driving field accumulates the propagation phases $\phi$ and $\theta$ between coupling points along the waveguide.
    \textbf{e)} Extensible quantum architecture concept for all-to-all connectivity using many giant atoms coupled sequentially to a waveguide. Here, one drive entangles the blue and pink pairs of giant atoms, as determined by the qubit frequency detunings. The driven-dissipative protocol supports configurable, pairwaise, and simultaneous entanglement generation between non-local quantum processing units (QPUs) interfacing with giant atoms.
}
    \label{fig:fig1}
\end{figure*}

In the field of superconducting quantum circuits, coherent, reversible interactions in closed systems underpin many deterministic remote entanglement schemes (Fig.~\ref{fig:fig1}a). Coherent coupling to a resonator interconnect mediates entanglement generation between distant qubits in a point-to-point fashion~\cite{Zhong2019, Zhong2021, Hensen2015, Pita_Vidal_2024}. In contrast, alternative interconnects utilize the physics of open quantum systems, where qubits dissipate into a mode continuum in a waveguide (Fig.~\ref{fig:fig1}b). These approaches combine coherent, tunable interactions with microwave photon emission and absorption, using the waveguide as a data bus to connect qubits across arbitrary distances~\cite{Kurpiers2018,Magnard2020, Storz2023, kannan2023, Almanakly2025, Almanakly2026}. To date, achieving high-fidelity entanglement across such interconnects has required precise timing and calibration of these underlying reversible interactions.

While dissipation must be carefully managed to avoid loss in schemes using propagating photons, it can instead be used to generate steady-state entanglement~\cite{PoyatosQuantumReservoir1996, PlenioCavitylossinducedGeneration1999,Gonzalez-TudelaEntanglementTwo2011,Pichler2015,KastoryanoDissipativePreparation2011,Vivas-VianaDissipativeStabilization2024,Vivas-VianaFrequencyResolvedPurcell2024,LingenfelterExactResults2024,Gonzalez-TudelaLightMatter2024,AgustiNonMarkovianThermal2026}. Qubits coupled to a common environment exhibit interference in their spontaneous emission, resulting in correlated dissipation (Fig.~\ref{fig:fig1}c). This interaction is manifest through the collective phenomena of Dicke superradiance and subradiance~\cite{Dicke1954,GrossSuperradianceEssay1982, VanLoo2013, Mlynek2014}, giving rise to bright and dark entangled states that are controlled through or protected from the environment~\cite{Lalumiere2013, Solano2017, EvansPhotonmediatedInteractions2018,Mirhosseini2019,TiranovCollectiveSuper2023}.

Theoretical work has proposed driven-dissipative entanglement generation, where continuously driving qubits through their environment gradually shelves population into the dark state~\cite{Pichler2015,Guimond2020}. This process is commonly referred to as the autonomous stabilization of entanglement, running without the calibration overhead of coherent protocols. Because only the dark state is protected from decay, extracting this entanglement for use in quantum algorithms is an outstanding challenge. Operations that attempt to access the dark state generally will alter its isolation condition, triggering~dissipation.

Demonstrations of driven-dissipative schemes in superconducting and trapped-ion systems have used cavity-mediated dissipation to entangle neighboring qubits~\cite{Shankar2013, Brown2022, Lin2013, Chen2025, guo2026}. These schemes require the coordination of several drive tones and energy transitions, often relying on heralding or post-selection to characterize the state. In waveguide-quantum electrodynamical (wQED) implementations, distant superconducting qubits are driven directly through the waveguide. In the absence of an isolation mechanism, the challenge of characterization---for example, by studying the multi-photon correlations of the subsequent emission into the waveguide---limits the generation and storage of entanglement~\cite{YangEntanglementPhotonic2025,Zanner2022, Shah2024, Irfan2025, Andresjuanes2025}. These driven-dissipative wQED schemes have achieved entanglement fidelity up to \SI{55}{\percent}.


Driven-dissipative entanglement schemes are plagued by the same dissipation that enables the entanglement generation. Using the entanglement requires rotating out of the protected state, so we must first suppress the dissipation used to generate the state. This tunable dissipation can be realized using giant atoms~\cite{Kannan2020, FriskKockum2014, Kockum2021review}. In these wQED systems, an artificial atom couples to a waveguide at multiple points separated by distances comparable to the interacting wavelength of light. Interference between these coupling points yields a frequency-dependent qubit-relaxation rate, enhancing or suppressing decay.

In this work, we realize a wQED system of sequential giant atoms each coupled twice to a common waveguide (Fig.~\ref{fig:fig1}d). By engineering the phases between coupling points, we stage interference effects that enable both tunable single-qubit and collective dissipation. Exploiting collective superradiance and subradiance, we entangle the distant qubits using a single, continuous waveguide drive. Upon reaching steady state, we suppress single-qubit dissipation by tuning both qubit frequencies such that destructive interference protects the two-qubit subspace, providing sufficient time to perform quantum-state tomography. We achieve Bell-state fidelity $F = 0.89 \pm 0.02$ and concurrence $C = 0.86 \pm 0.02$ using the deterministic driven-dissipative protocol. This scheme extends straightforwardly to entangle many distant giant atoms with essentially all-to-all connectivity for distributed quantum computing and~sensing (Fig.~\ref{fig:fig1}e).

\begin{figure*}[t!]
    \centering
    \includegraphics[width=\textwidth]{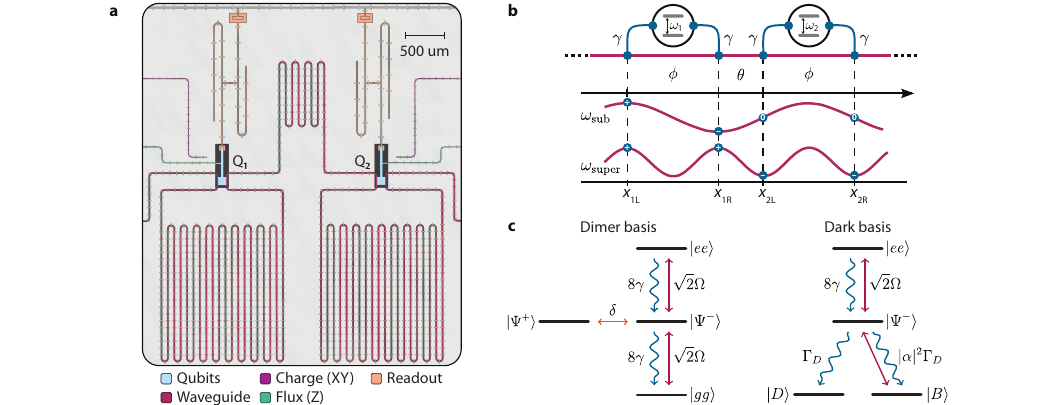}
    \caption{
    \textbf{Sequential giant atoms device and experimental set up.} 
    \textbf{a)} False-colored optical micrograph of the device. Giant atoms are realized using transmon qubits (light blue) coupled twice with strength $\gamma$ to a common waveguide (red). XY (purple) and Z (green) control lines and dispersive readout circuitry (orange) enable \textit{in situ} qubit tuning and state characterization. 
    \textbf{b)} Schematic of the sequential giant-atom configuration. The phase difference between the coupling points of an individual atom is $\phi(\omega) = \omega (x_\mathrm{1R} - x_\mathrm{1L})/\nu = \omega (x_\mathrm{2R} - x_\mathrm{2L})/\nu$, while the phase separating the atoms is $\theta(\omega) = \omega (x_\mathrm{2L} - x_\mathrm{1R})/\nu$, where $\nu$ is the speed of light in the waveguide. For qubit frequency $\omega = \omega_\mathrm{sub}$ [$\phi(\omega_\mathrm{sub}) = \pi (2n + 1)$, $n \in\mathbb{N}$], the atoms decouple from the waveguide via destructive interference (individual subradiance), highlighted by the signs of the phases of an example wave at each coupling point (blue circles).  For qubit frequency $\omega =\omega_\mathrm{super}$ [$\phi(\omega_\mathrm{super}) = 2\pi n$], constructive interference enhances dissipation (individual superradiance). Setting $\theta(\omega_\mathrm{super}) = (2m+1)\pi$ ($m \in\mathbb{N}$) at $\omega_\mathrm{super}$ maximizes correlated dissipation, giving rise to bright and dark entangled states.
    \textbf{c)} Energy level diagram of the two-qubit subspace. Operating the qubits near $\omega_\mathrm{super}$, the interplay between drive amplitude $\Omega$, waveguide dissipation $\gamma$, and qubit-drive detuning $\delta$ populates the subradiant Bell state $|\Psi^+\rangle = (|ge\rangle + |eg\rangle)/\sqrt{2}$. While $|\Psi^+\rangle$ is protected by symmetry from waveguide dissipation, the superradiant Bell state $|\Psi^-\rangle$ faces enhanced decay. Subsequently tuning the qubit frequencies from $\omega_\mathrm{super}$ to $\omega_\mathrm{sub}$ protects the generated entanglement. On the right, the level diagram highlights the dark $|D\rangle$ and bright $|B\rangle$ states, where $\alpha = i\Omega/\sqrt{2}\delta$. The dark state is decoupled from coherent dynamics, interacting with the system exclusively through dissipation from $|\Psi^-\rangle$ at rate $\Gamma_D$.
}
    \label{fig:fig2}
\end{figure*}

\begin{figure}[t!]
    \centering
    \includegraphics[width=3.45in]{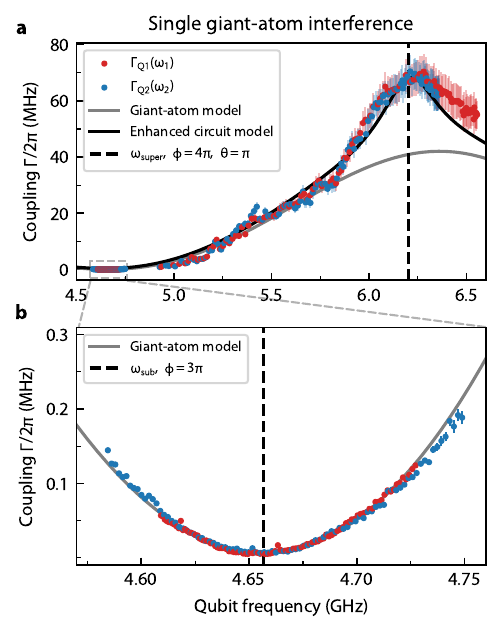}
    \caption{
    \textbf{Single giant-atom superradiance and subradiance.} 
    \textbf{a)} Individual waveguide coupling rates $\Gamma_\mathrm{Q1}(\omega_1)$ and $\Gamma_\mathrm{Q2}(\omega_2)$ across the qubit spectrum. For large couplings, decay rates are extracted from waveguide transmission spectroscopy while the other qubit idles at the subradiant frequency $\omega_\mathrm{sub}/2\pi = 4.656$ GHz. We observe superradiance, or maximal waveguide coupling, of the individual giant atoms at $\omega_\mathrm{super}/2\pi = 6.208$ GHz and determine the value of the individual waveguide coupling points $\Gamma(\omega_\mathrm{super})/8\pi = \gamma(\omega_\mathrm{super})/2\pi \approx 14.7$ MHz. We fit the $\Gamma_1(\omega_1)$ data to an enhanced circuit model that combines the expected giant-atom dissipation spectrum in Eq.~\eqref{eq:giant_atom_coupling} with a Lorentzian response to a low-Q mode arising from the meandered waveguide architecture and local impedance mismatches (black line---see Supplementary Information Eq.~\eqref{eq:SM-Effective-decay-rate}).
    \textbf{b)} Individual decay rates near the subradiant frequency $\omega_\mathrm{sub}$ extracted from qubit lifetime ($T_1$) measurements. The minimal dissipation, or non-radiative decay rate $\gamma_\mathrm{nr}/2\pi \approx 5.4$ kHz, corresponds to a maximal $T_1\approx$~\SI{29.5}{\micro\s}. We fit the data to the expected giant-atom model in Eq.~\eqref{eq:giant_atom_coupling} (gray), and plot this fit above in part \textbf{a} to illustrate the deviation of the data near the superradiant frequency. Note that maximum dissipation occurs at a qubit frequency slightly larger than $\omega_\mathrm{super}$, which we attribute to the quadratic frequency dependence of $\gamma(\omega)$.
}
    \label{fig:fig3}
\end{figure}

\section{Experiment}

Our device comprises two nominally identical flux-tunable transmon artificial giant atoms Q$_1$ and Q$_2$ (blue) coupled sequentially to a waveguide (red)~\cite{Koch2007} (Fig.~\ref{fig:fig2}a). Each giant atom is coupled with strength $\gamma$ to two physically separated points along the waveguide. This separation $\Delta x_i = x_\mathrm{\textit{i}R} - x_\mathrm{\textit{i}L}$ (qubit index $i = 1,2$), the frequency $\omega$,  and the speed of light in the waveguide $\nu$ determine the phase difference between the coupling points of an individual giant atom $\phi(\omega) = \omega\Delta x_i/\nu$ (Fig.~\ref{fig:fig2}b).

At each coupling point, each individual giant atom sees a distinct phase of the same wave propagating in the waveguide. Because of interference between the coupling points, the overall dissipation to the waveguide tunes with qubit frequency according to
\begin{equation}
    \Gamma_{\mathrm{Q}i}(\omega_i) = 2\gamma(\omega_i)\mleft[1 + \cos{\phi(\omega_i)}\mright] + \gamma_\mathrm{nr},
    \label{eq:giant_atom_coupling}
\end{equation}
where $\omega_i$ is the frequency of Q$_i$, and $\gamma_\mathrm{nr}$ is the rate of non-radiative decay to channels other than the waveguide~\cite{FriskKockum2014, Kockum2018,Soro2022}. The strength of each coupling point $\gamma(\omega_i) = (\omega_i/\omega_0)^2\gamma_0$ is defined in reference to the superradiant frequency ($\omega_0 = \omega_\mathrm{super} = 2 \pi \times 6.208 $ GHz) and decay rate $\gamma_0/2\pi = 14.7$ MHz.

Setting the qubit frequency to the subradiant frequency ($\omega_i = \omega_\mathrm{sub}$) such that $\phi(\omega_\mathrm{sub})$ is an odd multiple of $\pi$ results in destructive interference, decoupling the atom from the waveguide $[\Gamma_\mathrm{Q1}(\omega_\mathrm{sub}) = \Gamma_\mathrm{Q2}(\omega_\mathrm{sub}) = \gamma_\mathrm{nr}]$. Because the coupling points are completely out of phase, the qubit exhibits individual subradiance, or protection from decay, when tuned to $\omega_\mathrm{sub}/2\pi = \SI{4.66}{\giga\hertz}$.

In contrast, setting the qubit frequency to the superradiant frequency ($\omega_i=\omega_\mathrm{super}$) such that $\phi(\omega_\mathrm{super})$ is an integer multiple of $2\pi$ gives rise to constructive interference, maximally enhancing dissipation $[\Gamma_\mathrm{Q1}(\omega_\mathrm{super}) = \Gamma_\mathrm{Q2}(\omega_\mathrm{super}) \approx 4\gamma(\omega_\mathrm{super}) + \gamma_\mathrm{nr}]$. Because the coupling points are in phase, the qubit experiences individual superradiance when operated at $\omega_\mathrm{super}$. The enhanced coupling mediates qubit control through the waveguide.

To extract the dissipation spectrum of each qubit $\Gamma_\mathrm{Qi}(\omega_i)$, we measure the transmission of a weak coherent tone through the waveguide as a function of qubit frequency. During the measurement, the other qubit frequency is set to $\omega_\mathrm{sub}$, where it decouples from the waveguide. We extract the decay rates using Lorentzian fits and plot the dissipation spectra in Fig.~\ref{fig:fig3}a. The individual dissipation is maximum at qubit frequency $\omega_\mathrm{super}$, with a Lorentzian enhancement due to the meandered structure of the waveguide and resulting impedance mismatches (black line---see Supplementary Information).

Using qubit-lifetime ($T_1$) measurements, we extract the dissipation spectrum $\Gamma_\mathrm{Qi}(\omega_i)$ near the subradiant frequency $\omega_\mathrm{sub}$, shown in Fig.~\ref{fig:fig3}b. The local minimum in dissipation at $\omega_\mathrm{sub}$ indicates the non-radiative decay rate is $\gamma_\mathrm{nr}/2\pi \approx 5.4$ kHz and the maximum lifetime is $T_1\approx$~\SI{29.5}{\micro\s}. To characterize the qubits, we bias them at the subradiant frequency $\omega_\mathrm{sub}$, where the increased qubit lifetime allows for dispersive readout.

In addition to these individual interference effects, the sequential giant atoms also experience collective superradiance and subradiance controlled by the interference between the coupling points of both atoms. The nature of the interference is determined by the phase separating the atoms $\theta(\omega) = \omega (x_\mathrm{2L} - x_\mathrm{1R})/\nu$. If both qubits are tuned to $\omega_\mathrm{super}$ such that $\theta(\omega_\mathrm{super})$ is an integer multiple of $\pi$, the qubits interact with the same or opposite phase of the wave, as drawn in Fig.~\ref{fig:fig2}b~\cite{Soro2022}. As a result, the system features dark and bright entangled states that are protected from or controlled through the~waveguide.

Here, we set $\theta(\omega_\mathrm{super}) = \pi$, such that the waveguide coupling points of the different atoms are completely out of phase when the qubit-drive detuning $\delta$ is zero. This geometry determines the assignment of the Bell state $|\Psi^-\rangle = (|ge\rangle - |eg\rangle)/\sqrt{2}$ as the superradiant state with enhanced dissipation and $|\Psi^+\rangle = (|ge\rangle + |eg\rangle)/\sqrt{2}$ as the subradiant state with suppressed dissipation. The coherent superposition of the superradiant and subradiant states with the joint ground state $|gg\rangle$ state form the bright $|B\rangle$ and dark $|D\rangle$ states (Fig.~\ref{fig:fig2}c). The dark state
\begin{equation}
    |D\rangle = \frac{1}{\sqrt{1+ |\alpha|^2}}(\alpha|\Psi^+\rangle +|gg\rangle)
\end{equation}
approaches the subradiant state $|\Psi^+\rangle$ in the limit of strong drive $\Omega$ and small, non-zero detuning $\delta$, governed by $\alpha = i\Omega/\sqrt{2}\delta$ (see Supplementary Information).
%
%


To generate steady-state entanglement, the two giant atoms are coherently driven through the waveguide at frequency $\omega_\mathrm{d}$ and drive amplitude $\Omega$, illustrated in Fig.~\ref{fig:fig1}c. In the absence of detuning, the Bell state $|\Psi^+\rangle$ is fully decoupled from the driven manifold and therefore remains unpopulated. Introducing a small qubit-qubit detuning $2\delta$ coherently couples the bright and dark manifolds, allowing population transfer from superradiant state $|\Psi^-\rangle$ into the subradiant state $|\Psi^+\rangle$ during the drive (Fig.~\ref{fig:fig2}c). In the regime $\delta \ll \Omega, \gamma$, the subradiant state remains only weakly coupled to the dissipative manifold, while the drive and dissipation dominate the dynamics of the superradiant state. Consequently, population progressively accumulates in $|\Psi^+\rangle$, ultimately leading to stabilized entanglement~\cite{Pichler2015,Soro2022,Vivas-VianaDissipativeStabilization2024,Vivas-VianaUnconventionalMechanism2022}.

\begin{figure}[t!]
    \centering
    \includegraphics[width=3.45in]{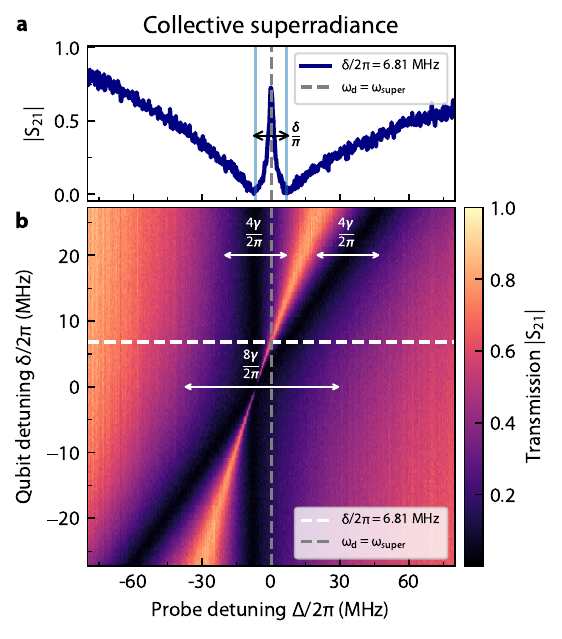} 
    \caption{
    \textbf{Collective superradiance.} 
    \textbf{a)} Waveguide transmission spectroscopy of the giant atoms each detuned by $\delta/2\pi = 6.81$~MHz from the superradiant frequency $\omega_\mathrm{super} = 6.208$~GHz, where $\Delta$ is the detuning of the transmission probe frequency from $\omega_\mathrm{super}$. To begin the driven-dissipative entanglement generation protocol, we operate the qubits in this frequency configuration and input a coherent drive at frequency $\omega_\mathrm{d} = \omega_\mathrm{super}$ through the waveguide. 
    \textbf{b)} Collective superradiant crossing. We measure transmission as a function of qubit detuning $\delta$ near $\omega_\mathrm{super}$. At zero detuning ($\delta = 0$), the qubits exhibit collective superradiance, characterized by an enhanced linewidth of $8\gamma(\omega_\mathrm{super})$.
}
    \label{fig:fig4}
\end{figure}

To demonstrate collective superradiance, we perform transmission spectroscopy as a function of qubit-qubit detuning $2\delta$ near $\omega_\mathrm{super}$, shown in Fig.~\ref{fig:fig4}. The detuned qubits couple to the waveguide with approximate strength $4\gamma(\omega_\mathrm{super})$ due to individual superradiance. At zero detuning, the system exhibits maximal collective superradiance, indicated by the broadening of the Lorentzian feature to a width of $8\gamma(\omega_\mathrm{super})$. We highlight the 1D transmission trace for qubit-drive detuning $\delta/2\pi = 6.81$~MHz in Fig.~\ref{fig:fig4}a. We initialize the qubits into this frequency configuration, centered around $\omega_\mathrm{super}$, driving through the waveguide at frequency $\omega_\mathrm{d} = \omega_\mathrm{super}$. 
\begin{figure*}[t!]
    \centering
    \includegraphics[width=\textwidth]{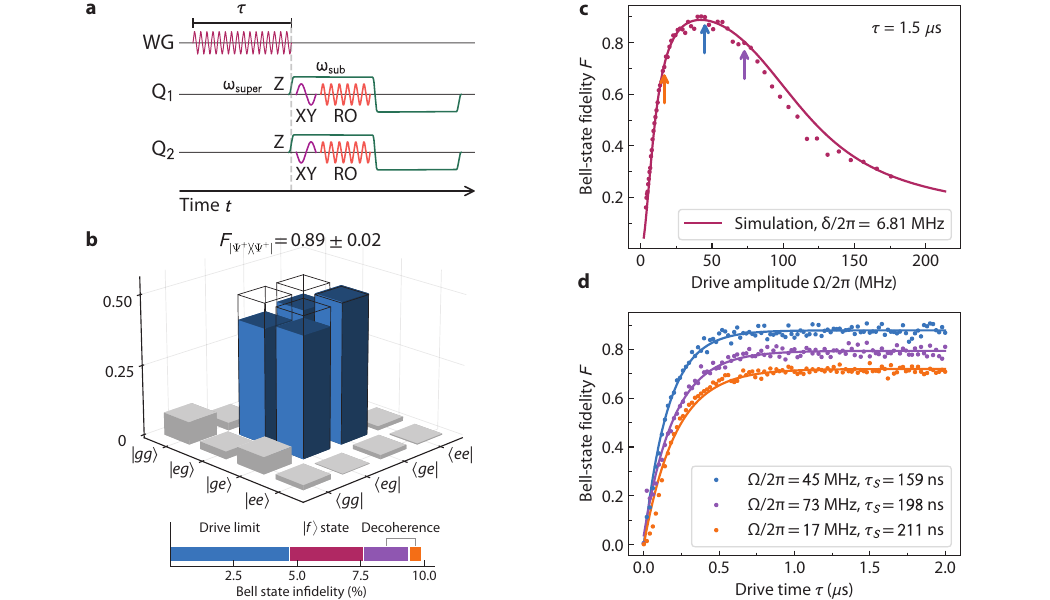}
    \caption{
    \textbf{Driven-dissipative entanglement generation using giant atoms.} 
    \textbf{a)} Experimental pulse scheme. Starting with qubits near $\omega_\mathrm{super}$, a coherent drive of amplitude $\Omega$ is applied at frequency $\omega_\mathrm{d} = \omega_\mathrm{super}$ for time duration $\tau$. Upon reaching steady state, the qubits are flux-tuned (Z, green) to $\omega_\mathrm{sub}$ to decouple the atoms from the waveguide, protecting the generated entangled state. We perform qubit state tomography using XY pulses (purple) and dispersive readout (orange) to quantify the Bell-state fidelity. To mitigate flux transients, we implement net-zero predistorted flux pulses.
    \textbf{b)} Average Bell-state fidelity. For $\Omega/2\pi = 45$~MHz and $\tau =$~\SI{1.5}{\micro\s}, we measure an average fidelity $F = 0.89 \pm 0.02$ and concurrence $C = 0.86 \pm 0.02$ (100 tomography repetitions of 10,000 shots). The reconstructed density matrix (blue) is plotted against the ideal Bell state $|\Psi^+\rangle$ (black wires). The population imbalance originates from a slight asymmetry in the AC Stark shift imposed by the drive. The sources of Bell-state infidelity are broken down beneath the density matrix.
    \textbf{c)} Bell-state fidelity as a function of drive amplitude $\Omega$. Given a particular qubit-drive detuning $\delta$, the fidelity reaches a local maximum before the cross-coherences of the density matrix begin to degrade. Master-equation simulations (solid line) confirm that the optimal drive amplitude, and thus the maximum fidelity, is limited by the transmon anharmonicity (see Supplementary Information).
    \textbf{d)} Entanglement stabilization. The dark-state stabilization time $\tau_S$, extracted from exponential fits, reaches a local minimum at the optimal drive amplitude.
}
    \label{fig:fig5}
\end{figure*} 

\section{Entanglement Generation}
We begin the driven-dissipative entanglement generation protocol by biasing the qubits near the superradiant frequency $\omega_\mathrm{super}$. We simultaneously drive the qubits through the waveguide with amplitude $\Omega$ for a time duration $\tau$ until the system reaches steady state and entanglement is stabilized. As discussed above, the protection of the dark state presents a challenge for the utilization of entanglement as a resource. Unitary operations, which are necessary for state characterization and in the execution of quantum algorithms, rotate the system out of the protected state. This transition into a radiative state triggers dissipation, ultimately corrupting the fidelity of the entanglement.

After stabilizing the entangled dark state, we suppress dissipation by flux-tuning the entangled qubits to the subradiant frequency $\omega_\mathrm{sub}$ such that each qubit individually decouples from the waveguide, as illustrated in Fig.~\ref{fig:fig5}a. As a result, the whole two-qubit subspace is protected from waveguide dissipation. This individual giant-atom subradiance isolates the qubits from the waveguide, granting sufficient time to apply the single-qubit XY rotations required for state tomography using local drive lines. 

We extract an average Bell-state fidelity of $F = 0.89 \pm 0.02$ from 100 two-qubit state tomography measurements of 10,000 shots each. The concurrence $C = 0.86 \pm 0.02$ quantifies the correlation between the states of the two qubits. The ideal density matrix (black wireframes) is overlaid atop an example reconstructed density matrix in Fig.~\ref{fig:fig5}b. In the ideal case of lossless two-level systems, this approach stabilizes a perfect Bell state in the limit of an increasingly strong drive ($\Omega \gg \delta$)~\cite{Pichler2015, Soro2022, Vivas-VianaDissipativeStabilization2024}. However, the presence of the transmon $|f\rangle$ state prevents arbitrarily strong drive amplitudes, resulting in $4.7\%$ infidelity due to residual joint ground state population ($|gg\rangle$). Similarly, leakage outside the two-qubit subspace in addition to an imbalance in the drive-induced AC Stark shift contributes $2.9\%$ infidelity. Estimates of qubit decoherence, including pure dephasing and non-radiative decay, account for another $2.4\%$ infidelity (see Supplementary Information Sec.~\ref{sec-SM:conditins-entanglement}).


To study the nature of the driven-dissipative entanglement process, we measure the Bell-state fidelity as a function of drive amplitude for a drive time of $\tau =$~\SI{1.5}{\micro\s} (Fig.~\ref{fig:fig5}c). As predicted by master-equation simulations (solid red line), the fidelity reaches a maximum at drive amplitude $\Omega/2\pi = 45$ MHz. Beyond this maximum, the strong drive produces a mixed state, diminishing the two-qubit coherences and corrupting the fidelity~\cite{WallsQuantumOptics2008}.

Next, we measure the fidelity as a function of drive time $\tau$ for various drive amplitudes. We observe the expected exponential time dynamics as the system stabilizes into the entangled steady state. As illustrated in Fig.~\ref{fig:fig5}d, the stabilization time is shortest for the optimal drive amplitude. The choice of detuning $\delta/2\pi=\SI{6.81}{\mega\hertz}$ determines the stabilization time in competition with decoherence while also optimizing the entanglement fidelity (see Supplementary Information, Sec.~\ref{sec-SM:conditins-entanglement}). After reaching steady state, the Bell-state fidelity does not depend on drive duration, reflecting the autonomous stabilization of entanglement via driven dissipation. 

\section{Conclusions}
Here, we generate driven-dissipative entanglement using a wQED system of two distant giant atoms coupled twice to the same waveguide. Leveraging collective superradiance and subradiance, we drive the qubits into a protected entangled state. We then tune the qubit frequencies such that destructive interference protects the individual giant atoms from decay. We measure an average Bell-state fidelity $F = 0.89 \pm 0.02$ using two-qubit state tomography. This autonomous entanglement stabilization scheme requires only a single continuous drive.

Using giant atoms, we exploit quantum interference to suppress the always-on dissipation that has historically limited the utility of driven-dissipative entanglement protocols. This approach enables both the generation of entanglement as well as its subsequent utility, e.g., as indicated by the tomography measurements performed while the qubits were individually protected from radiative decay into the waveguide. While that protection was used to characterize fidelity in this work, it would also allow the entanglement to be swapped into a processor and used, for example, to execute remote unitary operations.

While existing schemes generate high-fidelity remote entanglement using carefully calibrated coherent interactions, the driven-dissipative approach is a steady-state alternative with essentially all-to-all connectivity. 
The protocol performance here approaches the fidelity threshold required for interfacing logical qubits, which is more lenient than the corresponding surface-code threshold~\cite{ramette2023,haug2025}. Future iterations can improve the fidelity by increasing the anharmonicity, thereby reducing the influence of the next transmon energy level. For example, one could increase the anharmonicity using capacitively-shunted flux qubits~\cite{Yan2016} to reach an estimated fidelity of roughly 95\%, assuming the decoherence estimates presented here (see Supplementary Information). Adapting active cancellation techniques to negate drive-induced shifts could further increase the fidelity past the fault-tolerance threshold for quantum channels~\cite{Chiaro2025}.

An extension of the driven-dissipation entanglement protocol involves coupling many giant atoms sequentially to the same waveguide (Fig.~\ref{fig:fig1}e). Each giant atom serves as an interface between the waveguide bus and a quantum processing unit (QPU). By configuring pairs of qubit-frequency detunings relative to the drive, we can simultaneously generate Bell pairs with many-to-many or even all-to-all connectivity. This steady-state remote entanglement resource enables gate teleportation~\cite{Jiang2007, Chou2018}, a critical operation in distributed quantum computing and sensing. Such distributed architectures are likely key to reaching the scale and quality of quantum computers that will enable quantum advantage. The giant-atom platform also facilitates the quantum simulation of many-body physics and exotic light-matter interaction using configurable entanglement networks~\cite{Pichler2015,Lodahl2017}.


\section*{Author Contributions}
AA designed the experiment procedure and the device, conducted the measurements, analyzed data, and wrote the manuscript.
AS performed theoretical calculations and simulations to aid in the experiment design.
AS, AVV, and CG performed numerical simulations to model the experimental data. AVV and CG developed effective theoretical models used for the analytical derivations and interpretation of the experimental results. AA, AVV, CG, and DP wrote the Supplementary Information with input from all authors.
DP designed the readout circuit on the device.
JA assisted with predistortion calibration.
BY and GC assisted with experimental infrastructure.
MG, BMN, and HS fabricated the devices with coordination from RDP, MES, and KES.
MH, JAG, AFK, and WDO supervised the project.
All authors discussed the results and commented on the manuscript.

\section*{Acknowledgments}
This research was funded in part by the Army Research Office under Award No.~W911NF-23-1-0045 and in part under Air Force Contract No.~FA8702-15-D-0001. AS, AVV, CG, and AFK acknowledge support from the Swedish Foundation for Strategic Research (Grant No.~FFL21-0279), the Swedish Research Council (Grant No.~2019-03696), and the Knut and Alice Wallenberg Foundation through the Wallenberg Centre for Quantum Technology (WACQT). AFK is also supported by the Swedish Foundation for Strategic Research (Grant No.~FUS21-0063) and the Horizon Europe programme HORIZON-CL4-2022-QUANTUM-01-SGA via the project 101113946 OpenSuperQPlus100. The views and conclusions contained herein are those of the authors and should not be interpreted as necessarily representing the official policies or endorsements, either expressed or implied, of the U.S. Government. 

\section*{Data Availability}
The data that support the findings of this study are available from the corresponding author upon reasonable request.

\section*{Code Availability}
The code used for numerical simulations and data analyses is available from the corresponding author upon reasonable request.

\bibliography{AVV_Refs,main,supp}

\include{supp}

\end{document}

%% file: supp.tex
\onecolumngrid
\newpage

\begin{center}
    \textbf{\large SUPPLEMENTARY INFORMATION}
\end{center}
\begin{center}
    \textbf{\large Driven-dissipative entanglement of distant giant atoms}
\end{center}
\vspace{0.5cm}

\setcounter{section}{0}
\setcounter{figure}{0}
\setcounter{table}{0}
\setcounter{equation}{0}

\makeatletter
\renewcommand{\thefigure}{S\@arabic\c@figure}
\renewcommand{\thetable}{S\@arabic\c@table}
\renewcommand{\theequation}{S\arabic{equation}}
\makeatother

\newcommand{\tocnumberbox}[1]{\makebox[3em][l]{#1}}

\titlecontents{section}
  [3em] 
  {\addvspace{2pt}}
  {\contentslabel{3em}} 
  {\hspace*{-3em}\let\numberline\tocnumberbox} 
  {\titlerule*[0.8pc]{.}\contentspage}

\titlecontents{subsection}
  [5em] 
  {}
  {\contentslabel{3em}} 
  {\hspace*{-3em}\let\numberline\tocnumberbox} 
  {\titlerule*[0.8pc]{.}\contentspage}

\startcontents[supp]
\begin{center}
    \textbf{Table of Contents}
\end{center}

\printcontents[supp]{}{1}{}

\vspace{1cm}


\newpage

\section{Device and experimental setup}
\begin{figure*}[h]
    \centering
    \includegraphics[width=7in]{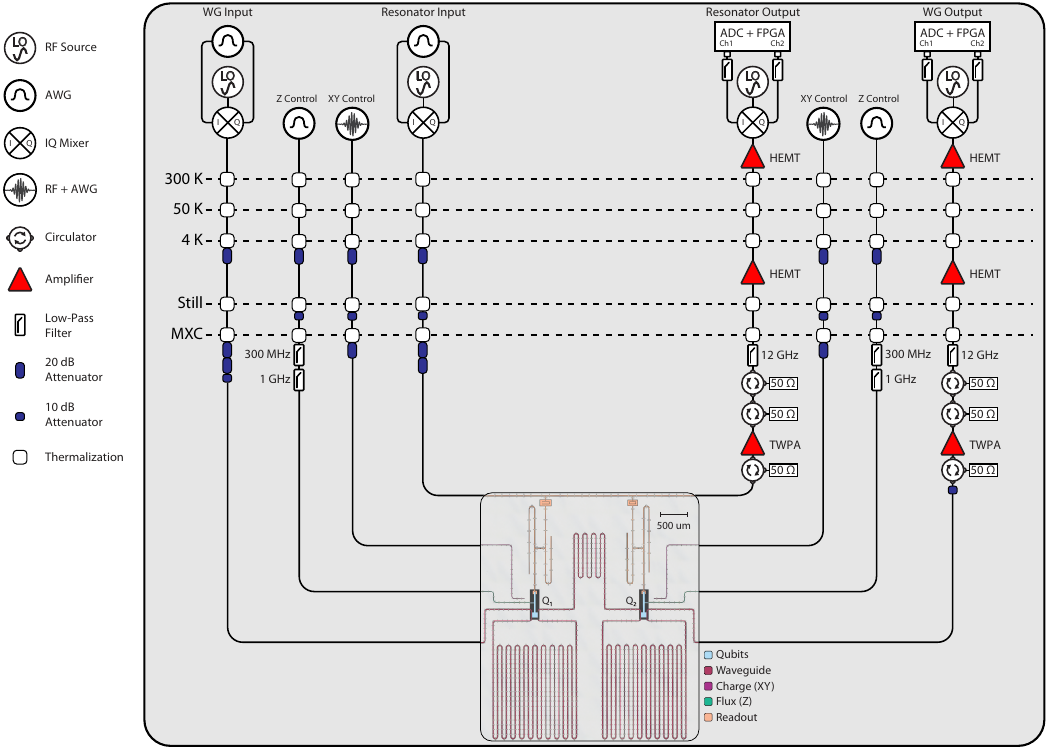}
    
    \caption{\textbf{Experimental setup.} Wiring schematic of the device and all electronics used to perform the experiment. Each qubit is coupled to an individual charge line (purple) and a flux line (green) with separate, but identical, control electronics. The same device micrograph from Fig.~\ref{fig:fig2}a is shown here.}
    \label{fig:setup}
\end{figure*} 

This experiment was conducted in a Bluefors XLD1000 dilution refrigerator, which operates at base temperature of around \SI{15}{\milli\kelvin} throughout the experiment.
The experimental setup is shown in Fig.~\ref{fig:setup}.
The device is protected from ambient magnetic fields by superconducting and Cryoperm-10 shields below the mixing chamber (MXC) stage.
To minimize thermal noise from higher temperature stages, the inputs are attenuated by \SI{20}{\decibel} at the \SI{4}{\kelvin} stage, \SI{10}{\decibel} at the Still stage, and \SI{60}{\decibel} (\SI{40}{\decibel} for resonator readout input) at the MXC stage.
The output signals are filtered with \SI{12}{GHz} low-pass filters.
Two additional isolators are placed after the circulator at the MXC stage to prevent noise from higher-temperature stages from traveling back into the sample. Traveling-wave parametric amplifiers (TWPA) are used at the MXC stage and high-electron-mobility transistor (HEMT) amplifiers are used at the \SI{4}{K} and room-temperature stages of the measurement chain to amplify the outputs from the device.
The signals are then down-converted to an intermediate frequency using an IQ mixer, after which they are filtered, digitized, and demodulated.
Both qubits are also equipped with their own radio-frequency (RF) control bias lines, which are attenuated by \SI{20}{\decibel} at the \SI{4}{K} stage, and by \SI{10}{\decibel} at the \SI{1}{\kelvin} stage.
The qubits are both equipped with a local charge line for independent single-qubit XY gates.
The specific control and measurement equipment used throughout the experiment is summarized in Table~\ref{tab:equipment}.
The relevant parameters of the device used in the experiment are summarized in Table~\ref{tab:params}.

\newpage
\begin{table}[h!]
\centering
\begin{tabular}{lll} 
\hline
\hline
Component & Manufacturer & Model \\
\hline
Dilution Refrigerator & Bluefors & XLD1000 \\
RF Source & Rohde \& Schwarz & SGS100A \\
Control Chassis & Keysight & M9019A \\
AWG & Keysight & M3202A \\
ADC & Keysight & M3102A \\
\hline
\hline
\end{tabular}
\caption{\textbf{Summary of control equipment.} The manufacturers and model numbers of experimental control equipment.}
\label{tab:equipment}
\end{table}

\begin{table}[h!]
\centering
\begin{tabular}{lll} 
\hline
\hline
Parameter & $\textrm{Q}_1$ & $\textrm{Q}_2$ \\
\hline
Superradiant Frequency $\omega_\mathrm{super}/2\pi$ & \SI{6.208}{\giga\hertz} & \SI{6.208}{\giga\hertz} \\ 
Subradiant Frequency $\omega_\mathrm{sub}/2\pi$ & \SI{4.656}{\giga\hertz} & \SI{4.662}{\giga\hertz} \\ 
Superradiant Coupling $\Gamma(\omega_\mathrm{super})/2\pi$ ($4\gamma(\omega_\mathrm{super})/2\pi$)~~~~~~~~& \SI{58.3}{\mega\hertz}~~~~~~~~~~& \SI{59.6}{\mega\hertz}\\
Subradiant Coupling $\Gamma(\omega_\mathrm{sub})/2\pi$ ($\gamma_\mathrm{nr}/2\pi$)& \SI{5.4}{\kilo\hertz} & \SI{6.8}{\kilo\hertz} \\
Anharmonicity ($\Lambda/2\pi$) & \SI{207}{\mega\hertz} & \SI{206}{\mega\hertz}\\
$T_1(\omega_\mathrm{sub})$ & \SI{29.5}{\micro\s} & \SI{23.4}{\micro\s} \\
$T_2^*(\omega_\mathrm{sub})$ & \SI{6.4}{\micro\s} & \SI{2.3}{\micro\s} \\
\hline
\hline
\end{tabular}
\caption{\textbf{Summary of device parameters.} The operational qubit frequencies, qubit–waveguide coupling strengths, transmon anharmonicity, and coherence times $T_1$ and $T_2^*$.}
\label{tab:params}
\end{table}

\section{Theoretical model}
\label{sec-SM:model}

In this section, we present the theoretical framework used to describe the experimental setup: two distant flux-tunable transmon qubits coupled at two spatially separated points to a common waveguide in a giant-atom configuration, as depicted in Fig.~\ref{fig:SM-Setup-Driven-Waveguide}a. 
For completeness, and in order to provide a self-contained theoretical description, we summarize the master equation obtained within the SLH formalism~\cite{CombesSLHFramework2017}. The corresponding SLH network representation is shown in Fig.~\ref{fig:SM-Setup-Driven-Waveguide}b, while the full derivation can be found in Refs.~\cite{Kockum2018,Soro2022}.  
In Table~\ref{tab:parameters}, we summarize the definitions of the main parameters used in this work.

\begin{table}[h!]
	\centering
\begin{tabular}{lll}
		\hline
		\hline
		\textbf{Symbol} & \textbf{Meaning} & \textbf{Equation} \\
		\hline
		$\bar \omega$ & Average frequency of the giant atoms & $(\omega_1+\omega_2)/2$ \\
		$\delta$ & Energy detuning between giant atoms & $(\omega_1-\omega_2)/2$ \\
        $\Lambda$ & Anharmoncity & $\omega_{ge} - \omega_{ef}>0$ \\
		$\omega_\mathrm{d}$ & Drive frequency & - \\
		$\Delta$ & Average detuning between the giant atoms and drive  & $\bar \omega - \omega_\text{d}$ \\
        $\Delta x$ & Distance between two coupling points & - \\
        $\Delta \bar x$ & Distance between adjacent coupling points of different emitters& - \\
        $\phi(\omega)$ & Frequency-dependent phase between two coupling points of the $i$th emitter  & $\omega \Delta x/v$ \\
        $\theta(\omega)$ & Frequency-dependent phase between adjacent coupling points of different emitters & $\omega \Delta \bar x/v$ \\
		$\gamma(\omega_i)$ & Coupling strength at each connection point & $(\omega_i/\omega_0)^2\gamma_0$ \\
        $\Gamma_i(\omega_i)$ & Total waveguide-induced decay rate & $2\gamma(\omega_i)[1+\cos \phi(\omega_i)]$ \\
        $\Gamma_{\text{Q}i}(\omega_i)$ & Total relaxation rate & $\Gamma_i(\omega_i)+\gamma_{\text{nr}}$ \\
        $g$ & Collective coupling & Eq.~\eqref{eq:SM-collective-coupling} \\
        $\Gamma_{\text{coll}}$ & Collective decay rate & Eq.~\eqref{eq:SM-collective-dissipator} \\
        $\delta\omega_i$ & Giant-atom induced Lamb shift & $\gamma(\omega_i)\sin \phi(\omega_i)$ \\
        $\omega_{\text{super/sub}}$ & Individual super-/subradiant frequency  & Eq.~\eqref{eq-SM:individual-sup-sub} \\
		$|\beta|^2$ & Photon flux (where $P$ is the drive power) &  $P/\omega_\text{d}$\\ 
		$\Omega$ & Rabi frequency of the drive (under triplet dark state conditions) &  $2\sqrt{2\gamma}\beta$\\ 
        $\alpha$ & Control parameter  &  $i\Omega/\sqrt{2}\delta$\\
        $|\Psi^\pm\rangle$ & Triplet/singlet state  &  $1/\sqrt{2}(|ge\rangle\pm|eg\rangle)$\\ 
        $|D^\pm\rangle$ & Triplet/singlet dark state  &  $1/\sqrt{1+|\alpha|^2}(\alpha|\Psi^\pm\rangle+|gg\rangle)$\\ 
        $\hat \rho$ & Density matrix describing both giant atoms as transmon qubits &  Eq.~\eqref{eq-SM:Full-masterEq}\\ 
        $\hat \rho_\text{eff}$ & Density matrix describing both giant atoms as effective two-level systems  &  Eq.~\eqref{eq-SM:Effective-masterEq}\\ 
        $\hat \rho_\text{TLS}$ & Density matrix describing both giant atoms as ideal two-level systems  &  Eq.~\eqref{eq-SM:GA-masterEq}\\ 
		\hline
        \hline
        \end{tabular}
	\caption{\textbf{Definitions of the notation and main parameters used in this work.}}
	\label{tab:parameters}
\end{table}

\subsection{Master equation for two distant giant atoms with two coupling points}

\begin{figure*}[t!]
    \centering
\includegraphics[width=\textwidth]{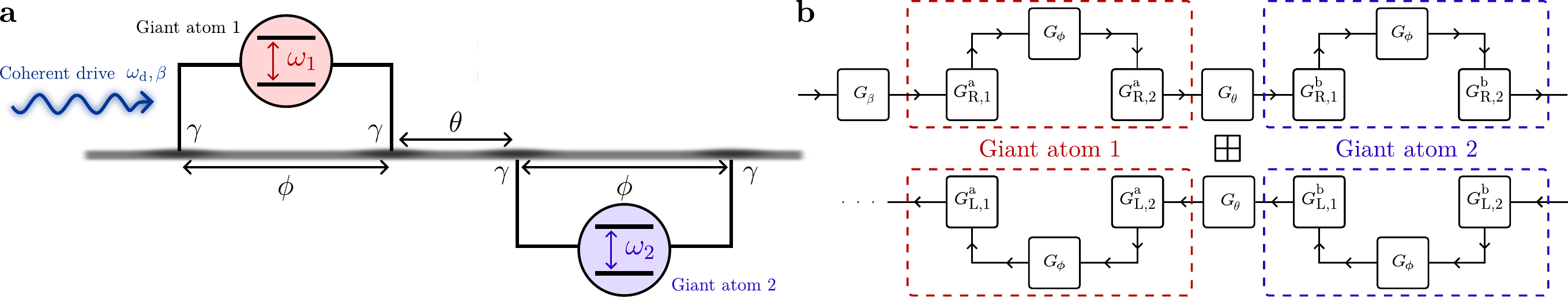}
    \caption{\textbf{Schematic representation of the device. }
\textbf{a)} Illustration of the system: two distant giant atoms, each coupled at two spatially separated points to a common waveguide. The atoms are coherently driven from the left end of the waveguide with frequency $\omega_{\mathrm{d}}$ and amplitude $\beta$, giving rise to effective drive amplitudes $\Omega_1$ and $\Omega_2$, which depend on the corresponding propagation phases and coupling strengths of each emitter. The separation between coupling points defines the phase $\phi$, while the distance between atoms introduces an additional propagation phase $\theta$. 
\textbf{b)}
SLH representation of the setup. The blocks $G^{a}_{R/L,j}$ and $G^{b}_{R/L,j}$ describe the coupling of the first and second giant atoms, respectively, to right- and left-propagating modes at coupling point $j=1,2$. The blocks $G_\phi$ and $G_\theta$ account for the propagation phases within each giant atom and between the two giant atoms, respectively, while $G_\beta$ represents the incoming coherent drive. These elementary components are combined using the SLH rules~\cite{CombesSLHFramework2017}. A detailed derivation of the model within the SLH framework can be found in Refs.~\cite{Kockum2018,Soro2022}.
    }
    \label{fig:SM-Setup-Driven-Waveguide}
\end{figure*} 

\noindent
\textbf{System Hamiltonian.}
We model the $i$th transmon qubit as a weakly anharmonic oscillator truncated to the three lowest energy levels $\{|g\rangle_i,|e\rangle_i,|f\rangle_i\}$, with bosonic annihilation operator $\hat a_i$. The corresponding two-emitter Hilbert space is spanned by states of the form $\{|gg\rangle, |ge\rangle, |gf\rangle, |eg\rangle, |fg\rangle, |ee\rangle, \ldots\}$, where $|gg\rangle \equiv |g\rangle_1 \otimes |g\rangle_2$, and similarly for the other states.
The bare $\ket{g} \leftrightarrow \ket{e}$ transition frequencies of the emitters are given by $\omega_1=\bar \omega+\delta$ and $\omega_2=\bar \omega-\delta$, where $\bar \omega\equiv (\omega_1+\omega_2)/2$ is the average frequency and $2\delta\equiv (\omega_1-\omega_2)$ is the detuning between the emitters. The anharmonicity $\Lambda_i$ leads to level spacings $\Delta E_i=\{\omega_i,\,\omega_i-\Lambda_i\}$, and we assume approximately identical nonlinearities, $\Lambda_1\approx \Lambda_2\approx \Lambda$.
Both emitters are driven from the left end of the waveguide by a coherent microwave field of frequency $\omega_\mathrm{d}$ and amplitude $\beta$, giving rise to effective drive amplitudes $\Omega_1$ and $\Omega_2$.
In the rotating frame of the drive, setting $\hbar=1$ henceforth, and within the rotating-wave approximation, the Hamiltonian of the two driven giant atoms reads
\begin{equation}
    \hat H=  \mleft[(\Delta+\delta) \hat a_1^\dagger \hat a_1 -\frac{\Lambda }{2} \hat a_1^{\dagger2} \hat a_1^2\mright]+\mleft[(\Delta-\delta) \hat a_2^\dagger \hat a_2    -\frac{\Lambda }{2} \hat a_2^{\dagger2} \hat a_2^2  \mright] 
    -\frac{i}{2}\mleft(\Omega_1\hat a_1^\dagger +\Omega_2\hat a_2^\dagger  - \text{H.c.}\mright)+g(\hat a_1^\dagger \hat a_2+\text{H.c.}),
    \label{eq-SM:Full-Hamiltonian}
\end{equation}
where $\Delta\equiv \bar \omega-\omega_\mathrm{d}$ denotes the qubit–drive detuning and $\text{H.c.}$ stands for Hermitian conjugate. The last term describes coherent interactions between the emitters mediated by the waveguide.
In the experiment, the drive frequency is chosen as $\omega_{\mathrm{d}}=\bar \omega$, such that $\Delta=0$. In this rotating frame, the emitter energies are therefore determined by the qubit-qubit detuning $2\delta$ and the anharmonicity $\Lambda$.
\newline

\noindent
\textbf{Dissipative dynamics and giant-atom geometry.}
Both transmon qubits interact with a common waveguide, giving rise to dissipative processes through spontaneous emission at each coupling point. In the giant-atom configuration depicted in Fig.~\ref{fig:SM-Setup-Driven-Waveguide}a, each transmon interacts with the waveguide at two spatially separated points. 
This geometry gives rise to interference effects characterized by local, frequency-dependent phases
\begin{equation}
    \phi(\omega_i)=\omega_i \Delta x_i / v,
    \label{eq-SM:local-phase}
\end{equation}
associated with each emitter, as well as a propagation phase 
\begin{equation}
    \theta(\omega)=\omega \Delta \bar x/v,
    \label{eq-SM:propagating-phase}
\end{equation}
accumulated by the field between adjacent coupling points of the two emitters. 
Here, $v$ denotes the phase velocity of the waveguide modes, $\Delta x_i$ is the separation between the two coupling points of the $i$th emitter, and $\Delta \bar x$ is the distance between adjacent coupling points of different emitters.
Strictly speaking, the propagation phase $\theta(\omega)$ depends on the frequency of the propagating mode and should be evaluated at the relevant atomic transition frequencies. 
However, when the phase variation over the qubit-qubit detuning is negligible, $\theta(\omega_2)-\theta(\omega_1)=2\delta\Delta\bar x/v\ll 1$, the propagation phase can be treated as approximately constant over this frequency range, such that $\theta(\omega_1)\approx\theta(\omega_2)$.

In the Markovian regime, the dissipative dynamics of the reduced density matrix of the two distant giant atoms is described by a Lindblad master equation~\cite{BreuerTheoryOpen2007,RivasOpenQuantum2012} given by~\cite{Kockum2018}:
\begin{equation}
    \frac{d \hat \rho}{dt}=-i[\hat H,\hat \rho]+\frac{\Gamma_{\mathrm{Q}1}(\omega_1)}{2}\mathcal{D}[\hat a_1]\hat \rho +\frac{\Gamma_{\mathrm{Q}2}(\omega_2)}{2}\mathcal{D}[\hat a_2]\hat \rho + \frac{\Gamma_{\text{coll}}}{2}(\mathcal{D}[\hat a_1,\hat a_2]\hat \rho+\mathcal{D}[\hat a_2,\hat a_1]\hat \rho) +\frac{\gamma_\phi}{4}(\mathcal{D}[2\hat a_1^\dagger\hat a_1]\hat \rho+\mathcal{D}[2\hat a_2^\dagger\hat a_2]\hat \rho),
 \label{eq-SM:Full-masterEq}   
\end{equation}
where $\mathcal{D}[\hat X](\cdot)\equiv 2 \hat X (\cdot) \hat X^\dagger-\{\hat X^\dagger \hat X,(\cdot)\}$ and $\mathcal{D}[\hat A,\hat B](\cdot)\equiv 2 \hat A (\cdot) \hat B^\dagger-\{\hat B^\dagger \hat A,(\cdot)\}$ denote the individual and collective Lindblad superoperators, respectively. 
Each emitter experiences individual decay processes due to its coupling to the waveguide, with rates given by~\cite{Kockum2018}
\begin{equation}
    \Gamma_i(\omega_i)\equiv 2\gamma(\omega_i)[1+\cos \phi(\omega_i)],
    \label{eq:giant_atom_dissipation-wo-non-radiative}
\end{equation}
%
where $\gamma(\omega_i)\equiv (\omega_i/\omega_0)^2 \gamma_0$ denotes the coupling strength at each connection point, defined with respect to a reference frequency $\omega_0$ and rate $\gamma_0$. The quadratic frequency dependence of the decay rate is derived from the circuit model of a qubit capacitively coupled to the waveguide~\cite{Kannan2020, Almanakly2025thesis}. 
In addition, each emitter is subject to non-radiative decay processes, including relaxation ($\gamma_{\mathrm{nr}}$) and pure dephasing ($\gamma_\phi$). We define the total relaxation rate of each emitter as
\begin{equation}
    \Gamma_{\mathrm{Q}i}(\omega_i)\equiv \Gamma_i(\omega_i)+\gamma_{\mathrm{nr}}=2\gamma(\omega_i)[1+\cos \phi(\omega_i)]+\gamma_{\mathrm{nr}}.
    \label{eq:giant_atom_dissipation}
\end{equation}
We note that incoherent excitation by thermal photons can be neglected in superconducting-circuit platforms operating at cryogenic temperatures~\cite{GarciaRipollQuantumInformation2022}.
In addition, the interference between the emission from the two coupling points of each giant atom gives rise to a frequency-dependent contribution to the Lamb shift~\cite{Kockum2018},
\begin{equation}
    \delta \omega_i=\gamma(\omega_i)\sin \phi(\omega_i),
    \label{eq:SM-Lamb-shift}
\end{equation}
as well as effective driving amplitudes given by
\begin{equation}
    \Omega_1=\sqrt{2\gamma(\omega_1)}[1+e^{i\phi(\omega_1)}]\beta, \quad \text{and}\quad \Omega_2=\sqrt{2\gamma(\omega_2)}[1+e^{i\phi(\omega_2)}]e^{i[\phi(\omega_1)+\theta(\omega)]}\beta.
    \label{eq:SM-Driving-amplitudes}
\end{equation}
Both the Lamb shifts and the driving amplitudes depend explicitly on the giant-atom geometry (see Fig.~\ref{fig:SM-Setup-Driven-Waveguide}a) through the local phases $\phi(\omega_i)$ and the phase $\theta(\omega)$ acquired by the field during propagation between the two emitters.
Furthermore, the waveguide mediates both coherent and dissipative interactions between the emitters. These are quantified by the coherent coupling $g$ and the collective decay rate $\Gamma_{\text{coll}}$, given by~\cite{Kockum2018}
\begin{subequations}
    \begin{align}
    g&=\frac{1}{2}\sqrt{\gamma(\omega_1)\gamma(\omega_2)}\mleft\{ \sin[\phi(\omega_1)+\phi(\omega_2)+\theta(\omega)]+\sin[\phi(\omega_2)+\theta(\omega)]+\sin[\phi(\omega_1)+\theta(\omega)]+\sin[\theta(\omega)]\mright\}, 
    \label{eq:SM-collective-coupling}
    \\
 \Gamma_{\text{coll}}&=\frac{1}{2}\sqrt{\gamma(\omega_1)\gamma(\omega_2)}\mleft\{ 
\cos[\phi(\omega_1)+\phi(\omega_2)+\theta(\omega)]+\cos[\phi(\omega_2)+\theta(\omega)]+\cos[\phi(\omega_1)+\theta(\omega)]+\cos[\theta(\omega)]
    \mright\}.
    \label{eq:SM-collective-dissipator}
    \end{align}
\end{subequations}
From these two expressions, it follows that the coherent and dissipative couplings are intrinsically determined by the interference phases. In particular, tuning the phases to suppress the exchange interaction ($g=0$) does not eliminate either the individual or collective decay channels. Conversely, setting the collective decay rate to zero ($\Gamma_{\text{coll}}=0$) does not suppress the exchange interaction or the individual decay processes. Only in the limiting case where the individual decay rates vanish can both $g$ and $\Gamma_{\text{coll}}$ be simultaneously suppressed. 
Therefore, aside from the trivial situation in which all waveguide-mediated interactions are switched off, the two-emitter giant-atom system necessarily exhibits collective effects arising from the coupling of both emitters to the common waveguide environment~\cite{Kockum2018}.

\subsection{Individual superradiance and subradiance} 
As shown in Eq.~\eqref{eq:giant_atom_dissipation}, the total decay rates experienced by the quantum emitters $\Gamma_{\mathrm{Q}i}(\omega_i)$ depend explicitly on the geometry of the giant-atom configuration through the local accumulated phases $\phi(\omega_i)$.
Consequently, depending on the value of these phases, the effective decay rates can be either suppressed or enhanced. 
This interference-induced modification of the spontaneous emission rates constitutes one of the defining hallmarks of giant-atom architectures~\cite{FriskKockum2014,Kockum2021review}.
In principle, the emission properties can be engineered by modifying the emitter frequencies, the number of coupling points, and the distances between them. However, in a given experimental device, the number and spatial arrangement of the coupling points are fixed by the fabricated design, such that only the emitter frequencies can be tuned \textit{in situ}.
Under these conditions, one can define the frequencies at which the interference between coupling two points becomes perfectly constructive or destructive:
\begin{subequations}
\label{eq-SM:individual-sup-sub}
    \begin{align}
        \omega_{\text{super}}:&\quad \phi(\omega_\text{super})=\frac{\omega_\text{super}\Delta x}{v}=2\pi m, \quad \hspace{0.3cm}  \Rightarrow\quad \Gamma_{\text{Q}i}(\omega_{\text{super}})=4\gamma(\omega_{\text{super}})+\gamma_{\text{nr}}, 
        \label{eq-SM:individual-superradiance}
        \\
        \omega_{\text{sub}}:&\quad \phi(\omega_\text{sub})=\frac{\omega_\text{sub}\Delta x}{v}=(2n+1)\pi,   \quad  \Rightarrow\quad \Gamma_{\text{Q}i}(\omega_{\text{sub}})=\gamma_{\text{nr}} ,
        \label{eq-SM:individual-subradiance}
    \end{align}
\end{subequations}
with $m, n\in\mathbb{N}$. Here, $\omega_{\text{super}}$ and $\omega_{\text{sub}}$ define the \textit{individual superradiant} and \textit{individual subradiant} frequencies, respectively, at which the waveguide-mediated decay is maximally enhanced or suppressed due to interference between the coupling points.
Within this framework, perfect constructive interference corresponds to individual superradiance, whereas perfect destructive interference corresponds to individual subradiance.
Importantly, the terminology \textit{individual superradiance/subradiance} is inspired by the well-known collective superradiant and subradiant effects---also known as Dicke super- and subradiance---emerging in many-body quantum systems~\cite{Dicke1954,GrossSuperradianceEssay1982}. In the present case, however, the enhancement and suppression of spontaneous emission originate from single-emitter interference effects associated with the giant-atom geometry, rather than from cooperative interactions between multiple emitters.
Therefore, this distinction allows us to clearly differentiate between individual and collective superradiance/subradiance, depending on whether the modification of the emission properties originates from the giant-atom geometry or from cooperative many-body effects.%
\newline

\begin{figure*}[b!]
    \centering
\includegraphics[width=1.0\textwidth]{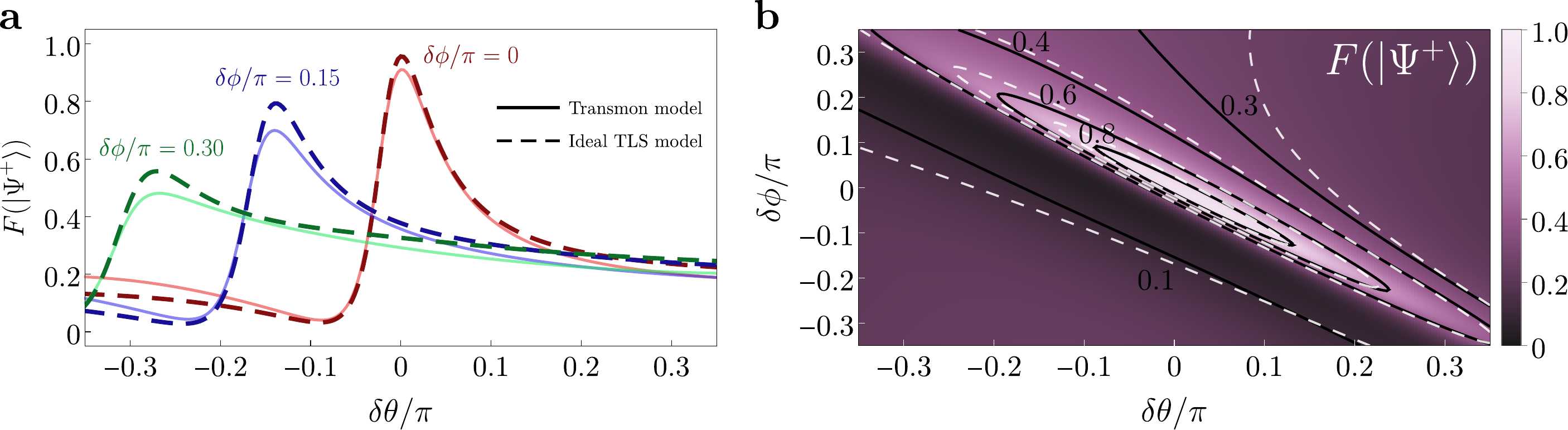}
    \caption{
\textbf{Triplet-state fidelity under phase mismatches.}
\textbf{a)} Fidelity $F(|\Psi^+\rangle)$ as a function of the propagation-phase mismatch $\delta\theta$ for several values of the internal-phase mismatch $\delta\phi/\pi=\{0,0.15,0.30\}$. Solid lines correspond to numerical simulations of the full transmon model [Eq.~\eqref{eq-SM:Full-masterEq}], while dashed lines show the effective TLS model [Eq.~\eqref{eq-SM:Effective-masterEq}]. The fidelity is maximized at the ideal superradiant condition, $\delta\theta=\delta\phi=0$.
\textbf{b)} Fidelity $F(|\Psi^+\rangle)$ as a function of both phase mismatches $(\delta\theta,\delta\phi)$ obtained from the full transmon model. Black contours denote constant-fidelity lines from numerical simulations of the full master equation [Eq.~\eqref{eq-SM:Full-masterEq}], while gray dot-dashed contours correspond to the effective TLS model [Eq.~\eqref{eq-SM:Effective-masterEq}]. The high-fidelity region is concentrated along a narrow diagonal band in phase space, revealing the sensitivity of the protocol to deviations from the ideal dark-state conditions.
Parameters: $\{\Omega,\delta,\gamma\}/2\pi=\{45,6.81,14.7\}\,\mathrm{MHz}$.
 }
    \label{fig:SM-PhaseMismatch-AVV}
\end{figure*}

\noindent
\textbf{Phase mismatches.}
In the experiment, both giant atoms are operated at their superradiant frequencies [see Eq.~\eqref{eq-SM:individual-superradiance}], where the driven-dissipative entanglement-generation protocol is optimal. To assess the robustness of the scheme against calibration errors and frequency offsets, we consider deviations from the ideal phases,
\begin{equation}
    \delta\phi \equiv\phi(\omega)-\phi(\omega_{\text{super}}),\quad \text{and} \quad \delta\theta \equiv\theta(\omega)-\theta(\omega_{\text{super}}),
\end{equation}
and evaluate the fidelity~\cite{JozsaFidelityMixed1994} between the steady state and the target triplet Bell state,
\begin{equation}
    F(|\Psi^+\rangle)\equiv \langle \Psi^+| \hat \rho |\Psi^+\rangle.
\end{equation}
As shown in Fig.~\ref{fig:SM-PhaseMismatch-AVV}, the fidelity remains high for small phase deviations around the ideal superradiant conditions, $\delta\theta\approx\delta\phi \approx0$.
This high-fidelity region lies along a narrow diagonal band in the $(\delta\theta,\delta\phi)$ plane, revealing that opposite deviations of the internal and propagation phases can partially compensate each other and preserve the interference conditions required for triplet-state stabilization (panel b). Away from this region, once $|\delta\phi/\pi|$ or $|\delta \theta/\pi|$ exceed $5\  \%$, the fidelity decreases rapidly due to the breakdown of the triplet dark-state conditions, which introduces geometry-induced Lamb shifts and coherent interactions [see Eqs.~\eqref{eq:SM-Lamb-shift} and~\eqref{eq:SM-collective-coupling}] and thereby suppresses the dissipative mechanism responsible for stabilizing $|\Psi^+\rangle$.
In the experiment, however, the uncertainty in the calibrated qubit frequencies corresponds to phase deviations much smaller than the tolerance window shown in Fig.~\ref{fig:SM-PhaseMismatch-AVV}, ensuring operation close to the optimal superradiant point and enabling the observed triplet-state fidelity of $0.89\pm0.02$.%
\newline

\noindent
\textbf{Circuit-induced shift of the superradiant frequency.}
Due to the quadratic frequency dependence of the decay rate, $\gamma(\omega_i)=(\omega_i/\omega_0)^2\gamma_0$, arising from the circuit model of a qubit capacitively coupled to the waveguide~\cite{Kannan2020,Almanakly2025thesis}, the superradiant frequency is slightly shifted with respect to its ideal theoretical value based on the giant-atom model.
As a consequence, the maximum individual dissipation, $\Gamma_{\mathrm{max}}(\omega_{\mathrm{super}}')=4\gamma(\omega_{\mathrm{super}}')$, does not occur exactly at
$\omega_{\mathrm{super}}$, but rather at a shifted frequency $\omega_{\mathrm{super}}'=\omega_{\mathrm{super}}+\delta\omega$, with $\delta\omega\ll\omega_{\mathrm{super}}$. 
This effect can be observed in Fig.~\ref{fig:fig3}a, where the ideal superradiant frequency is $\omega_{\mathrm{super}}/2\pi=\SI{6.208}{\giga\hertz}$, while the theoretically optimal dissipation occurs at
$\omega_{\mathrm{super}}'/2\pi=\SI{6.36}{\giga\hertz}$.
We attribute this deviation to details of the circuit design. Nevertheless, it does not qualitatively affect the underlying physics or the conclusions of the present work.

\subsection{Geometric enhancement of the dissipation rate}
In the experiment, we observe deviations of the measured total dissipation rate $\Gamma_{\mathrm{Q}i}(\omega_i)$ from the giant-atom prediction in Eq.~\eqref{eq:giant_atom_dissipation} when the qubit frequencies are tuned near the superradiant frequency $\omega_{\mathrm{super}}$, as illustrated in Fig.~\ref{fig:fig3}. 
This effect likely originates from impedance mismatches due to the geometry of the on-chip waveguide, as similar features appear in the measured dissipation spectra of both qubits. Because the spacing between coupling points corresponds to $\lambda$ or $\lambda/2$ near the superradiant frequency, we hypothesize that impedance mismatches at the site of the other, decoupled qubit create a low-Q mode that enhances dissipation.
We model this effect phenomenologically by introducing two independent low-$Q$ background modes, each coupled to one of the emitters. The reduced density matrix of the combined system, $\hat\rho'$, evolves according to
\begin{equation}
    \frac{d\hat \rho'}{dt}= \mleft(\mathcal{\hat L}_0+\mathcal{\hat L}_{\text{modes}}+ \mathcal{\hat L}_{\text{modes,int}} \mright) \hat  \rho',
    \label{eq:SM-Nakajima-Zwanzig}
\end{equation}
where $\hat{\mathcal L}_0$ describes the giant-atom dynamics introduced in Eq.~\eqref{eq-SM:Full-masterEq}. The Liouvillian associated with the low-$Q$ modes is
\begin{equation}
     \mathcal{\hat L}_{\text{modes}} \hat \rho'=\sum_{j=1}^2 \mleft( -i[\omega_{\text{super}} \hat c_j^\dagger \hat c_j,\hat  \rho']+\frac{\kappa}{2}\mathcal{D}[\hat c_j]\hat  \rho' \mright),
\end{equation}
where $\hat c_i$ is the annihilation operator of the $i$th background mode, whose resonance frequencies are centered around the superradiant frequency, and $\kappa$ denotes their decay rates. Finally, $\mathcal{\hat L}_{\text{modes,int}}$ describes the coherent interaction between the emitters and the background modes, 
\begin{equation}
    \mathcal{\hat L}_{\text{modes,int}}\hat  \rho'=-i\sum_{i=1}^2[\eta(\hat c_i^\dagger \hat a_i+\hat a_i^\dagger\hat c_i ),\hat  \rho'],
\end{equation}
where $\eta$ is the emitter-mode coupling strength.
In the regime where the modes are able to resolve the atomic transition frequency even in a lossy regime (i.e., $\kappa \gg \eta$ and $\kappa < \omega_i$), the low-$Q$ modes can be adiabatically eliminated using the Nakajima–Zwanzig formalism~\cite{NakajimaQuantumTheory1958,ZwanzigEnsembleMethod1960,Gonzalez-BallesteroTutorialProjector2024}. Following a derivation similar to those in Refs.~\cite{Vivas-VianaDissipativeStabilization2024,Vivas-VianaNonclassicalDrivenDissipative2025}, we obtain the following effective master equation for the reduced dynamics of the giant atoms:
$
   d\hat \rho/dt= \left(\mathcal{\hat L}_0+\delta\mathcal{\hat L}_{\text{modes}} \right)\hat \rho,
$
where $\delta\mathcal{\hat L}_{\text{modes}}$ represents the effective Bloch--Redfield contribution arising from the adiabatic elimination of the low-$Q$ modes:
\begin{equation}
   \delta \mathcal{\hat L}_{\text{modes}} \hat \rho\equiv 
    \sum_{i=1}^2 \mleft( \frac{\eta^2}{\kappa/2+i(\omega_{\text{super}}-\omega_i)}[\hat a_i\hat \rho,\hat a_i^\dagger]+\text{H.c.} \mright)=
    \sum_{i=1}^2 \mleft(
    - 
    i[\delta_{\text{modes}}^{(i)}\hat a_i^\dagger \hat a_i, \hat \rho]
    +\frac{\Gamma_{\text{modes}}^{(i)}}{2}\mathcal{D}[\hat a_i]\hat \rho \mright),
\end{equation}
where the effective energy shifts  and decay rates are given by
\begin{equation}
    \delta_{\text{modes}}^{(i)}\equiv \frac{\eta^2 (\omega_{\text{super}}-\omega_i) }{\kappa^2/4+(\omega_{\text{super}}-\omega_i)^2} \quad  \text{and}
    \quad  
    \Gamma_{\text{modes}}^{(i)}\equiv \frac{\eta^2 \kappa/2 }{\kappa^2/4+(\omega_{\text{super}}-\omega_i)^2}.
\end{equation}
Retaining only the dissipative contribution---the corresponding energy shifts are small in the vicinity of the superradiant frequency---we find that the effective low-$Q$ background modes introduce an additional Lorentzian decay channel. This allows us to model the enhancement of the individual dissipation rates of the qubits through the relation
\begin{equation}
    \Gamma_{\mathrm{Q}i}'(\omega_i) \equiv \Gamma_{\mathrm{Q}i}(\omega_i) + \Gamma_{\text{modes}}^{(i)}=2\gamma(\omega_i)[1+\cos \phi(\omega_i)]+ \frac{\eta^2 \kappa/2 }{\kappa^2/4+(\omega_{\text{super}}-\omega_i)^2}+\gamma_{\mathrm{nr}}.
    \label{eq:SM-Effective-decay-rate}
\end{equation}
Fitting the experimental data in Fig.~\ref{fig:fig3}a near the superradiant frequency, we extract the low-$Q$ mode parameters $\{\eta,\kappa\}/2\pi=\{\SI{27}{\kilo\hertz},\, \SI{350}{\mega\hertz}\}$.
We emphasize that this effective description, based on an additional Lorentzian decay channel, is only intended to capture the mismatch in the vicinity of the superradiant frequency and is therefore not expected to accurately describe other spectral regions.
As such, this model should be viewed as a phenomenological description rather than a microscopic circuit-QED explanation. Nevertheless, it provides a satisfactory quantitative description of the dynamics relevant for the generation and stabilization of entanglement between the giant atoms.   
\newline

\noindent
\textbf{Parameter extraction.}
Throughout this work, Eq.~\eqref{eq-SM:Full-masterEq} is used to perform numerical simulations and to fit the experimental data [see Sec.~\ref{subsec:qutrit_power_scan} and Sec.~\ref{subsec:collective_spec}]. 
The individual decay rates are extracted by fitting the transmission spectrum to the effective Lorentzian model introduced in Eq.~\eqref{eq:SM-Effective-decay-rate} as a function of the qubit frequency. 
The full set of dissipation parameters is listed in Table~\ref{tab:params}.

\subsection{Two-level-system approximation}
\label{sec-SM:AE-transmon}

The master equation in Eq.~\eqref{eq-SM:Full-masterEq} describes the giant atoms as transmon qubits, i.e., weakly anharmonic oscillators, rather than ideal two-level systems.
While the inclusion of the third level $|f\rangle$ is required for a quantitatively accurate description of the experiment, it takes a subdominant role in the dynamics because the anharmonicity is large compared to the drive strengths.
%
%
This separation of energy scales enables the derivation of an effective two-level description by adiabatically eliminating the $|f\rangle$ state, leading to a reduced model that is more transparent from a physical perspective.
\newline

\begin{figure*}[b!]
    \centering
\includegraphics[width=1.0\textwidth]{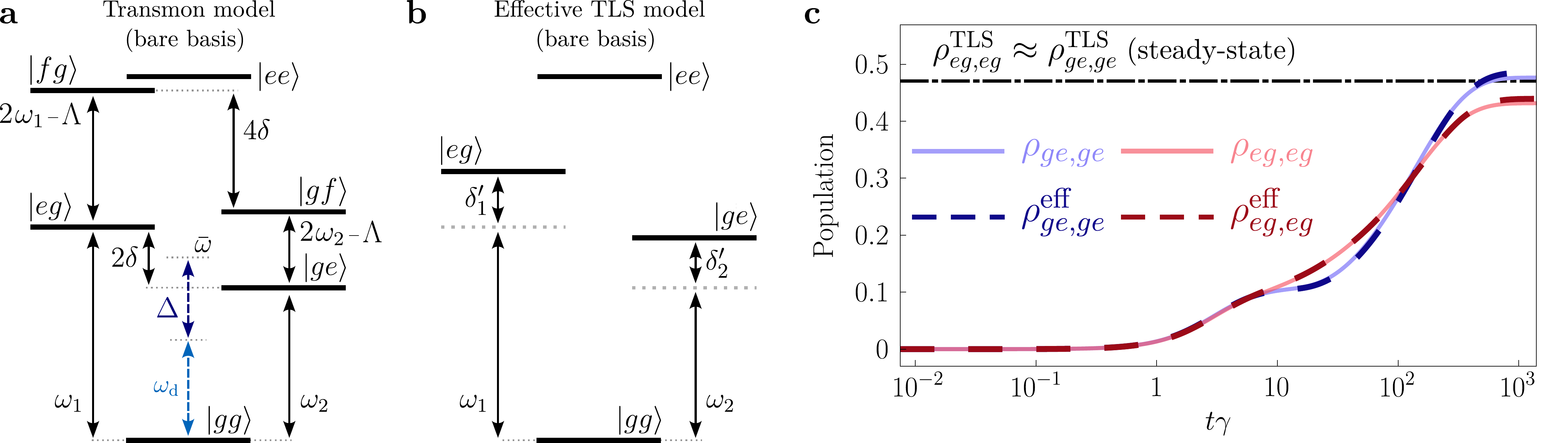}
    \caption{
\textbf{Effective two-level description of the transmon model.} 
\textbf{a)}
Energy-level structure of the full transmon model in the bare basis.
The system is coherently driven at frequency $\omega_d$, detuned by $\Delta=\bar{\omega}-\omega_d$ from the average transition frequency $\bar{\omega}=(\omega_1+\omega_2)/2$.
\textbf{b)} 
Effective energy diagram obtained after adiabatically eliminating the third transmon level and restricting the dynamics to the two-level subspace.
Virtual transitions involving the eliminated states induce effective energy corrections $\delta_1'$ and $\delta_2'$ within the single-excitation manifold. The diagrams are not drawn to scale and the energy splittings are exaggerated for visualization purposes.
\textbf{c)}
Time evolution of the single-excitation populations, $\rho_{ge,ge}(t)$ (blue) and $\rho_{eg,eg}(t)$ (red), under the triplet dark-state conditions given in Eq.~\eqref{eq-SM:dark-state-conditions-driven-triplet}.
Solid lines correspond to numerical simulations using the transmon model in Eq.~\eqref{eq-SM:Full-masterEq}, while dashed lines are obtained from the effective two-level system model in Eq.~\eqref{eq-SM:Effective-masterEq}. 
The horizontal dot-dashed line indicates the stationary-state population predicted by the ideal two-level model in Eq.~\eqref{eq-SM:GA-masterEq}.
These results were obtained using the parameters outlined in Table~\ref{tab:params}. For the simplified case, we consider the averaged parameters.
}
    \label{fig:SM-Model-Comparison}
\end{figure*}

\noindent
\textbf{Adiabatic elimination of the $|f\rangle$ state.}
The anharmonicity of the transmon qubit is defined as the difference between the lowest two energy transition frequencies $\Lambda = \omega_{ge} - \omega_{ef}>0$.
As a consequence, transitions between states within the two-level-system subspace, $\{|gg\rangle,|ge\rangle, |eg\rangle, |ee\rangle\}$, are detuned from those involving the third level $|f\rangle$.
Therefore, the subspace containing the states $\{|fg\rangle,|gf\rangle,\ldots\}$ interacts only weakly with the effective two-level subspace, with the anharmonicity providing the corresponding detuning, as illustrated in Fig.~\ref{fig:SM-Model-Comparison}a.

\begin{figure*}[t!]
    \centering
\includegraphics[width=1.0\textwidth]{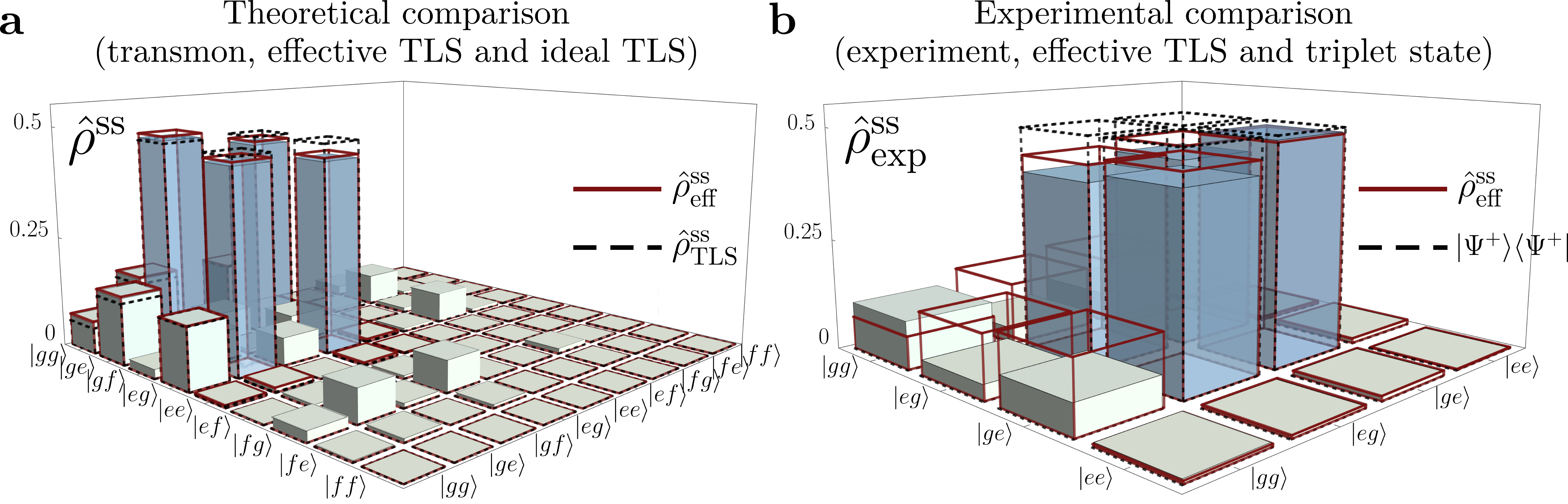}
\caption{
\textbf{Absolute value of the steady-state density matrix.} 
\textbf{a)}
Comparison between the steady-state density matrix $\hat\rho_{\mathrm{ss}}$ obtained from the full transmon model [Eq.~\eqref{eq-SM:Full-masterEq}] and the effective two-level model [Eq.~\eqref{eq-SM:Effective-masterEq}].
The colored bars correspond to the full transmon simulation, the red wireframe denotes the effective two-level prediction, and the dashed black wireframe corresponds to the simplified model in Eq.~\eqref{eq-SM:GA-masterEq}.
\textbf{b)}
Comparison between the experimentally reconstructed steady-state density matrix, $\hat\rho_{\mathrm{ss}}^{\mathrm{exp}}$, and the effective two-level prediction (red wireframe). The dashed black wireframe denotes the ideal triplet state.
}
    \label{fig:SM-Model-Comparison-AbsValues}
\end{figure*}

In the anharmonic regime (i.e., $|\Lambda| \gg |\Omega_1|,|\Omega_2|$), states involving $|f\rangle$ mainly mediate virtual processes within the two-level subspace, while their populations and coherences remain strongly suppressed.
Numerical simulations confirm that the total population of states involving the third level remains close to $\sim \SI{2}{\percent}$ given the optimal conditions for entanglement generation ($|\Omega_1|/2\pi = \SI{45}{MHz}$), as shown in Fig.~\ref{fig:SM-Model-Comparison-AbsValues}. For $|\Omega_1|/2\pi \gtrsim \SI{75}{MHz}$, this population exceeds $\sim\SI{5}{\percent}$, signaling the onset of deviations from the approximation that neglects the third-level population.
The higher excited states induce residual corrections to the system dynamics: in particular, they generate an asymmetry between the populations of $|ge\rangle$ and $|eg\rangle$, as observed experimentally in Fig.~\ref{fig:fig5}b.
This effect, which is absent in an ideal two-level description, reflects the influence of virtual transitions involving the $|f\rangle$ state.
Nevertheless, the dynamics can be accurately captured within an effective two-level description obtained by adiabatically eliminating the $|f\rangle$ state and restricting the dynamics to the two-level subspace.

For simplicity, in the following derivations, we employ the original notation introduced in Ref.~\cite{Kockum2018}, based on collective left- and right-propagating jump operators,%
\begin{equation}
    \hat L_\text{L/R}\equiv\sqrt{\Gamma_{1,\text{L/R}}}\hat a_1+\sqrt{\Gamma_{2,\text{L/R}}}\hat a_2.
\end{equation}
We note that this formulation is fully equivalent to the master equation presented in Eq.~\eqref{eq-SM:Full-masterEq}, which is recovered after expanding the Lindblad terms, $(\mathcal{D}[\hat L_\text{L}]+\mathcal{D}[\hat L_\text{R}])\hat \rho$.
Following Ref.~\cite{ReiterEffectiveOperator2012}, we partition the Hilbert space into two subspaces:

\begin{enumerate}[label=(\roman*)]
    \item 
        \textit{Relevant subspace}, $\mathcal{H}_0 \equiv \{|gg\rangle,|ge\rangle,|eg\rangle,|ee\rangle\}$, whose effective dynamics we aim to describe. The corresponding Hamiltonian is
        \begin{equation}
            \hat H_0\equiv(\Delta+\delta)|eg\rangle\langle eg|+(\Delta-\delta)|ge\rangle\langle ge|-i\frac{1}{2}(\Omega_1|ee\rangle\langle ge|+\Omega_1|eg\rangle\langle gg|-\text{H.c.})-i\frac{1}{2}(\Omega_2|ee\rangle\langle eg|+\Omega_2|ge\rangle\langle gg|-\text{H.c.}),
        \end{equation}
        with the directional collective jump operators
        \begin{equation}
        \hat L_{0,\text{L/R}}\equiv\sqrt{\Gamma_{1,\text{L/R}}}(|gg\rangle\langle eg|+|ge\rangle\langle ee|)+\sqrt{\Gamma_{2,\text{L/R}}}(|gg\rangle\langle ge|+|eg\rangle\langle ee|).
        \end{equation}
    \item 
    \textit{Virtual subspace}, 
    $\mathcal{H}_V\equiv \{|fg\rangle,|gf\rangle\}$, which is adiabatically eliminated. The dynamics are governed by the non-Hermitian Hamiltonian
    \begin{equation}
        \hat H_{V}=[2(\Delta+\delta)-\Lambda-i\Gamma_{1}(\omega_1)]|fg\rangle\langle fg|+[2(\Delta-\delta)-\Lambda-i\Gamma_{2}(\omega_2)]|gf\rangle\langle gf|.
    \end{equation}
    These are the lowest-energy states involving the $|f\rangle$ level that couple to the drive, so higher excited states can be safely neglected as long as $|\Lambda| \gg |\Omega_1|, |\Omega_2|$. 
    Note that here $\Gamma_i(\omega_i)$ refers to the waveguide-induced decay rate introduced in Eq.~\eqref{eq:giant_atom_dissipation-wo-non-radiative}, i.e., excluding the non-radiative contribution considered in Eq.~\eqref{eq:giant_atom_dissipation}. %
    \item The two subspaces are coupled both coherently and dissipatively via
    \begin{equation}
    \hat H_{\text{int}}=- \frac{i}{\sqrt{2}}\mleft(\Omega_1|fg\rangle\langle eg|+\Omega_2 |gf\rangle\langle ge|-\text{H.c.}\mright),
    \end{equation}
    and
    %
    %
        \begin{equation}
        \hat L_{\text{int},\text{L/R}}=\sqrt{2\Gamma_{1,\text{L/R}}}|eg\rangle\langle fg|+\sqrt{2\Gamma_{2,\text{L/R}}}|ge\rangle\langle gf|.
    \end{equation}
\end{enumerate}

Applying the effective operator formalism~\cite{ReiterEffectiveOperator2012}, the elimination of $\mathcal{H}_V$ yields coherent and dissipative corrections to the dynamics in $\mathcal{H}_0$, such that we obtain
$\hat H'\equiv\hat H_0+\delta\hat H $ and $\hat L_{\text{L/R}}'\equiv \hat L_{0,\text{L/R}}+\delta\hat L_{\text{L/R}}$, where:
\begin{subequations}
\label{eq:SM-anharmonic-corrections}
   \begin{align}
        \delta\hat H &\equiv-\frac{1}{2}\hat H_{\text{int}}^\dagger [\hat H_{V}^{-1}+(\hat H_{V}^{-1})^\dagger]\hat H_{{\text{int}}}  
        =\delta_1'|eg\rangle\langle eg|+\delta_2'|ge\rangle\langle ge|,\\
        \delta\hat L_\text{L/R}&\equiv-\hat L_{\text{int},\text{L/R}}\hat H_{V}^{-1}\hat H_{\text{int}}
        =
        \sqrt{\Gamma_{1,\text{L/R}}'}|eg\rangle\langle eg|+\sqrt{\Gamma_{2,\text{L/R}}'}|ge\rangle\langle ge|,
   \end{align}
\end{subequations}
where have defined
\begin{equation}
\delta_1'\equiv \frac{[-2(\Delta+\delta)+\Lambda]|\Omega_1|^2}{2\Gamma^2_{1}(\omega_1)+2[-2(\Delta+\delta)+\Lambda]^2} \quad \text{and}\quad \delta_2'\equiv \frac{[-2(\Delta-\delta)+\Lambda]|\Omega_2|^2}{2\Gamma^2_{2}(\omega_2)+2[-2(\Delta-\delta)+\Lambda]^2}, 
\end{equation}
and
\begin{equation}
    \sqrt{\Gamma_{1,\text{L/R}}'}\equiv \frac{i\Omega_1\sqrt{\Gamma_{1,\text{L/R}}}}{[2(\Delta+\delta)-\Lambda]-i\Gamma_{1}(\omega_1)} \quad \text{and}\quad  \sqrt{\Gamma_{2,\text{L/R}}'}\equiv\frac{i\Omega_2\sqrt{\Gamma_{2,\text{L/R}}}}{[2(\Delta-\delta)-\Lambda]-i\Gamma_{2}(\omega_2)}.
\end{equation}
We must note that an additional minus sign is included in the definition of $\delta\hat L_{\mathrm{L/R}}$ in order to recover the correct relative phase between $\hat L_{0,\mathrm{L/R}}$ and $\delta\hat L_{\mathrm{L/R}}$. 
This phase is not fixed unambiguously by the effective-operator prescription of Ref.~\cite{ReiterEffectiveOperator2012}. We determine it instead by comparing with a derivation based on the Nakajima--Zwanzig formalism~\cite{NakajimaQuantumTheory1958,ZwanzigEnsembleMethod1960,Gonzalez-BallesteroTutorialProjector2024}.
These corrections reveal two main effects. First, the single-excitation states acquire different energy shifts, which break the balance between $|ge\rangle$ and $|eg\rangle$ and thereby contribute to the population imbalance observed experimentally in Fig.~\ref{fig:fig5}b. Second, the corrections to the jump operators induce an effective state-selective pure-dephasing mechanism within the single-excitation manifold.
\newline

\noindent
\textbf{Effective master equation.}
We can now derive an effective master equation in terms of two-level quantum emitters, following the approach of Refs.~\cite{Kockum2018,Soro2022}. This equation is obtained by projecting the dynamics onto the two-level subspace and replacing the bosonic operators by Pauli operators, $\hat a_i \rightarrow \hat \sigma_i$ and $2\hat a_i^\dagger \hat a_i \rightarrow \hat \sigma_{z,i}$, where $\hat \sigma_i \equiv |g_i\rangle\langle e_i|$ and $\hat \sigma_{z,i} \equiv 2\hat \sigma_i^\dagger \hat \sigma_i - \mathbb{I}$.
After eliminating the $|f\rangle$ level, the resulting effective Hamiltonian, $\hat H_{\text{eff}}=\hat H_{0}+\delta \hat H$, is given by
\begin{subequations}
\label{eq-SM:Effective-Hamiltonian}
    \begin{align}
        \hat H_{0}&\equiv (\Delta+\delta)   \hat\sigma_1^\dagger\hat\sigma_1 
    +
    (\Delta-\delta) \hat\sigma_2^\dagger\hat\sigma_2
    -\frac{i}{2}(\Omega_1\hat\sigma_1^\dagger+\Omega_2\hat\sigma_2^\dagger-\text{H.c.}),
    \\
    \delta  \hat H&\equiv \delta'_1 \hat\sigma_1^\dagger\hat\sigma_1  (\mathbb{I}-\hat\sigma_2^\dagger\hat\sigma_2) +\delta'_2 \hat\sigma_2^\dagger\hat\sigma_2  (\mathbb{I}-\hat\sigma_1^\dagger\hat\sigma_1).
    \end{align}
\end{subequations}
Similarly, the effective directional jump operators, $\hat L_{L/R}^\text{eff}=\hat L_{0,\text{L/R}}+\delta \hat L_{\text{L/R}}$, are given by
\begin{subequations}
\label{eq-SM:Effective-Jump}
    \begin{align}
        \hat L_{0,\text{L/R}}&\equiv \sqrt{\Gamma_{1,\text{L/R}}}\hat \sigma_1+\sqrt{\Gamma_{2,\text{L/R}}}\hat \sigma_2,
    \\
   \delta \hat L_{\text{L/R}}&\equiv 
        \sqrt{\Gamma_{1,\text{L/R}}'}\hat\sigma_1^\dagger\hat\sigma_1  (\mathbb{I}-\hat\sigma_2^\dagger\hat\sigma_2)+\sqrt{\Gamma_{2,\text{L/R}}'}\hat\sigma_2^\dagger\hat\sigma_2  (\mathbb{I}-\hat\sigma_1^\dagger\hat\sigma_1),
    \end{align}
\end{subequations}
The terms proportional to $\delta_1'$ and $\delta_2'$ describe the effective AC Stark shifts induced by virtual transitions involving the $|f\rangle$ state (see Fig.~\ref{fig:SM-Model-Comparison}b), while those proportional to $\sqrt{\Gamma_{i,\text{L/R}}'}$ describe state-selective pure dephasing and correlated decoherence effects since the Lindblad term is of the form $ \sim \mathcal{D}[\hat L_{\text{L/R}}^{\text{eff}}]= \mathcal{D}[\hat L_{0,\text{L/R}}+\delta \hat L_{\text{L/R}}]$.
We have expressed the projectors onto the single-excitation states in terms of spin operators as $|eg\rangle\langle eg|
    =
    \hat\sigma_1^\dagger \hat\sigma_1
    (\mathbb{I}-\hat\sigma_2^\dagger \hat\sigma_2)$ and $|ge\rangle\langle ge|
    =
    \hat\sigma_2^\dagger \hat\sigma_2
    (\mathbb{I}-\hat\sigma_1^\dagger \hat\sigma_1)$.
The effective dynamics of the reduced density matrix of the two giant atoms is then described by
\begin{equation}
    \frac{d \hat \rho_{\text{eff}}}{dt}=(\mathcal{\hat L}_0 +\delta\mathcal{\hat L}) \hat \rho_{\text{eff}},
     \label{eq-SM:Effective-masterEq}   
\end{equation}
where $\mathcal{\hat L}_0$ is the Liouvillian superoperator analogous to that in Eq.~\eqref{eq-SM:Full-masterEq}, but restricted to the two-level subspace,
%
%
    %
    %
    %
    %
%
%
while the superoperator $\delta\mathcal{\hat L}$ contains the coherent and dissipative corrections induced by the elimination of the $|f\rangle$ level,
\begin{multline}
    \delta \mathcal{\hat L} \hat \rho_{\text{eff}}\equiv -i[\delta \hat H,\hat \rho_{\text{eff}}]
    \frac{1}{2}(\mathcal{D}[\delta\hat L_\text{L}]+\mathcal{D}[\delta\hat L_\text{R}])\hat \rho_\text{eff}
+
     \frac{1}{2}(\mathcal{D}[\hat L_\text{L},\delta\hat L_\text{L}]+\mathcal{D}[\delta\hat L_\text{L},\hat L_\text{L}])\hat \rho_\text{eff}
    \\
    +\frac{1}{2}(\mathcal{D}[\hat L_\text{R},\delta\hat L_\text{R}]+\mathcal{D}[\delta\hat L_\text{R},\hat L_\text{R}])\hat \rho_\text{eff}.
\end{multline}
The effective Liouvillian $\delta \mathcal{\hat L}$ contains dissipative corrections generated solely by the effective jump operators $\delta\hat L_{\mathrm{L/R}}$.
Since these operators are diagonal in the single-excitation manifold, they generate state-selective dephasing between $|eg\rangle$ and $|ge\rangle$. More explicitly, they contain dissipators of the form
\begin{equation}
\mathcal{D}[\delta\hat L_\text{L/R}]\sim \mathcal{D}[\hat\sigma_i^\dagger\hat\sigma_i (\mathbb{I}-\hat\sigma_j^\dagger\hat\sigma_j),\hat\sigma_m^\dagger\hat\sigma_m  (\mathbb{I}-\hat\sigma_n^\dagger\hat\sigma_n)],
\end{equation}
with $i,j,m,n\in \{1,2\}$ ($i\neq j, m\neq n$). 
These terms suppress coherences involving the single-excitation states and therefore act as an effective collective pure-dephasing mechanism.
%
%
The mixed dissipators involving both $\hat L_{\mathrm{L/R}}$ and $\delta\hat L_{\mathrm{L/R}}$ arise because these two contributions are correlated noise channels, as they correspond to different components of the same directional collective decay process. 

Figures~\ref{fig:SM-Model-Comparison}c and~\ref{fig:SM-Model-Comparison-AbsValues}a show that the effective master equation in Eq.~\eqref{eq-SM:Effective-masterEq} reproduces the dynamics and steady state of the full transmon model in Eq.~\eqref{eq-SM:Full-masterEq}. In particular, it captures the imbalance between the single-excitation states $\{|eg\rangle,|ge\rangle\}$ induced by the anharmonicity.
In Fig.~\ref{fig:SM-Model-Comparison-AbsValues}a, we compare the steady-state density matrix from the transmon model with the effective two-level system model (red wireframes), and the ideal two-level model of Eq.~\eqref{eq-SM:GA-masterEq} (black dashed wireframes). The effective model agrees well with the exact calculation, reproducing the redistribution of populations and coherences within the single-excitation manifold.
Figure~\ref{fig:SM-Model-Comparison-AbsValues}b shows the experimental stationary density matrix from the main text (see Fig.~\ref{fig:fig5}b), compared with the effective model and the ideal triplet state. The effective description captures the populations and coherences within the single-excitation manifold, though the model is less accurate for coherences involving the ground state.

%
%


\section{Properties of the emitter-emitter system: dark-state conditions}
\label{sec:dark-State}

In this section, we analyze the energy structure of the two-giant-atom system (both in the presence and absence of coherent driving), and derive the conditions under which dark states arise.  
For simplicity, we restrict the description to the two-level approximation---we set to zero the energy and dissipative corrections derived in the previous section, i.e., $\delta \hat H,\delta \mathcal{\hat L}= 0$---and neglect non-radiative decay processes ($\gamma_{\mathrm{nr}},\gamma_\phi=0$). Within this approximation, the system is described by~\cite{Soro2022}
\begin{equation}
    \frac{d \hat \rho_{\text{TLS}}}{dt}=-i[\hat H_{\text{TLS}},\hat\rho_{\text{TLS}}]+\frac{\Gamma_{\mathrm{Q}1}(\omega_1)}{2}\mathcal{D}[\hat \sigma_1]\hat \rho_{\text{TLS}} +\frac{\Gamma_{\mathrm{Q}2}(\omega_2)}{2}\mathcal{D}[\hat \sigma_2]\hat \rho_{\text{TLS}} 
    + \frac{\Gamma_{\text{coll}}}{2}\mleft(\mathcal{D}[\hat \sigma_1,\hat \sigma_2]+\mathcal{D}[\hat \sigma_2,\hat \sigma_1]\mright)\hat \rho_{\text{TLS}} ,
 \label{eq-SM:GA-masterEq}   
\end{equation}
where the Hamiltonian $\hat H_{\text{TLS}}$ reads
\begin{equation}
    \hat H_{\text{TLS}}= 
(\Delta+\delta)\hat\sigma_1^\dagger\hat\sigma_1 
    +
    (\Delta-\delta)\hat\sigma_2^\dagger\hat\sigma_2
    -\frac{i}{2}\mleft(\Omega_1\hat\sigma_1^\dagger+\Omega_2\hat\sigma_2^\dagger-\text{H.c.}\mright).
    \label{eq-SM:GA-Hamiltonian}
\end{equation}
The expressions for $\Gamma_{\mathrm{Q}i}(\omega_i)$,  $\Omega_i$, and $\Gamma_{\text{coll}}$ are given by Eqs.~\eqref{eq:giant_atom_dissipation}, \eqref{eq:SM-Driving-amplitudes}, and \eqref{eq:SM-collective-dissipator}, respectively.
These approximations provide a transparent analytical description of the eigenstructure of the system. Since the anharmonicity introduces only small corrections to the energy levels and dynamics, the two-level model is sufficient for the present analysis. The resulting energy-level structure in the bare basis is equivalent to one shown in Fig.~\ref{fig:SM-Model-Comparison}a but neglecting higher states.
We note that the following discussion was already derived in Ref.~\cite{Soro2022}; here, we provide a more detailed analysis for the particular case of two distant giant atoms.

\subsection{Formal definition of a dark state}

Collective light–matter systems can exhibit, under special conditions, pure states $|D\rangle$ that are protected from decoherence, preventing decay during the evolution towards the steady state. 
These states are referred to as \textit{dark states} and are defined by the following conditions~\cite{KrausPreparationEntangled2008,DiehlQuantumStates2008}:
(i) $|D\rangle$ is an eigenstate of the system Hamiltonian; and  
(ii) $|D\rangle$ lies in the kernel of the dissipative dynamics, i.e., it is annihilated by all quantum jump operators $\{\hat J_k\}$.
Equivalently,
\begin{subequations}
\begin{align}
    \hat H |D\rangle &= E_D |D\rangle.\\
    \hat J_k |D\rangle &= 0 \quad \forall k,
\end{align}
\end{subequations}
If the dark state is unique and no other invariant subspaces exist, it defines the steady state of the system~\cite{DiehlQuantumStates2008}, that is,  $\hat \rho_{\text{ss}}=|D\rangle \langle D|$.

In the present system, dark states can be engineered by exploiting the giant-atom geometry. In particular, by tuning the phases $\{\phi(\omega_i),\theta(\omega)\}$, it is possible to stabilize entangled states in the long-time limit, as shown in Ref.~\cite{Soro2022}.
Operationally, this procedure consists of: (i) diagonalizing the system Hamiltonian, (ii) evaluating the action of the phase-dependent jump operators on its eigenstates, and (iii) identifying the phase conditions under which a given eigenstate is annihilated by all jump operators, thus becoming a dark state.

\subsection{Bare emitter system}

We begin by considering the simplest scenario, in which the coherent drive is absent, i.e., $\Omega_1=\Omega_2=0$. 
Although the Hamiltonian in Eq.~\eqref{eq-SM:GA-Hamiltonian} is already diagonal in the bare basis, it is more convenient to express it in the \textit{dimer basis}, $\{|gg\rangle, |\Psi^+\rangle, |\Psi^-\rangle, |ee\rangle\}$, where
\begin{equation}
    |\Psi^\pm\rangle \equiv \frac{1}{\sqrt{2}}\mleft(|ge\rangle \pm |eg\rangle\mright),
\end{equation}
are the symmetric and antisymmetric combinations of the single-excitation bare states, i.e., \textit{Bell states}. These states are also commonly referred to as the triplet and singlet states, respectively. Throughout the text, we use these terminologies interchangeably.
In this basis, the Hamiltonian becomes
\begin{equation}
    \hat H^{(\Omega=0)}_{\text{TLS}}= 
(\Delta+\delta)\hat\sigma_1^\dagger\hat\sigma_1 
    +
    (\Delta-\delta)\hat\sigma_2^\dagger\hat\sigma_2 
    =
    2\Delta |ee\rangle \langle ee|+\Delta(|\Psi^+\rangle \langle \Psi^+|+|\Psi^-\rangle \langle \Psi^-|)+\delta (|\Psi^+\rangle \langle \Psi^-|+\text{H.c.}).
    \label{eq-SM:hamiltonian-no-drive}
\end{equation}
The collective decay rates associated with the symmetric and antisymmetric sectors are
\begin{equation}
    \Gamma_{\pm}\equiv \frac{1}{2}\mleft[ \Gamma_{\mathrm{Q}1}(\omega_1)+\Gamma_{\mathrm{Q}2}(\omega_2)\pm 2\Gamma_{\text{coll}}\mright].
        \label{eq-SM:dimer-decay-rates}
\end{equation}
\begin{figure*}[t!]
    \centering
\includegraphics[width=1.0\textwidth]{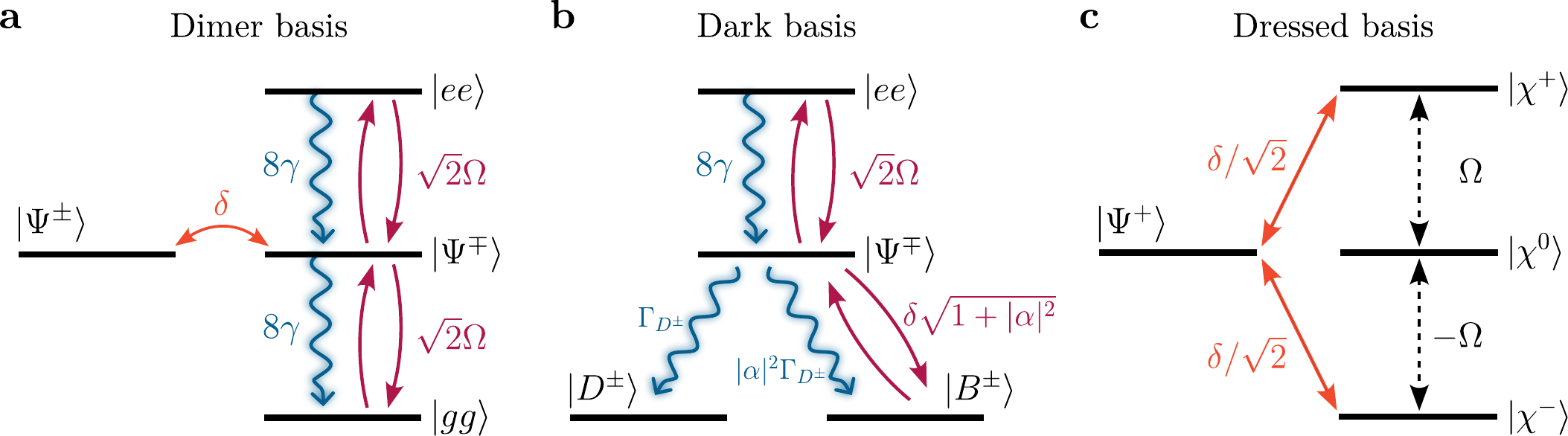}
    \caption{
    \textbf{Energy-level diagrams under dark-state conditions}. 
\textbf{a)} 
Dimer basis $\{|gg\rangle,|\Psi^+\rangle,|\Psi^-\rangle,|ee\rangle\}$, where $|\Psi^\pm\rangle=(|ge\rangle\pm|eg\rangle)/\sqrt{2}$ are the triplet and singlet states. The bright channel $|gg\rangle\leftrightarrow|\Psi^\mp \rangle\leftrightarrow|ee\rangle$ is coherently driven and dissipatively connected with decay rate $8\gamma$. In contrast, the state $|\Psi^\pm\rangle$ is only coherently coupled through the qubit-qubit detuning $\delta$.
\textbf{b)} Dark-state representation $\{|D^\pm\rangle,|B^\pm\rangle,|\Psi^\mp\rangle,|ee\rangle\}$. The dark state is coherently decoupled from the dynamics and only connected dissipatively through the residual decay channel $\Gamma_{D^\pm}$. The remaining states form a coherently connected ladder,
$|ee\rangle \leftrightarrow |\Psi^\mp\rangle \leftrightarrow |B^\pm\rangle$,
with coherent coupling strengths $\sqrt{2}\Omega$ and $\delta\sqrt{1+|\alpha|^2}$, respectively. 
\textbf{c)} Dressed-basis representation $\{|\Psi^+\rangle,|\chi^+\rangle,|\chi^0\rangle,|\chi^-\rangle\}$ under the triplet dark-state conditions. 
The triplet state $|\Psi^+\rangle$ is coupled to the dressed states $|\chi^\pm\rangle$ with strength $\delta/\sqrt{2}$, while remaining detuned by an energy $\pm\Omega$. In the regime $\Omega\gg\delta$, the triplet state contributes only perturbatively to the dynamics, effectively behaving as a virtual state. Incoherent processes are not shown in this panel.%
    }
    \label{fig:SM-Energy-Diagrams}
\end{figure*}

\noindent
\textbf{Dark-state conditions.}
In the original formulation of the problem~\cite{Kockum2018}, the dissipative dynamics are naturally expressed in terms of directional left- and right-collective jump operators as a consequence of applying the SLH formalism~\cite{CombesSLHFramework2017}. These jump operators are given by
\begin{subequations}
    \begin{align}
        \hat L_{\text{L}}&\equiv 
        \sqrt{\Gamma_{1,\text{L}}}\hat \sigma_1+\sqrt{\Gamma_{2,\text{L}}}\hat \sigma_2
        =
        \sqrt{\frac{\gamma_1(\omega_1)}{2}}\mleft[ 1+ e^{i \phi_1} \mright]\hat \sigma_1+\sqrt{\frac{\gamma_2(\omega_2)}{2}}\mleft[ e^{i(\phi_1+\theta)}+ e^{i( \phi_1+\theta+\phi_2)} \mright]\hat \sigma_2 , 
        \label{eq-SM:left-jump}
        \\
     \hat L_{\text{R}}&\equiv \sqrt{\Gamma_{1,\text{R}}}\hat \sigma_1+\sqrt{\Gamma_{2,\text{R}}}\hat \sigma_2=\sqrt{\frac{\gamma_1(\omega_1)}{2}}\mleft[  e^{i(\theta+\phi_2)}+ e^{i( \phi_1+\theta+\phi_2)} \mright]\hat \sigma_1+\sqrt{\frac{\gamma_2(\omega_2)}{2}}\mleft[ 1+ e^{i\phi_2} \mright]\hat \sigma_2,
    \label{eq-SM:right-jump}
    \end{align}
\end{subequations}
where we have introduced the compact notation $\phi_i\equiv\phi(\omega_i)$ and $\theta\equiv\theta(\omega)$ to lighten the discussion.
For clarity, we emphasize that the different decay rates introduced above are conceptually different: $\gamma_i(\omega_i)$ denotes the transmon–waveguide coupling strength at each individual coupling point, $\Gamma_{\mathrm{Q}i}(\omega_i)$ is the effective geometry-dependent decay rate experienced by the $i$th transmon due to the giant-atom configuration, and $\Gamma_{i,\text{L/R}}$ are the directional decay rates associated with left- and right-propagating waveguide modes.
We also note that expanding the dissipators associated with the directional jump operators in Eqs.~\eqref{eq-SM:left-jump} and \eqref{eq-SM:right-jump}, $(\mathcal{D}[\hat L_{\text{L}}]+\mathcal{D}[\hat L_{\text{R}}])\hat \rho$, one recovers the master equation given in Eq.~\eqref{eq-SM:GA-masterEq}.
Expressed in the dimer basis, the directional jump operators take the form
\begin{equation}
            \hat L_{\text{L/R}}= 
        \frac{1}{\sqrt{2}}
        \mleft[\sqrt{\Gamma_{1,\text{L/R}}}-\sqrt{\Gamma_{2,\text{L/R}}}\mright](|\Psi^- \rangle \langle ee|-|gg\rangle \langle \Psi^-|)
        +
      \frac{1}{\sqrt{2}}
        \mleft[\sqrt{\Gamma_{1,\text{L/R}}}+\sqrt{\Gamma_{2,\text{L/R}}}\mright](|\Psi^+ \rangle \langle ee|+|gg\rangle \langle \Psi^+|).
        \label{eq-SM:left-right-jump-dimer-basis}
\end{equation}
This representation is particularly useful for identifying dark-state conditions, since it allows us to directly determine the phase configurations for which the Bell states $|\Psi^\pm\rangle$ are annihilated by both jump operators.
In particular, the undriven system exhibits dark states whenever $\hat L_{\text{L/R}} |\Psi^\pm\rangle = 0$, which leads to the following conditions:
\begin{subequations}
\label{eq-SM:dark-state-conditions-undriven}
    \begin{align}
        |D\rangle=|\Psi^-\rangle:& \quad \sqrt{\Gamma_{1,\text{L/R}}}=\sqrt{\Gamma_{2,\text{L/R}}}, \quad \ \ \Rightarrow\quad \mleft\{\phi_{1/2},\theta\mright\}= 2\pi m,\\
         |D\rangle=|\Psi^+\rangle:& \quad \sqrt{\Gamma_{1,\text{L/R}}}=-\sqrt{\Gamma_{2,\text{L/R}}}, \quad  \Rightarrow\quad \mleft\{\phi_{1/2},\theta\mright\}= \mleft\{2\pi m ,(2m+1)\pi \mright\},
    \end{align}
\end{subequations}
with $m\in\mathbb{N}$, assuming $\Gamma_1=\Gamma_2$ and $\delta=0$ (so that the Bell states become actual eigenstates of the Hamiltonian), in agreement with Ref.~\cite{Soro2022}.
Using Eq.~\eqref{eq-SM:dimer-decay-rates}, one finds that when the dark state is $|\Psi^+\rangle$, the symmetric decay channel is suppressed, $\Gamma_+=0$, while the antisymmetric decay channel is enhanced, $\Gamma_-=8\gamma$. Conversely, when the dark state is $|\Psi^-\rangle$, the antisymmetric decay channel is suppressed, $\Gamma_-=0$, while the symmetric decay channel is enhanced, $\Gamma_+=8\gamma$.
Note that, as a consequence of the collective structure of the directional dissipators, cross-Lindblad terms emerge of the form $\sim (\mathcal{D}\mleft[|\Psi^\mp\rangle\langle ee|,|gg\rangle\langle \Psi^\mp\mright]+\mathcal{D}\mleft[|gg\rangle\langle \Psi^\mp|,|\Psi^\mp\rangle\langle ee|\mright])$. These terms do not describe independent decay channels, but rather dissipative interference between the two transitions forming the cascade $|ee\rangle \rightarrow |\Psi^\mp\rangle \rightarrow |gg\rangle$.
Figure~\ref{fig:SM-Energy-Diagrams}a shows the resulting energy-level structure in the dimer basis under the dark-state conditions, together with the corresponding coherent and dissipative couplings.

\subsection{Dressed drive-emitter system}

In contrast to the undriven case, the presence of a coherent driving considerably enriches the structure of the system. Although general analytical expressions for the eigenvalues and eigenstates can still be obtained, they become cumbersome and offer limited physical insight.
Nevertheless, it is still convenient to express the Hamiltonian in the dimer basis introduced previously. Using Eq.~\eqref{eq-SM:GA-Hamiltonian}, the Hamiltonian can be written as
\begin{multline}
    \hat H_{\text{TLS}}= 
2\Delta |ee\rangle \langle ee|+\Delta(|\Psi^+\rangle \langle \Psi^+|+|\Psi^-\rangle \langle \Psi^-|)+\delta (|\Psi^+\rangle \langle \Psi^-|+|\Psi^-\rangle \langle \Psi^+|)\\
-\frac{i}{2}\mleft[\frac{1}{\sqrt{2}}(\Omega_1+\Omega_2)(|ee\rangle \langle \Psi^+|+|\Psi^+\rangle \langle gg|)+\frac{1}{\sqrt{2}}(\Omega_1-\Omega_2)(|ee\rangle \langle \Psi^-|-|\Psi^-\rangle \langle gg|)-\text{H.c.}\mright].
    \label{eq-SM:GA-Hamiltonian-dimer}
\end{multline}
\noindent
\textbf{Dark-state conditions.}
The coherent drive modifies only the Hamiltonian sector of the dynamics, while the dissipative contributions remain unchanged. Therefore, the directional jump operators derived in Eq.~\eqref{eq-SM:left-right-jump-dimer-basis} can still be employed to determine the dark-state conditions.
In the general case, the eigenstates of the driven Hamiltonian are nontrivial superpositions of the dimer states. To identify possible dark states, we therefore consider a general state
$|\xi\rangle=a|gg\rangle+b| \Psi^+\rangle+c|\Psi^-\rangle+d|ee\rangle$,
and impose the condition $\hat L_{\text{L/R}} |\xi \rangle = 0$, such that:
\begin{multline}
   \hat  L_{\text{L/R}} |\xi \rangle = \frac{1}{2}\mleft[
(b+c)\sqrt{\Gamma_{1,\text{L/R}}}+
(b-c)\sqrt{\Gamma_{2,\text{L/R}}}\mright] |gg\rangle 
\\
+
\frac{1}{2}d \mleft(\sqrt{\Gamma_{1,\text{L/R}}}+\sqrt{\Gamma_{2,\text{L/R}}}\mright)|\Psi^+\rangle 
+\frac{1}{2}d \mleft(-\sqrt{\Gamma_{1,\text{L/R}}}+\sqrt{\Gamma_{2,\text{L/R}}}\mright)|\Psi^-\rangle=0. 
\end{multline}
From this expression, it follows that the dark states must be superpositions of the ground state $|gg\rangle$ and either the singlet or triplet Bell state. The corresponding conditions are $\sqrt{\Gamma_{1,\text{L/R}}}=\sqrt{\Gamma_{2,\text{L/R}}}$ or $\sqrt{\Gamma_{1,\text{L/R}}}=-\sqrt{\Gamma_{2,\text{L/R}}}$, respectively, which reproduce the same phase constraints obtained in the undriven case [see Eq.~\eqref{eq-SM:dark-state-conditions-undriven}].
Under these conditions, and imposing resonant driving $\Delta=0$, the dark states take the form
\begin{equation}
\label{eq-SM:dark-state}
    |D^\pm\rangle=\frac{1}{\sqrt{1+|\alpha|^2}}(\alpha |\Psi^\pm \rangle+|gg\rangle),
\end{equation}
where 
\begin{equation}
    \alpha\equiv i\frac{\sqrt{2}\Omega}{2\delta}
\end{equation}
is the control parameter that weights the degree of entanglement between the emitters, since the dark state is a coherent superposition of the corresponding Bell state and the ground state. In particular, in the limit $|\alpha|\rightarrow\infty$---reached either for strong driving, $\Omega\gg\delta$, or for vanishingly small but finite detuning, $\delta\rightarrow0$---, the dark state approaches the Bell state,
\begin{equation}
    \lim_{|\alpha|\rightarrow\infty}|D^\pm\rangle=|\Psi^\pm\rangle,
\end{equation}
thereby yielding maximal stationary entanglement generated autonomously by the dissipative dynamics.
%
%
The corresponding dark-state conditions are
\begin{subequations}
\label{eq-SM:dark-state-conditions-driven}
    \begin{align}
        |D\rangle=|D^-\rangle:& \quad \sqrt{\Gamma_{1,\text{L/R}}}=\sqrt{\Gamma_{2,\text{L/R}}}, \quad \ \ \Rightarrow\quad \mleft\{\phi_{1/2},\theta\mright\}= 2\pi m \quad \quad \quad \hspace{1.5cm}    \text{and} \quad \Omega\equiv\Omega_1=\Omega_2,
        \label{eq-SM:dark-state-conditions-driven-singlet}
\\
         |D\rangle=|D^+\rangle:& \quad \sqrt{\Gamma_{1,\text{L/R}}}=-\sqrt{\Gamma_{2,\text{L/R}}}, \quad  \Rightarrow\quad \mleft\{\phi_{1/2},\theta\mright\}= \mleft\{2\pi m ,(2m+1)\pi \mright\}  \quad \text{and} \quad \Omega\equiv\Omega_1=-\Omega_2,
         \label{eq-SM:dark-state-conditions-driven-triplet}
    \end{align}
\end{subequations}
with $m\in\mathbb{N}$. Since the driving amplitudes depend explicitly on the giant-atom phases, the phase constraints simultaneously determine the symmetry of the coherent driving~\cite{Pichler2015}.
Finally, the states orthogonal to the dark states can be identified as the corresponding bright states,
\begin{equation}
    |B^\pm \rangle\equiv \frac{1}{\sqrt{1+|\alpha|^2}}(|\Psi^\pm\rangle -\alpha^* |gg\rangle).
\end{equation}
Then, defining the \textit{dark basis}, $\{|D^\pm\rangle,|B^\pm\rangle,|\Psi^\mp\rangle,|ee\rangle\}$, the Hamiltonian under the dark-state conditions associated with $|D^\pm\rangle$ takes the form

\begin{equation}
     \hat H_{\text{TLS}}^{(\pm)}= -\frac{i}{2}[\sqrt{2}\Omega (|ee\rangle \langle \Psi^\mp|-|\Psi^\mp\rangle \langle ee|)]+\delta\sqrt{1+|\alpha|^2} (|B^\pm\rangle\langle \Psi^\mp\rangle |+|\Psi^\mp\rangle\langle B^\pm \rangle |).
     \label{eq-SM:Hamiltonian-DarkBasis}
\end{equation}
%
%
%
%
%
%
%
%
%

%
We emphasize that the dark basis does not diagonalize the Hamiltonian. Rather, it provides a convenient representation that makes the coherent and dissipative structure of the system more transparent for analyzing the present problem. 
The Hamiltonian in Eq.~\eqref{eq-SM:Hamiltonian-DarkBasis} shows that, when the dark-state conditions for $|D^\pm\rangle$ are satisfied, the corresponding dark state becomes coherently decoupled from the dynamics. In contrast, the remaining states form a coherently connected ladder,  $|ee\rangle \leftrightarrow|\Psi^\mp\rangle\leftrightarrow|B^\pm\rangle$ with coupling strengths $\sqrt{2}\Omega$ and $\delta\sqrt{1+|\alpha|^2}$, respectively, as illustrated in Fig.~\ref{fig:SM-Energy-Diagrams}b.   
Expressed in the dark basis, the directional jump operators take the form
\begin{multline}
    \hat L_{R/L}^{(\pm)}=\frac{\alpha (\sqrt{\Gamma_{1,R/L}}\pm\sqrt{\Gamma_{2,R/L}})}{\sqrt{2}(1+|\alpha|^2)}(|D^\pm\rangle\langle D^\pm|-|B^\pm\rangle\langle B^\pm|)
    +\frac{\sqrt{\Gamma_{1,R/L}}\pm\sqrt{\Gamma_{2,R/L}}}{\sqrt{2}(1+|\alpha|^2)} (|D^\pm\rangle \langle B^\pm|-\alpha^2 |B^\pm\rangle \langle D^\pm|)
    \\
   + \frac{\sqrt{\Gamma_{1,R/L}}\mp\sqrt{\Gamma_{2,R/L}}}{\sqrt{2}\sqrt{1+|\alpha|^2}} (|D^\pm\rangle \langle \Psi^\mp\rangle|-\alpha |B^\pm\rangle \langle \Psi^\mp|-\sqrt{1+|\alpha|^2}|\Psi^\mp\rangle \langle ee|)
    +\frac{\pm\sqrt{\Gamma_{1,R/L}}+\sqrt{\Gamma_{2,R/L}}}{\sqrt{2}\sqrt{1+|\alpha|^2}} (\alpha^*|D^\pm\rangle \langle ee\rangle|+ |B^\pm\rangle \langle ee|).
    \label{eq-SM:Collective-Jump-Op-DarkBasis}
\end{multline}
Then, using the Hamiltonian and the directional collective jump operators in Eqs.~\eqref{eq-SM:Hamiltonian-DarkBasis} and~\eqref{eq-SM:Collective-Jump-Op-DarkBasis}, we construct the master equation in the dark basis under the corresponding dark-state conditions:
\begin{equation}
        \frac{d \hat \rho_{\text{TLS}}^{(\pm)}}{dt}=-i[\hat H^{(\pm)},\rho^{(\pm)}_{\text{TLS}}]+\frac{1}{2}\mleft(\mathcal{D}[\hat L_{R}^{(\pm)}]+\mathcal{D}[\hat L_{L}^{(\pm)}]\mright)\hat \rho^{(\pm)}_{\text{TLS}}.
\end{equation}
By analyzing the associated rate equations for $\langle D^\pm| \dot {\hat \rho}_{\text{TLS}}^{(\pm)}|D^\pm \rangle$ and $\langle B^\pm| \dot {\hat \rho}_{\text{TLS}}^{(\pm)}|B^\pm \rangle$, we extract the effective decay rates from $|\Psi^\mp\rangle$ into $|D^\pm\rangle$ and $|B^\pm\rangle$:
\begin{equation}
    \Gamma_{D^\pm}\equiv \frac{\Gamma_{\text{Q}1}(\omega_1)+\Gamma_{\text{Q}2}(\omega_2)}{1+|\alpha|^2},\quad \text{and}\quad \Gamma_{B^\pm}\equiv |\alpha|^2\Gamma_{D^\pm}.
\end{equation}
The resulting energy-level structure in the dark basis, including the corresponding coherent and dissipative couplings, is shown in Fig.~\ref{fig:SM-Energy-Diagrams}b.

\subsection{Dark-state operating regime}
In the experiment, we operate under the triplet dark-state conditions stablished in Eq.~\eqref{eq-SM:dark-state-conditions-driven-triplet}, namely
\begin{equation}
    \phi(\omega_1)=\phi(\omega_2)=2\pi m \quad \text{and}\quad \theta(\omega)=(2m+1)\pi, \quad \text{with $m\in\mathbb{N}$.}
\end{equation}
Under these conditions, the geometry-induced Lamb shifts and coherent exchange interaction from Eqs.~\eqref{eq:SM-Lamb-shift} and ~\eqref{eq:SM-collective-coupling} vanish, i.e., 
$
    \delta\omega_1,\delta\omega_2,g=0.
$
The remaining parameters from Eqs.~\eqref{eq:giant_atom_dissipation}, \eqref{eq:SM-Driving-amplitudes}, and \eqref{eq:SM-collective-dissipator} then simplify to
\begin{equation}
        \Gamma\equiv \Gamma_{\text{Q}{1/2}}(\omega_{1/2})\approx4\gamma, \quad \Gamma_{\text{coll}}=-2\gamma,\quad \text{and}\quad
        \Omega\equiv \Omega_1=-\Omega_2=2\sqrt{2\gamma}\beta,
        \label{eq:SM-GammaOmega-definition}
\end{equation}
where we assume $\gamma(\omega_1)\approx\gamma(\omega_2)\approx\gamma$.
Following the notation introduced in Sec.~\ref{sec-SM:model}, the triplet dark-state conditions require the qubits to be tuned to the \textit{individual superradiant} frequencies [see Eq.~\eqref{eq-SM:individual-superradiance}]. In this regime, the fields emitted from the internal coupling points of each giant atom interfere constructively, leading to an enhancement of the spontaneous emission rate.

\section{Mechanism of entanglement generation}
\label{Sec-SM-Unconventional}

In this section, we further analyze the mechanism underlying the generation of entanglement between the two distant giant atoms from two different and complementary approaches. 

\subsection{Dark-state approach}

As mentioned in the main text, the emergence of large steady-state entanglement between two coherently driven, non-interacting quantum emitters has previously been reported in several theoretical contexts, including chiral waveguides~\cite{Pichler2015}, giant atoms~\cite{Soro2022}, and more recently cavity QED systems~\cite{Vivas-VianaDissipativeStabilization2024}.
In particular, Refs.~\cite{Pichler2015,Soro2022} explain the stabilization of entanglement by expressing the dynamics in the dark basis, as depicted in Fig.~\ref{fig:SM-Energy-Diagrams}b. In this representation, it becomes clear that the system relaxes toward the dark state $|D^\pm\rangle$ in the long-time limit, $\hat \rho_{\text{ss}}=|D^\pm\rangle \langle D^\pm|$, since this state is protected against decoherence and therefore cannot decay during the dissipative evolution toward the steady state~\cite{KrausPreparationEntangled2008,DiehlQuantumStates2008}. Consequently, population progressively accumulates in the dark state until the stationary regime is reached.
While this picture in terms of the dark-state basis is mathematically transparent, it can be complemented by a more physically intuitive description of the entanglement-generation mechanism based on quantum trajectories.

\subsection{Unconventional mechanism of entanglement generation}

An alternative interpretation of this effect was introduced in Ref.~\cite{Vivas-VianaUnconventionalMechanism2022}, where it was shown that a state exhibiting the properties of a \textit{virtual state}---namely, energetically off-resonant and effectively decoupled from dissipative channels~\cite{Cohen-TannoudjiAtomPhotonInteractions1998}---can nevertheless accumulate a sizable population in the long-time limit.
This phenomenon was termed \textit{unconventional population of virtual states}, as it challenges the usual intuition that a virtual state should remain essentially unpopulated.
As discussed in Refs.~\cite{Vivas-VianaUnconventionalMechanism2022,Vivas-VianaDissipativeStabilization2024}, and further detailed in Ref.~\cite{Vivas-VianaNonclassicalDrivenDissipative2025}, the virtual state gradually accumulates population during the non-Hermitian evolution between quantum jumps. The key observation is that, conditioned on the absence of quantum jumps, the system evolves preferentially toward the virtual state. In other words, the longer a trajectory evolves without a jump, the more likely it is that the system occupies the virtual state. This information gained from the absence of jumps feeds back into the state of the system, and the effect accumulates over time, ultimately leading to a finite steady-state population of the virtual state.

In the present system---two distant and coherently driven giant atoms~\cite{Soro2022}---this mechanism underlies the generation of entanglement since, under the triplet dark-state conditions, the Bell state $|\Psi^+\rangle$ exhibits precisely the properties of such a virtual state. 
This becomes evident after diagonalizing the driving term in the Hamiltonian from Eq.~\eqref{eq-SM:GA-Hamiltonian-dimer}. The resulting eigenstates can be expressed in terms of the two-excitation dressed states---$|\Phi^\pm\rangle\equiv1/\sqrt{2}(|gg\rangle\pm |ee\rangle)$---as 
\begin{equation}
    |\chi^\pm\rangle =\frac{1}{\sqrt{2}}(|\Psi^-\rangle\mp i|\Phi^+\rangle),\quad \text{and}\quad |\chi^0\rangle=|\Phi^-\rangle,
\end{equation}
with eigenenergies $\lambda_\pm =\pm \Omega$ and $\lambda_0=0$. 
In this dressed basis $\{|\Psi^+\rangle,|\chi^+\rangle,|\chi^0\rangle,|\chi^-\rangle\}$, illustrated in Fig.~\ref{fig:SM-Energy-Diagrams}c, the triplet state $|\Psi^+\rangle$ couples to $|\chi^\pm\rangle$ with strength $\delta/\sqrt{2}$, while remaining detuned by $\pm\Omega$. Therefore, in the regime $\Omega\gg\delta$, the coupling is strongly off-resonant and the triplet state only affects the dynamics perturbatively, effectively behaving as a virtual state~\cite{Cohen-TannoudjiAtomPhotonInteractions1998}.
The same feature appears in the original model of coherently driven emitters coupled to a chiral waveguide~\cite{Pichler2015}, as well as in its cavity-QED counterpart~\cite{Vivas-VianaDissipativeStabilization2024}: in both cases, the entangled state behaves as a virtual state that ultimately acquires a sizable population through this unconventional mechanism~\cite{Vivas-VianaUnconventionalMechanism2022}.
\newline

\begin{figure*}[b!]
    \centering
\includegraphics[width=1.0\textwidth]{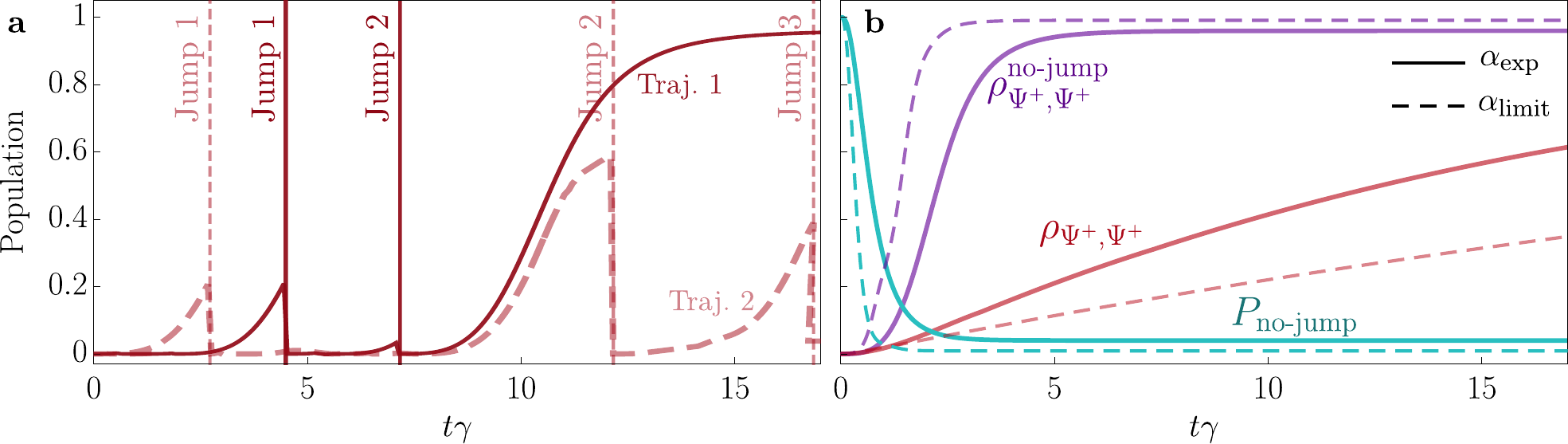}
    \caption{
\textbf{Analysis of the unconventional mechanism of virtual-state population.}
\textbf{a)} 
Population of the triplet state $\rho_{\Psi^+,\Psi^+}(t)$  during the non-Hermitian evolution between quantum jumps, obtained from Eq.~\eqref{eq-SM:Non-Hermitian-Hamiltonian}. Two quantum trajectories are shown: trajectory 1 (solid dark red) and trajectory 2 (dashed light red). Vertical lines indicate the occurrence of quantum jumps.
\textbf{b)}
Conditional no-jump dynamics. 
Purple lines show the conditioned population of the triplet state,
$\rho^{\text{no-jump}}_{\Psi^+,\Psi^+}(t)$ using Eq.~\eqref{eq-SM:Conditional-master-eq}, 
while red lines correspond to the unconditional triplet-state population,  $\rho_{\Psi^+,\Psi^+}(t)$, computed with Eq.~\eqref{eq-SM:GA-masterEq}.
Blue lines denote the probability of no quantum jumps, $P_{\text{no-jump}}(t)$.
Solid lines correspond to numerical simulations using the experimental parameters $\{\Omega,\delta,\gamma\}/2\pi=\{45,6.81,14.7\}\,\mathrm{MHz}$, while dashed lines show the limiting regime $|\alpha|\gg1$, with $\{\Omega,\delta,\gamma\}/2\pi=\{100,6.81,14.7\}\,\mathrm{MHz}$. %
In the limiting regime ($\alpha_{\text{limit}}$, second case), the conditioned triplet-state population approaches unity in the long-time limit,
 $\rho^{\text{no-jump}}_{\Psi^+,\Psi^+}\rightarrow1$ while the no-jump probability vanishes, $P_{\text{no-jump}}\rightarrow0$, 
 in agreement with the unconventional mechanism discussed in Refs.~\cite{Vivas-VianaUnconventionalMechanism2022}.
    }
    \label{fig:SM-Unconventional-Mechanism}
\end{figure*}

\noindent
\textbf{Quantum-trajectory perspective. }
To illustrate this mechanism explicitly, we perform quantum-trajectory simulations using the quantum jump method~\cite{CarmichaelPhotoelectronWaiting1989,MolmerMonteCarlo1993,MolmerMonteCarlo1996,PlenioQuantumjumpApproach1998}.
We consider the effective non-Hermitian Hamiltonian in the dimer basis
\begin{equation}
    \tilde H=\hat H
   - \frac{i}{2}
   \hat J^\dagger \hat J,
    \label{eq-SM:Non-Hermitian-Hamiltonian}
\end{equation}
where $\hat H$ is the Hamiltonian defined in Eq.~\eqref{eq-SM:GA-Hamiltonian-dimer} under the triplet dark-state conditions, i.e., $\Delta=0$ and $\Omega_1=-\Omega_2$, while the collective jump operator is given by
\begin{equation}
    \hat J\equiv\sqrt{8\gamma}\, (|\Psi^-\rangle\langle ee|-|gg\rangle\langle\Psi^-|).
\end{equation}
Throughout this analysis, we consider the same assumptions introduced in the previous section: (i) the two-level approximation, neglecting the energy and dissipative corrections associated with the third transmon level, and (ii) negligible non-radiative processes, i.e., $\gamma_{\mathrm{nr}},\gamma_\phi=0$.
Inspection of individual trajectories in Fig.~\ref{fig:SM-Unconventional-Mechanism}a shows that the virtual state---i.e., the triplet state---becomes populated during the non-Hermitian evolution between quantum jumps, while the occurrence of a jump strongly suppresses its population.
Interestingly, a single trajectory can rapidly stabilize into the triplet state after only a few quantum jumps, as illustrated by trajectory 1 (solid red line in Fig.~\ref{fig:SM-Unconventional-Mechanism}a). 
Due to the \textit{dark} nature of the steady state~\cite{KrausPreparationEntangled2008,DiehlQuantumStates2008}, there exists the possibility that the two-giant-atom system reaches the dark state during the non-Hermitian evolution. Once this occurs, the system becomes protected against further dissipative evolution and no additional quantum jumps take place.%
\newline

\noindent
\textbf{Conditional evolution.}
We can further formalize this intuition by considering the particular conditional trajectory in which no jumps occur during the evolution, as shown in Fig.~\ref{fig:SM-Unconventional-Mechanism}b.
The corresponding conditioned master equation reads
\begin{equation}
    \frac{d\hat \rho_{\text{no-jump}}}{dt}=-i[\hat H,\hat \rho_{\text{no-jump}}]-
    \frac{1}{2}
    \mleft\{
    \hat J^\dagger \hat J,
    \hat\rho_{\text{no-jump}}
    \mright\},
    \label{eq-SM:Conditional-master-eq}
\end{equation}
which corresponds to the standard Lindblad evolution after removing the jump term, $\sim \hat J \hat \rho \hat J^\dagger$.
As a consequence, the dynamics is described by an unnormalized density matrix, $\hat \rho_{\text{no-jump}}$,  conditioned on the absence of jumps during a time interval $t$. Its trace gives the probability that no jump has occurred up to time $t$:
\begin{equation}
    P_{\text{no-jump}}(t)\equiv \text{Tr}[\hat \rho_{\text{no-jump}}(t)].
\end{equation}
To compare this conditional state with the physical density matrix obtained from Eq.~\eqref{eq-SM:GA-masterEq}, the conditional density matrix must be renormalized according to
\begin{equation}
    \hat \rho_{\text{no-jump}}(t) \longrightarrow \frac{\hat \rho_{\text{no-jump}}(t)}{\text{Tr}[\hat \rho_{\text{no-jump}}(t)]}.
\end{equation}

From the dark-state definition in Eq.~\eqref{eq-SM:dark-state}, the stationary population of the triplet Bell state is given by
$ \rho_{\Psi^+,\Psi^+}^{\text{ss}}
    =
    |\alpha|^2/(1+|\alpha|^2),$
while the remaining population occupies the ground state,
$
    \rho_{gg,gg}^{\text{ss}}
    =
    1/(1+|\alpha|^2).
$
Under the dark-state conditions, this stationary regime is reached both in the unconditional and conditioned dynamics.
However, as shown in Fig.~\ref{fig:SM-Unconventional-Mechanism}b, the conditioned no-jump evolution stabilizes into the triplet state on a much faster timescale, $\sim1/\gamma$, than the actual one [see Section~\ref{sec:HAE}], while the no-jump probability simultaneously decreases towards its asymptotic value,
$
    P_{\text{no-jump}}(t\rightarrow\infty)=  \rho_{gg,gg}^{\text{ss}}
.
$
In the limiting case $|\alpha|\gg1$, corresponding to optimal entanglement generation, the no-jump probability vanishes,  $P_{\text{no-jump}}\rightarrow 0$, while the conditioned triplet-state population approaches unity, $\rho_{\Psi^+,\Psi^+}(t\rightarrow \infty)\approx 1.$

This analysis confirms that the unconventional mechanism proposed in Ref.~\cite{Vivas-VianaUnconventionalMechanism2022} applies directly to the present system. In this alternative approach, the non-Hermitian evolution conditioned on the absence of quantum jumps provides the mechanism through which the triplet state becomes populated, thereby enabling the emergence of high steady-state entanglement.

\section{Timescale of entanglement stabilization}
\label{sec:HAE}

\begin{figure*}[b!]
    \centering
\includegraphics[width=1.0\textwidth]{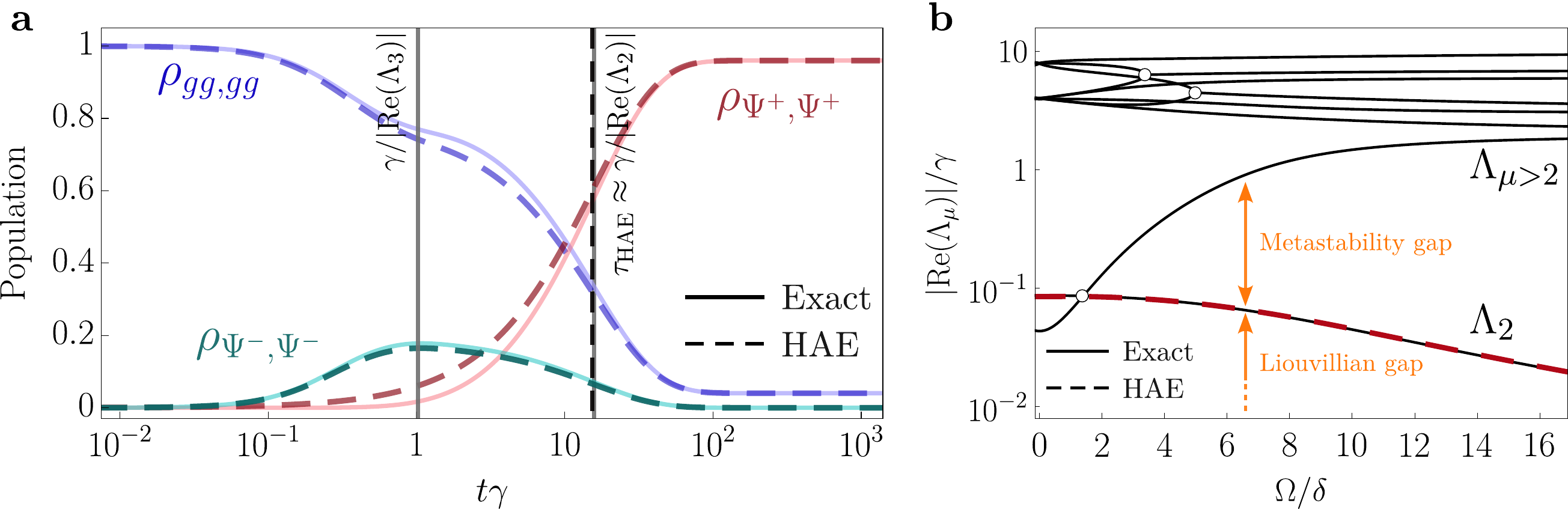}
    \caption{
\textbf{Validity of the hierarchical adiabatic elimination (HAE).}
\textbf{a)}
Time evolution of the populations $\rho_{gg,gg}(t)$ (blue), $\rho_{\Psi^-,\Psi^-}(t)$ (light blue), and $\rho_{\Psi^+,\Psi^+}(t)$ (red).
Solid lines correspond to numerical simulations of Eq.~\eqref{eq-SM:Effective-HAE-masterEq}, while dashed lines denote the analytical predictions obtained from the HAE method.
The vertical lines indicate the characteristic dissipative timescales: the fast relaxation timescale $\sim \gamma/|\mathrm{Re}(\Lambda_3)|$ and the relaxation timescale $\tau_{\mathrm{HAE}}\sim \gamma/|\mathrm{Re}(\Lambda_2)|$.
\textbf{b)}
Absolute value of the real part of the Liouvillian spectrum.
The spectral separation between $\Lambda_2$ and $\Lambda_{\mu>2}$ defines the metastability timescale, while the distance between $\Lambda_2$ and the steady-state eigenvalue $\Lambda_1=0$ (not shown) determines the Liouvillian gap.
Solid lines correspond to the exact Liouvillian eigenvalues obtained from direct diagonalization, while the red dashed line denotes the analytical prediction $\Gamma_{\mathrm{eff}}$ from Eq.~\eqref{eq-SM:HAE-LiouvGap}.
The white circles correspond to
exceptional points (EP), which represent bifurcations, convergence, or crossings of the Liouvillian eigenvalues~\cite{MullerExceptionalPoints2008,OzdemirParityTime2019,MiriExceptionalPoints2019,MingantiQuantumExceptional2019}.
Parameters: $\{\Omega,\delta,\gamma\}/2\pi=\{45,6.81,14.7\}\,\mathrm{MHz}$.
}
    \label{fig:SM-HAE}
\end{figure*}

Under the conditions established in the previous section for the stabilization of entanglement [see Eq.~\eqref{eq-SM:dark-state-conditions-driven-triplet}], the triplet state $|\Psi^+\rangle$ behaves as a virtual state, i.e., a state that is both energetically off-resonant and effectively decoupled from dissipative channels, as illustrated in Fig.~\ref{fig:SM-Energy-Diagrams}c.
As a consequence, the system becomes metastable and develops two well-separated dissipative timescales [see Fig.~\ref{fig:SM-HAE}a]: (i) a fast timescale governed by the collective decay rate, and (ii) a slow timescale associated with an effective relaxation rate, $\sim \Gamma_{\mathrm{eff}}^{-1}$.
This metastable behavior is illustrated in the Liouvillian spectrum~\cite{MacieszczakTheoryMetastability2016,MacieszczakTheoryClassical2021,BrownUnravelingMetastable2024}, with eigenvalues $\{ \Lambda_\mu,\ \mu=1,2,\ldots\}$ ordered by their real values, so that
$\text{Re}(\Lambda_\mu)\geq \text{Re}(\Lambda_{\mu+1}) $. 
The steady state corresponds to $\Lambda_1=0$ by Evans' theorem~\cite{EvansIrreducibleQuantum1977,EvansGeneratorsPositive1979}, while the second eigenvalue defines the Liouvillian gap~\cite{KesslerDissipativePhase2012}, which determines the relaxation timescale, $\tau_{\text{ss}}=1/|\text{Re}(\Lambda_2)|$. 
Metastability emerges when a large spectral separation exists between adjacent Liouvillian eigenvalues~\cite{MacieszczakTheoryMetastability2016,MacieszczakTheoryClassical2021,BrownUnravelingMetastable2024}, i.e.,  $|\text{Re}(\Lambda_m)|\gg |\text{Re}(\Lambda_{m+1})|$, as it occurs here between $\Lambda_2$ and $\Lambda_{\mu>2}$ [see Fig.~\ref{fig:SM-HAE}b].

This hierarchy of timescales enables the application of the \textit{hierarchical adiabatic elimination} (HAE) method introduced in Ref.~\cite{Vivas-VianaUnconventionalMechanism2022}. 
This framework is based on the successive application of adiabatic elimination procedures, allowing one to systematically derive effective reduced dynamics across different timescales and, when possible, obtain analytical expressions for both the dynamics and the Liouvillian gap, which are generally difficult to obtain by direct diagonalization of the Liouvillian.
The HAE method has been successfully applied to analyze the entanglement stabilization in coherently driven emitters coupled to chiral waveguides~\cite{Pichler2015}, where it provided analytical access to the dynamics and Liouvillian gap~\cite{Vivas-VianaUnconventionalMechanism2022,Vivas-VianaNonclassicalDrivenDissipative2025}. More recently, it was extended to a similar cavity-QED configuration~\cite{Vivas-VianaDissipativeStabilization2024}.
In this section, we apply the HAE method to obtain an analytical estimate of the entanglement-stabilization timescale in the present giant-atom system.

\subsection{Hierarchical adiabatic elimination (HAE)}

Here, we show step-by-step the application of the HAE technique. 
As assumed in the previous sections, we restrict the description of the system to the effective two-level approximation and neglect non-radiative decoherence processes, i.e., $\gamma_{\mathrm{nr}},\gamma_\phi=0$.
Under these assumptions, we consider the master equation introduced in Eq.~\eqref{eq-SM:GA-masterEq}. In the dimer basis and under the triplet dark-state conditions, it takes the form
\begin{equation}
    \frac{d \hat \rho^{(+)}_{\mathrm{TLS}}}{dt}=-i[\hat H^{(+)}_{\mathrm{TLS}},\hat \rho^{(+)}_{\mathrm{TLS}}]+\sum_{i,j}\frac{2\Gamma}{2}\mathcal{D}[\tilde \sigma_i,\tilde \sigma_j]\hat \rho_{\text{TLS}}^{(+)},
 \label{eq-SM:Effective-HAE-masterEq}   
\end{equation}
where we defined the ladder operators $\tilde \sigma_1\equiv |gg\rangle\langle\Psi^-|$ and  $\tilde \sigma_2\equiv - |\Psi^-\rangle\langle ee|$,  and the Hamiltonian
\begin{equation}
    \hat H^{(+)}_{\mathrm{TLS}}\equiv\frac{i\sqrt{2}\Omega}{2} (\tilde \sigma_1^\dagger+\tilde \sigma_2^\dagger-\text{H.c.})+\delta (|\Psi^+\rangle \langle \Psi^-|+\text{H.c.} ).
    \label{eq-SM:HAE-Hamiltonian}
\end{equation}
Here, we use $\Gamma\equiv 4\gamma$, as defined in Eq.~\eqref{eq:SM-GammaOmega-definition}, corresponding to the waveguide-induced decay rate of an individual giant atom. With this convention, the collective decay rate associated with the dissipative ladder $|ee\rangle \rightarrow |\Psi^-\rangle \rightarrow |gg\rangle$ is $2\Gamma$.
We note that Eq.~\eqref{eq-SM:Effective-HAE-masterEq} contains crossed dissipative contributions arising from the collective nature of the directional decay channels.
Within the terminology introduced by the HAE framework~\cite{Vivas-VianaUnconventionalMechanism2022}, we identify the states $\{|gg\rangle, |\Psi^-\rangle,|ee\rangle\}$ as the \textit{real states}, while the triplet state, $|\Psi^+\rangle$, acts as the \textit{virtual state} mediating the effective dynamics.
The method proceeds as follows:
\newline

\noindent
\textbf{First adiabatic elimination.} 
The first step of the hierarchical adiabatic elimination consists of adiabatically eliminating the virtual state under the assumption that the dynamics is effectively confined to the real subspace, $\mathcal{H}_R=\{|gg\rangle, |\Psi^-\rangle,|ee\rangle\}$, since, on a timescale $\sim 1/\Gamma$, the triplet state plays the role of a virtual state.
As a consequence, $|\Psi^+\rangle$ remains weakly populated during this stage of the evolution, and its main effect is to induce effective corrections within the real subspace. Therefore, we adiabatically eliminate the coherence terms involving the triplet state by imposing
    \begin{equation}
        \dot \rho_{i,\Psi^+}=0, \quad \text{with}\quad i=\{gg,\Psi^-,ee\},
    \end{equation}
and replacing the corresponding coherences by their stationary values. 
This procedure leads to the following reduced set of differential equations:
    \begin{subequations}
    \label{eq-SM:First-HAE}
        \begin{align}
            \dot \rho_{gg,\Psi^-}&\approx
            -\Gamma(\rho_{gg,\Psi^-}+2\rho_{\Psi^-,ee})
            +\frac{\Omega}{\sqrt{2}}(1+\rho_{gg,ee}-\rho_{\Psi^+,\Psi^+}-2\rho_{\Psi^-,\Psi^-}-\rho_{ee,ee})
            \notag\\
            &\hspace{1.1cm}+\delta^2\mleft[\frac{\sqrt{2}}{\Omega}(\rho_{\Psi^-,\Psi^-}-\rho_{\Psi^+,\Psi^+})+\frac{1}{\Gamma}(-\rho_{gg,\Psi^-}+\rho_{ee,\Psi^-})-\frac{2\Gamma}{\Omega^2}\rho_{gg,\Psi^-} \mright]
            , \\
            \dot \rho_{gg,ee}&\approx-\Gamma\rho_{gg,ee}-\frac{\Omega}{\sqrt{2}}(\rho_{gg,\Psi^-}+\rho_{\Psi^-,ee}), \\
            \dot \rho_{\Psi^-,\Psi^-}&\approx-\frac{2\sqrt{2}\delta^2}{\Omega}\text{Re}[\rho_{gg,\Psi^-}] + \sqrt{2}\Omega\,  \text{Re}[\rho_{gg,\Psi^-}+\rho_{\Psi^-,ee}] -2\Gamma (\rho_{\Psi^-,\Psi^-}-\rho_{ee,ee}), \\
            \dot \rho_{\Psi^-,ee}&\approx \frac{\delta^2}{\Gamma}( \rho_{\Psi^-,gg}- \rho_{\Psi^-,ee})+\frac{\Omega}{\sqrt{2}}(\rho_{gg,ee}+\rho_{ee,ee}-\rho_{\Psi^-,\Psi^-})-2\Gamma\rho_{\Psi^-,ee}, \\
            \dot \rho_{ee,ee}&\approx-\sqrt{2}\Omega \,\text{Re}[\rho_{\Psi^-,ee}]-2\Gamma \rho_{ee,ee},
        \end{align}
    \end{subequations}
where $\mathrm{Re}[(*)]$ and $\mathrm{Im}[(*)]$ denote real and imaginary parts, respectively. 
    The evolution equation governing the population of the triplet state after the first adiabatic elimination becomes
\begin{equation}
    \dot \rho_{\Psi^+,\Psi^+}\approx \frac{2\sqrt{2}\delta^2}{\Omega}\text{Re}[\rho_{gg,\Psi^-}].
    \label{eq-SM:HAE-Triplet-Diff-Eq}
\end{equation}
%
%

\noindent
\textbf{Second adiabatic elimination.} 
The longest characteristic timescale is governed by the slow evolution of the triplet state  $\rho_{\Psi^+,\Psi^+}(t)$, associated with the actual relaxation dynamics toward the steady state.
On this timescale, the \textit{real-state} variables relax almost instantaneously on the much faster dissipative timescale $\sim 1/\Gamma$.
Thus, the hierarchy of slow and fast variables is effectively reversed with respect to the first adiabatic elimination step: the real-subspace variables now act as fast dissipative degrees of freedom that adiabatically follow the slow evolution of $\rho_{\Psi^+,\Psi^+}(t)$.
Mathematically, the real-subspace variables relax toward a time-dependent quasistationary state parametrized by the triplet-state population.
Starting from the equations obtained after the first adiabatic elimination in Eq.~\eqref{eq-SM:First-HAE}, together with the dynamical equation for the virtual-state population in Eq.~\eqref{eq-SM:HAE-Triplet-Diff-Eq}, one obtains an equation of the form
   $\dot \rho_{\Psi^+,\Psi^+}(t)=f[\rho_{\Psi^+,\Psi^+}(t);\rho_{gg,\Psi^-}(t)]$.
The second adiabatic elimination consists of replacing $\rho_{gg,\Psi^-}(t)$ by its quasistationary value, parametrized by the slow variable $\rho_{\Psi^+,\Psi^+}(t)$. This yields a closed dynamical equation for the triplet-state population,
\begin{equation}
    \dot \rho_{\Psi^+,\Psi^+}(t)=f[\rho_{\Psi^+,\Psi^+}(t);\rho^{\text{ss}}_{gg,\Psi^-}[\rho_{\Psi^+,\Psi^+}(t)]]. 
\end{equation}%
To obtain the quasistationary solution  $\rho^{\text{ss}}_{gg,\Psi^-}[\rho_{\Psi^+,\Psi^+}(t)]$, we set to zero the equations of motion for the real-sector variables in Eq.~\eqref{eq-SM:First-HAE}, while treating $\rho_{\Psi^+,\Psi^+}(t)$ as a slowly varying parameter.
Equivalently, this amounts to solving the linear system $M.\vec{\rho}+\vec{b}=0$, where $M$ and $\vec b$ are determined by the coefficients of the reduced equations of motion (their explicit expressions are omitted here for brevity).
Substituting the resulting quasistationary expression for  $\rho_{gg,\Psi^-}^{\text{ss}}[\rho_{\Psi^+,\Psi^+}(t)]$ back into Eq.~\eqref{eq-SM:HAE-Triplet-Diff-Eq} leads to the effective differential equation
\begin{equation}
    \dot \rho_{\Psi^+,\Psi^+}(t)\approx \Gamma_{\text{eff}}\left[ \rho_{\Psi^+,\Psi^+}^{\text{ss}}-\rho_{\Psi^+,\Psi^+}(t)\right].
\end{equation}
Solving this equation yields the analytical expression for the evolution of the triplet-state population,%
\begin{equation}
    \rho_{\Psi^+,\Psi^+}(t)=\rho_{\Psi^+,\Psi^+}^{\text{ss}}\mleft(1-e^{-\Gamma_{\text{eff}}t}\mright),
    \label{eq:SM-HAE-Triplet-Pop}
\end{equation}
where $\rho_{\Psi^+,\Psi^+}^{\text{ss}}$ corresponds to the steady-state population of the triplet state
$   \rho_{\Psi^+,\Psi^+}^{\text{ss}}=|\alpha|^2/(1+|\alpha|^2),
$
and $\Gamma_{\text{eff}}$ defines the effective relaxation rate,
\begin{equation}
    \Gamma_{\text{eff}}=\frac{4\Gamma\delta^2(2\delta^2+\Omega^2)[2\Gamma^2+\delta^2+\Omega^2]}{4\delta^2(\Gamma^2+\delta^2)(2\Gamma^2+\delta^2) +(4\Gamma^4+6\Gamma^2\delta^2-4\delta^4)\Omega^2+(\Gamma^2+\delta^2)\Omega^4+3\Omega^6}.
    \label{eq-SM:HAE-LiouvGap}
\end{equation}
This quantity provides an analytical estimation of the Liouvillian gap and therefore determines the timescale of entanglement stabilization, $\tau_S$, through the relation
\begin{equation}
\tau_S\approx    \tau_{\text{HAE}}
    \sim
   \Gamma_{\mathrm{eff}}^{-1}.
   \label{eq:SM-Stab-HAE}
\end{equation}
This analytical expression for the stabilization timescale, in combination with the triplet-state population in Eq.~\eqref{eq:SM-HAE-Triplet-Pop}, provides a simple framework to determine the parameter region that simultaneously maximizes the degree of entanglement and minimizes the stabilization time.
Using the experimental parameters $\{\Omega,\delta,\gamma\}/2\pi=\{45,6.81,14.7\}\,\mathrm{MHz}$, this analysis predicts a dark-state generation time of $\tau_S \approx \Gamma_\mathrm{eff}^{-1} = \SI{167}{\nano\second}$, in good agreement with the experimentally measured $\tau_S = \SI{159}{\nano\second}$ shown in Fig.~\ref{fig:fig5}b.
We emphasize that the present analytical estimate of the relaxation timescale, $\tau_{\mathrm{HAE}}$, obtained using the HAE method~\cite{Vivas-VianaUnconventionalMechanism2022}, provides a more accurate prediction than those originally proposed in Refs.~\cite{Pichler2015,Soro2022}, as discussed in Ref.~\cite{Vivas-VianaNonclassicalDrivenDissipative2025}.

\begin{figure*}[b!]
    \centering
\includegraphics[width=1.0\textwidth]{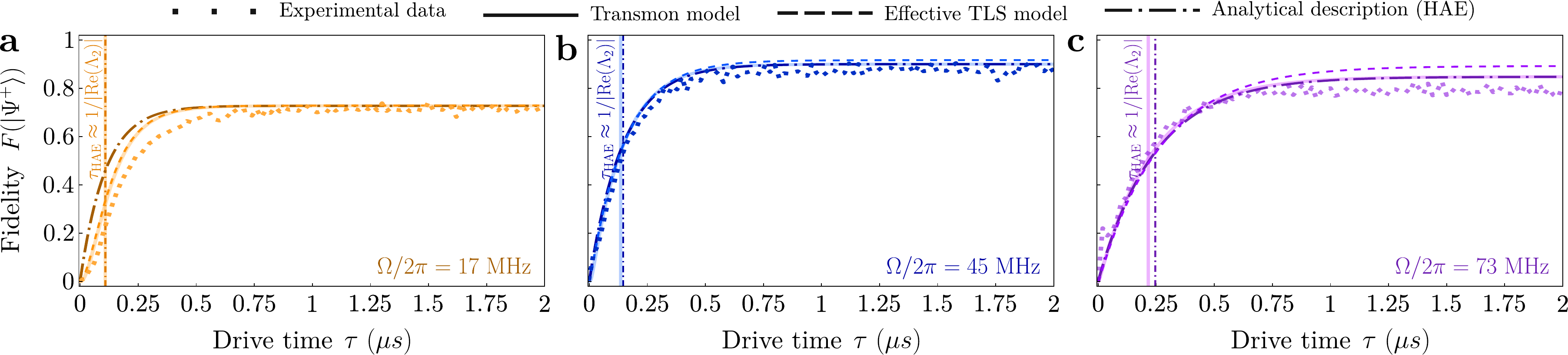}
    \caption{
\textbf{Entanglement stabilization. }
Time evolution of the fidelity to the triplet state, $F(|\Psi^+\rangle)$, for the three driving strengths shown in Fig.~\ref{fig:fig5}d, $\Omega/2\pi=\{17,45,73\}\,\mathrm{MHz}$, from left to right, respectively.
Solid lines correspond to numerical simulations of the full transmon model [Eq.~\eqref{eq-SM:Full-masterEq}], dashed lines to the effective two-level-system (TLS) model [Eq.~\eqref{eq-SM:Effective-masterEq}], and dot-dashed lines to the analytical prediction obtained from the HAE method [Eq.~\eqref{eq:SM-HAE-Triplet-Pop}]. For the analytical curves, the steady-state fidelity is taken from the full transmon simulations.
Symbols denote the experimental data.
The vertical solid lines indicate the characteristic stabilization timescale extracted from the full transmon model via the Liouvillian gap, while the vertical dot-dashed lines show the analytical prediction $\tau_{\mathrm{HAE}}$ obtained from Eq.~\eqref{eq:SM-Stab-HAE}.
Parameters are given in Tab.~\ref{tab:parameters}.
}
    \label{fig:SM-Experimental-Fidelity-AVV}
\end{figure*}

Once the analytical expression for the triplet-state population is known, the remaining density-matrix elements can be straightforwardly obtained, since they are explicit functions of $\rho_{\Psi^+,\Psi^+}(t)$.
The validity of the HAE method is illustrated in Fig.~\ref{fig:SM-HAE}, which compares the analytical predictions with the exact dynamics for the populations $\{gg,\Psi^\pm\}$ (panel a) and the Liouvillian gap (panel b). Both cases show a good agreement, illustrating the accuracy of the effective description.
Although the present treatment neglects higher transmon levels and non-radiative decoherence processes, the resulting analytical expressions provide valuable physical insight into the metastable dynamics and the mechanism of entanglement stabilization. They are also expected to accurately describe the two–giant-atom system in the regime of larger anharmonicity, where the two-level approximation is well justified. 

Alternatively, we can use the analytical solution provided by the HAE method together with the numerical steady-state values obtained from the transmon model [Eq.~\eqref{eq-SM:Full-masterEq}]. This hybrid approach accounts for the reduction in the stationary triplet-state fidelity at larger drive strengths arising from the finite anharmonicity of the giant atoms and additional decoherence channels that are not captured by the ideal TLS model [Eq.~\eqref{eq-SM:GA-masterEq}]. This is illustrated in Fig.~\ref{fig:SM-Experimental-Fidelity-AVV}, where we compare the experimental time evolution of the triplet-state fidelity, $F(|\Psi^+\rangle)$, for the three driving strengths $\Omega/2\pi=\{17,45,73\}\,\mathrm{MHz}$.
For $\Omega\ll\Lambda$ (panels a and b), the effective TLS model [Eq.~\eqref{eq-SM:Effective-masterEq}] accurately reproduces the transmon dynamics. Moreover, the analytical prediction obtained from the HAE method [Eq.~\eqref{eq:SM-HAE-Triplet-Pop}], using the steady-state value from the transmon model, closely matches the full numerical simulations and experimental data for all three driving strengths.
In particular, the HAE captures both the transient dynamics and the relaxation timescale toward the steady state, demonstrating that the effective descriptions developed here provide a reliable framework for analyzing the experimental results and understanding the mechanism of entanglement stabilization.

\section{Dark-state lifetime}

The mechanism responsible for the generation of entanglement is intrinsically driven-dissipative, i.e., the entangled state is stabilized in the steady state under continuous coherent driving.
Consequently, once the driving field is switched off, the system eventually relaxes toward the ground state.
This occurs because the dark state is a superposition of $|\Psi^+\rangle$ and $|gg\rangle$, while the state $|\Psi^-\rangle$ remains coherently coupled to $|\Psi^+\rangle$ through the qubit-qubit detuning $\delta$, thereby forming an effective dissipative channel toward $|gg\rangle$.
Nevertheless, this relaxation process can be avoided by rapidly tuning the giant atoms into their decoherence-free subspace before the entanglement is lost~\cite{Kockum2018}.
To achieve this, the emitters must be tuned on a timescale shorter than the dark-state lifetime, $\tau_D$.
In this section, we derive an analytical expression for the dark-state lifetime in the absence of coherent driving.

\begin{figure*}[b!]
    \centering
\includegraphics[width=0.85\textwidth]{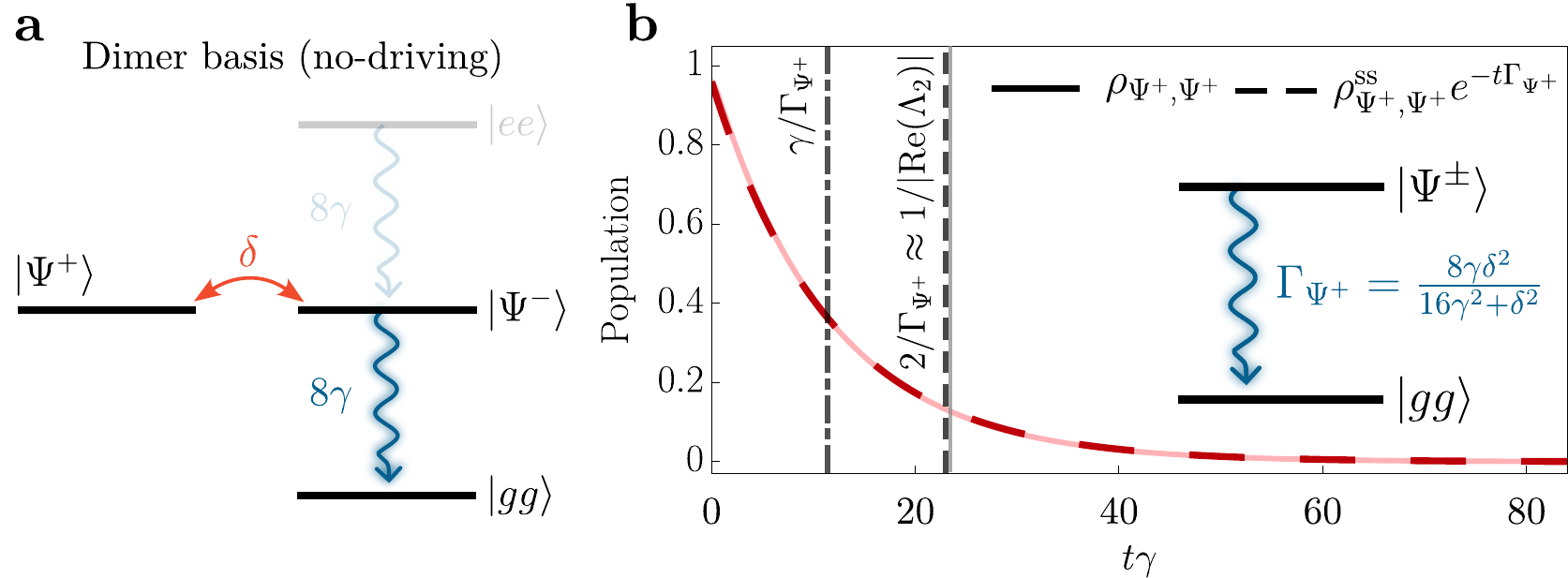}
    \caption{
\textbf{Dark-state lifetime. }%
\textbf{a)} 
Energy-level diagram in the dimer basis after switching off the coherent drive.
In this regime, the doubly excited state can be neglected and the dynamics becomes restricted to the subspace $\{|gg\rangle,|\Psi^+\rangle,|\Psi^-\rangle\}$.  
The triplet and singlet states are coherently coupled through the qubit-qubit detuning $\delta$, while dissipation occurs via the decay channel $|\Psi^-\rangle\rightarrow|gg\rangle$.
\textbf{b)}
Exponential decay of the undriven system initialized in the triplet dark state, $\hat \rho(0)=|D^+\rangle \langle D^+|$.
The solid black line corresponds to numerical simulations of Eq.~\eqref{eq-SM:AE-singlet-masterEQ}, while the dashed red line denotes the analytical prediction $\rho_{\Psi^+,\Psi^+}(t)\approx \rho_{\Psi^+,\Psi^+}^{\text{ss}}e^{-t\Gamma_{\Psi^+}}$, where $\rho_{\Psi^+,\Psi^+}^{\mathrm{ss}}$  is the initial triplet-state population and $\Gamma_{\Psi^+}$ is the effective decay rate.
The vertical dashed line indicates the Liouvillian gap obtained numerically (solid line) and analytically (dashed line).
The vertical dot-dashed line indicates the dark-state lifetime.
The inset illustrates the effective two-level description formed by $\{|gg\rangle,|\Psi^+\rangle\}$ with effective decay rate $\Gamma_{\Psi^+}$.
Parameters: $\{\Omega,\delta,\gamma\}/2\pi=\{45,6.81,14.7\}\ \text{MHz}$.
}
    \label{fig:SM-DarkState-Lifetime}
\end{figure*}

\subsection{Adiabatic elimination of the singlet state}

\begin{figure*}[t!]
    \centering
    \includegraphics[width=0.95\textwidth]{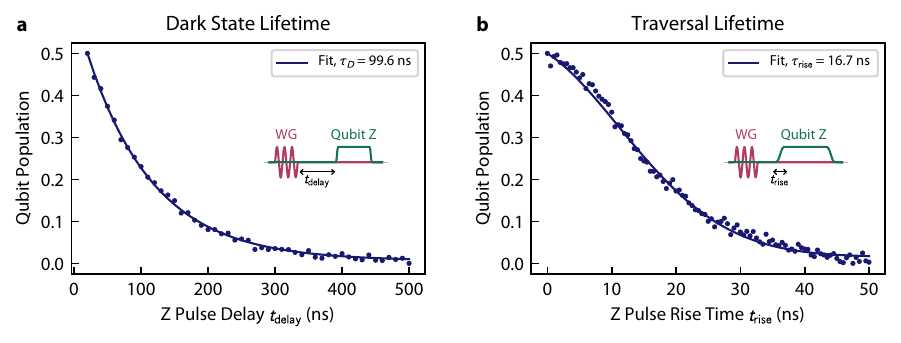}
    \caption{\textbf{Dark-state lifetime and traversal lifetime measurements. a)} Excited-state population as a function of time delay $t_\mathrm{delay}$ between the waveguide drive pulse and the Z pulse that decouples the qubit from the waveguide for readout. This measurement characterizes the exponential decay and lifetime ($\tau_D = \SI{99.6}{\nano\second}$) of the stabilized dark state in the absence of driving. \textbf{b)} Excited-state population as a function of the rise time $t_\mathrm{rise}$ of the fast flux Z pulse. At both $\omega_\mathrm{super}$ and $\omega_\mathrm{sub}$ the entanglement dissipation to the waveguide is suppressed by interference. However, the qubits must dynamically traverse across the spectrum during the Z pulse. The qubits experience significant dissipation at frequencies between $\omega_\mathrm{super}$ and $\omega_\mathrm{sub}$, as shown in Fig.~\ref{fig:fig3}a. In order to preserve the entanglement fidelity, the Z-pulse rise time $t_\mathrm{rise}$ must be as short as possible; in the experiment, we use a Gaussian ramp of width $t_\mathrm{rise} = \SI{2}{\nano\second}$. We characterize this traversal lifetime, fitting to an exponential-Gaussian function $P(t_\mathrm{rise}) \propto e^{-t_\mathrm{rise}/\tau_\mathrm{exp}} e^{-t_\mathrm{rise}^2/2\tau_\mathrm{rise}^2}$, extracting $\tau_\mathrm{rise} = \SI{16.7}{\nano\second}$ and $\tau_\mathrm{exp} = \SI{49.3}{\nano\second}$. 
    }
    \label{fig:dark_lifetime}
\end{figure*}

When the coherent drive is switched off and the initial state is prepared in the triplet dark state [Eq.~\eqref{eq-SM:dark-state}], the dynamics becomes restricted to the subspace $\{|gg\rangle,|\Psi^+\rangle,|\Psi^- \rangle\}$ since the doubly excited state $|ee\rangle$ can no longer be populated [see Fig.~\ref{fig:SM-DarkState-Lifetime}a].
Under these conditions, the reduced dynamics of the system, described by the density matrix $\hat \xi$, is governed by the master equation%
\begin{equation}
    \frac{d\hat \xi}{dt}=-i[\hat H_{\text{TLS},0}^{(+)},\hat \xi]+\frac{8\gamma}{2}\mathcal{D}[|gg\rangle\langle\Psi^-|]\hat \xi,
    \label{eq-SM:AE-singlet-masterEQ}
\end{equation}
where the Hamiltonian corresponds to Eq.~\eqref{eq-SM:HAE-Hamiltonian} with $\Omega=0$,
\begin{equation}
    \hat H_{\text{TLS},0}^{(+)}\equiv \delta (|\Psi^+\rangle \langle \Psi^-|+|\Psi^-\rangle \langle \Psi^+|).
\end{equation}
In this regime, the singlet state $|\Psi^-\rangle$ acts as an intermediate state mediating the relaxation from $|\Psi^+\rangle$ toward $|gg\rangle$ through the coherent coupling $\delta$.
Following the same strategy introduced in Sec.~\ref{sec:HAE}, we adiabatically eliminate the singlet state by imposing $\dot \xi_{i,\Psi^-}=0$ with $i=\{gg,\Psi^+,\Psi^-\}$, and and replacing the corresponding coherences and populations by their stationary values. 
This procedure yields the effective equations
\begin{equation}
        \dot \xi_{\Psi^+,\Psi^+}\approx-\frac{8\gamma\delta^2}{16\gamma^2+\delta^2}\xi_{\Psi^+,\Psi^+},\quad \text{and}\quad 
        \dot \xi_{gg,\Psi^+}\approx-\frac{\delta^2}{4\gamma}\xi_{gg,\Psi^+}.
\end{equation}
That is, after switching off the drive, the dynamics effectively reduces to a two-level system formed by $\{|\Psi^+\rangle,|gg\rangle\}$ [see inset in Fig.~\ref{fig:SM-DarkState-Lifetime}b], with effective relaxation and dephasing rates
\begin{equation}
    \Gamma_{\Psi^+}\equiv \frac{8\gamma\delta^2}{16\gamma^2+\delta^2} \quad \text{and}\quad \gamma_{\Psi^+}\equiv\frac{\delta^2}{4\gamma}.
\end{equation}
Since the dark state is a coherent superposition of $|\Psi^+\rangle$ and $|gg\rangle$, its decay is governed by the relaxation of the triplet component. Therefore, the dark-state lifetime can be identified with the lifetime of the triplet population,
\begin{equation}
    \tau_{D}\equiv \tau_{\Psi^+}
    \approx
    1/\Gamma_{\Psi^+}.
    \label{eq:SM-lifetime}
\end{equation}
For the parameters considered in this work, this effective relaxation rate dominates over the additional non-radiative decoherence channels, $\gamma_{\mathrm{nr}}$ and $\gamma_\phi$, and therefore sets the relevant timescale over which the qubit frequencies must be tuned into the decoherence-free subspace to preserve the entangled state.  
Using Eq.~\eqref{eq:SM-lifetime}, we estimate the dark-state lifetime to be $\tau_D= \SI{100}{\nano\second}$, in good agreement with the experimentally extracted value $\tau_D= \SI{99.6}{\nano\second}$ shown in Fig.~\ref{fig:dark_lifetime}a.

\section{Conditions for efficient entanglement stabilization}
\label{sec-SM:conditins-entanglement}

For the mechanism of entanglement generation presented in Secs.~\ref{Sec-SM-Unconventional} and~\ref{sec:HAE} to efficiently stabilize the triplet state in long-time limit, several physical conditions must be satisfied.
In this section, we analyze these conditions and discuss the main detrimental mechanisms that reduce the achievable degree of entanglement: 
(i) finite driving strength, (ii) leakage to higher transmon levels, and (iii) additional decoherence channels (non-radiative decay and pure dephasing).
In this section, to quantify the degree of entanglement, we use the fidelity~\cite{JozsaFidelityMixed1994} between the steady state and the triplet state
\begin{equation}
    F(|\Psi^+\rangle)\equiv \langle \Psi^+| \hat \rho |\Psi^+\rangle.
\end{equation}

\begin{figure*}[t!]
    \centering
\includegraphics[width=1.0\textwidth]{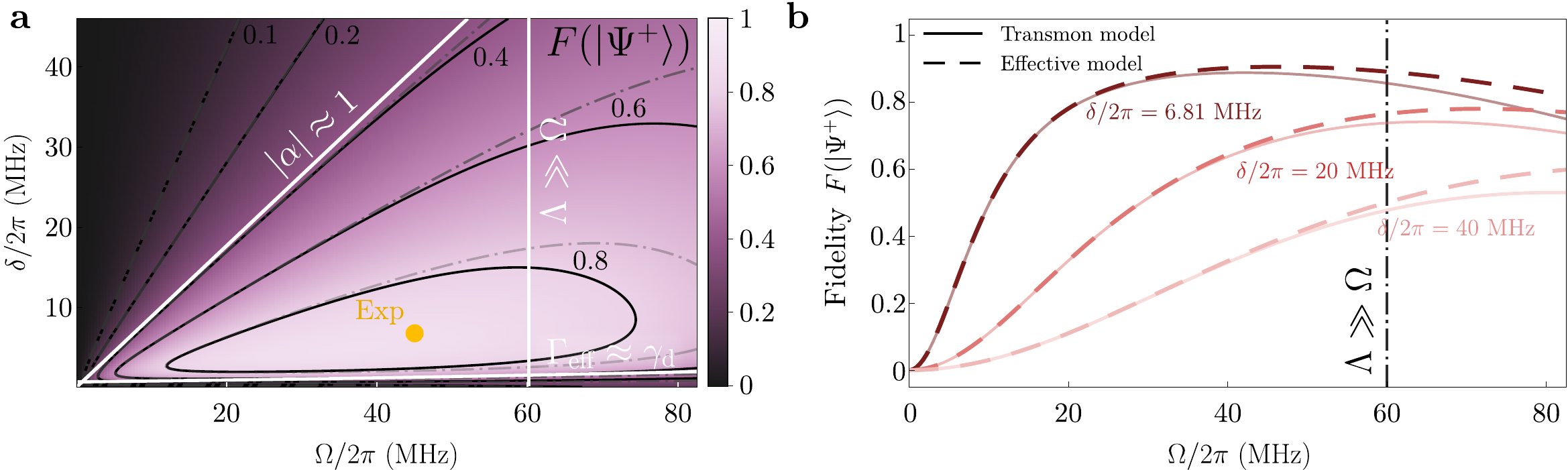}
    \caption{
    \textbf{Conditions for efficient entanglement stabilization. }
    \textbf{a)} 
    Steady-state triplet fidelity $F(|\Psi^+\rangle)$ as a function of the drive amplitude $\Omega$ and qubit-qubit detuning $\delta$. The black contours show lines of constant fidelity obtained from numerical simulations of the full master equation [Eq.~\eqref{eq-SM:Full-masterEq}], while the gray dot-dashed contours correspond to simulations using the effective TLS model [Eq.~\eqref{eq-SM:Effective-masterEq}]. The white solid lines mark the conditions outlined in the text and bound the regions of formation of entanglement.
    The yellow point marks the experimental operating point used for the measurements shown in Fig.~\ref{fig:fig5}.
    \textbf{b)}
    Triplet fidelity as a function of $\Omega$ for fixed detunings $\delta/2\pi=\{6.81,20,40\}~\mathrm{MHz}$. This panel corresponds to horizontal cuts of the fidelity map in panel a. 
 }
    \label{fig:SM-Validity-AVV}
\end{figure*}

Following a similar analysis to that of Ref.~\cite{Vivas-VianaDissipativeStabilization2024}, efficient triplet-state stabilization requires the following conditions:
\begin{enumerate}[label=(\roman*)]
    \item \textit{Virtualization of the triplet state.}    
    The target triplet state $|\Psi^+\rangle$ must be effectively protected from dissipation and energetically off-resonant from the other states [see Sec.~\ref{Sec-SM-Unconventional} and Fig.~\ref{fig:SM-Energy-Diagrams}c].
    Dissipative decoupling is achieved by satisfying the dark-state conditions in Eq.~\eqref{eq-SM:dark-state-conditions-driven-triplet}. In addition, the drive must be sufficiently strong compared with the qubit-qubit detuning,
    \begin{equation}
        \Omega\gg\delta.
    \end{equation}
    Equivalently, in terms of the control parameter, $\alpha=i\sqrt{2}\Omega/2\delta$, this condition becomes $|\alpha|\gg 1$. 
    As discussed previously, in the limit $|\alpha|\rightarrow\infty$, the dark state approaches the triplet state, $|D^+\rangle\approx|\Psi^+\rangle$.
    \newline

    The largest contribution to the total infidelity in the experiment comes from the fact that the drive strength is finite. A finite value of $\Omega$ implies a finite value of $|\alpha|$, so the dark state contains a non-negligible ground state population. From Eq.~\eqref{eq-SM:dark-state}, the corresponding infidelity is given, for identical qubits, by the ground-state weight in the dark state, $|\langle gg|D\rangle|^2=1/(1+|\alpha|^2)$.
    Increasing the drive strength reduces this contribution. However, arbitrarily large drives are not possible in practice, since strong driving also enhances leakage to higher transmon levels [see condition (iii)].
    \item \textit{Fast stabilization timescale.}
    The characteristic rate at which the system is dissipatively-driven into the entangled state must be larger than the decay rate of the triplet state derived in Eq.~\eqref{eq-SM:dimer-decay-rates},
    \begin{equation}
        \Gamma_{\text{eff}}\gg \Gamma_+.
    \end{equation}
    When the dark-state conditions are exactly fulfilled, this requirement is automatically satisfied because the target state is perfectly decoupled from dissipation. In practice, however, small phase mismatches in the giant-atom interference conditions might leave a residual decay rate $\Gamma_+$, in which case the above inequality becomes essential.
    This condition also imposes a lower bound on the qubit-qubit detuning. Indeed, from Eq.~\eqref{eq-SM:HAE-LiouvGap}, taking $\delta\rightarrow 0$ gives $\Gamma_{\mathrm{eff}}\rightarrow 0$, so that the entanglement-formation timescale diverges. Therefore, although the condition $\Omega/\delta\gg1$ is theoretically beneficial, the detuning cannot be made arbitrarily small.
    \newline
    
    When additional decoherence channels are included, such as non-radiative decay and pure dephasing, two further requirements must be satisfied:
\begin{equation}
      \Gamma_{\text{Q}i}(\omega_i)\gg \gamma_{\text{d}}, \quad \text{and }\quad \tau_{\text{HAE}} \ll \gamma_{\text{d}}^{-1}, \quad \text{with}\quad \gamma_{\text{d}}=\{\gamma_{\text{nr}},\gamma_\phi\}.
\end{equation}
    The first condition states that the engineered decay rate induced by the giant-atom configuration must dominate over the extra decoherence channels.
    The second condition requires the entanglement-stabilization timescale $\tau_{\mathrm{HAE}}$ to be shorter than the timescale associated with additional decoherence.
    In the present setup, these conditions are generally well satisfied: the non-radiative decay and pure-dephasing rates are in the kHz regime, whereas the engineered dissipative rates are in the MHz regime. Consequently, these additional noise channels give the smallest contribution to the total infidelity in the experiment. Since these noise channels are purely dissipative, they do not qualitatively alter the stabilization mechanism as long as entanglement is generated faster than the corresponding decoherence timescales.
\item \textit{Two-level-system approximation.}
    To justify the effective two-level description in which higher energy levels do not play any role, the dynamics must remain confined to the two-level subspace. This requires the transmon anharmonicity to be larger than the drive amplitude, $\Lambda\gtrsim\Omega$.
More stringently, achieving very high fidelities requires
\begin{equation}
    \Lambda\gg\Omega,
\end{equation}
so that transitions involving the third transmon level are far off-resonant.
\newline

    The second-largest contribution to the infidelity in the experiment comes from including the third transmon level $|f\rangle$ in the dynamics. This level reduces the triplet-state fidelity through two main mechanisms: (i) population can leak out of the computational subspace into $|f\rangle$, and (ii) virtual transitions through $|f\rangle$ generate energy shifts and collective pure-dephasing terms, as derived in Sec.~\ref{sec-SM:AE-transmon}.
    These effects can imbalance the two components of the triplet state and thereby reduce the overlap with $|\Psi^+\rangle$.
    Increasing the anharmonicity suppresses these effects by making transitions involving $|f\rangle$ more off-resonant. Thus, while condition (i) favors large $\Omega/\delta$, condition (iii) imposes an upper bound on the drive strength. The optimal regime is therefore determined by the available separation of scales between the qubit-qubit detuning $\delta$, the drive amplitude $\Omega$, and the anharmonicity $\Lambda$.
\end{enumerate}
Figure~\ref{fig:SM-Validity-AVV} illustrates how these three conditions delimit the region of high triplet-state fidelity. Conditions (i) and (ii) set the lower and upper constraints associated with the detuning $\delta$, while condition (iii) limits the maximum usable drive strength through leakage to higher transmon levels.
The individual contributions to the triplet-state infidelity in the experiment are summarized in Table~\ref{tab:infidelity}a. They are evaluated at the parameter point that yields the largest fidelity in Fig.~\ref{fig:fig5}d, namely $\Omega/2\pi= \SI{45}{\mega\hertz}$. The values were obtained from simulations with Eq.~\eqref{eq-SM:Full-masterEq} by introducing the detrimental effects successively, in the order shown from top to bottom in the table. Therefore, each entry quantifies the incremental reduction in fidelity caused by the corresponding additional physical effect.

 Capacitively-shunted flux qubits~\cite{Yan2016} could help us reach a larger anharmonicity ($\Lambda/2\pi=\SI{1}{\giga\hertz}$) and waveguide dissipation ($\gamma/2\pi=\SI{25}{\mega\hertz}$), which in turn would allow us to now benefit from larger driving amplitudes. Table~\ref{tab:infidelity}b shows how with these upgraded parameters, estimated fidelities well above \SI{90}{\percent} are reached, mainly via reduction of the first two infidelity sources when compared to Table~\ref{tab:infidelity}a.

\begin{table}[h!]
\centering

\begin{tabular}{cc}
\textbf{(a)} Experimental point
&
\hspace{1.5cm}\textbf{(b)} Improved parameters
\\[0.5em]

\begin{tabular}{lc}
\hline
\hline
Source & Triplet-state infidelity \\
\hline
Finite drive strength & $4.7\%$ \\
$|f\rangle$ state & $2.9\%$ \\
Pure dephasing ($\gamma_\phi$) & $1.8\%$ \\
Non-radiative decay ($\gamma_\text{nr}$) & $0.5\%$ \\
\hline
\hline
\end{tabular}
&
\hspace{1.5cm}
\begin{tabular}{lc}
\hline
\hline
Source & Triplet-state infidelity \\
\hline
Finite drive strength & $2.2\%$ \\
$|f\rangle$ state & $0.9\%$ \\
Pure dephasing ($\gamma_\phi$) & $1.9\%$ \\
Non-radiative decay ($\gamma_\text{nr}$) & $0.5\%$ \\
\hline
\hline
\end{tabular}
\end{tabular}
\caption{
\textbf{Summary of the main contributions to the triplet-state infidelity for the driven-dissipative stabilization protocol.}
\textbf{(a)} Experimental operating point, $\Omega/2\pi=\SI{45}{\mega\hertz}$.
\textbf{(b)} Improved parameter set, $\{\Omega,\gamma,\Lambda\}/2\pi=\{85,25,1000\}~\mathrm{MHz}$.
The listed values are obtained by successively adding each detrimental effect to the master-equation simulation, so that each entry represents the incremental loss of fidelity associated with that effect.
}
\label{tab:infidelity}
\end{table}

\section{Characterization of the degree of entanglement between giant atoms}

\subsection{Entanglement measure: concurrence}

In this work, in addition to looking at the fidelity with the triplet state, we also characterize the degree of entanglement between the giant atoms via the concurrence $\mathcal{C}$, a widely used measure of bipartite entanglement~\cite{WoottersEntanglementFormation1998,WoottersEntanglementFormation2001,PlenioIntroductionEntanglement2007,HorodeckiQuantumEntanglement2009}. 
For the particular case of two two-level systems, the concurrence admits a closed-form expression given by
\begin{equation}
    \mathcal{C}=\text{max}[0,\sqrt{\lambda_1}-\sqrt{\lambda_2}-\sqrt{\lambda_3}-\sqrt{\lambda_4}],
    \label{eq-SM:concurrence}
\end{equation}
%
%
where the $\lambda_i$ are the eigenvalues---in decreasing order---of the matrix
\begin{equation}
    \lambda=\text{eigenvalues}[\hat \rho (\hat \sigma_y\otimes \hat \sigma_y)\hat \rho^*(\hat \sigma_y\otimes \hat \sigma_y)],
\end{equation}
%
%
where $\hat \sigma_y$ is the Pauli Y matrix and $^*$ denotes complex conjugation. 
Since $\hat \sigma_y$ implements a spin-flip transformation~\cite{NielsenQuantumComputation2012,SakuraiModernQuantum2017}, the concurrence can be interpreted as quantifying the degree of invariance of the quantum state under simultaneous spin-flip of both qubits in their respective Bloch-sphere representations. Accordingly, $\mathcal{C}=0$ corresponds to separable states (e.g., vacuum or thermal states), while $\mathcal{C}=1$ corresponds to maximally entangled states (i.e., Bell states).

\subsection{Practical implementation}

As mentioned previously in Sec.~\ref{sec-SM:model}, the theoretical model [see Eq.~\eqref{eq-SM:Full-masterEq}] used to simulate the system and fit the experimental data considers the quantum emitters as transmon qubits with three energy levels. As a consequence, the expression of the concurrence in Eq.~\eqref{eq-SM:concurrence} cannot be directly applied, since the Hilbert space of each emitter has dimension larger than two. 
A possible alternative would be to employ other entanglement measures, such as the negativity~\cite{VidalComputableMeasure2002,PlenioLogarithmicNegativity2005,YangEntanglementPhotonic2025}, or the quantum state fidelity~\cite{JozsaFidelityMixed1994} to the triplet state $F(|\Psi^+\rangle)$, as previously used in Sec.~\ref{sec-SM:conditins-entanglement}.
Nevertheless, in order to have a direct comparison with previous experimental results~\cite{Shah2024, Irfan2025}, we employ the concurrence as our entanglement quantifier. This requires reducing the dimensionality of the density matrix.
As discussed in Sec.~\ref{sec-SM:AE-transmon}, the population of higher transmon levels is negligible, apart from small energy shifts and dissipative corrections. We therefore project the density matrix onto the two-level subspace and neglect the contribution of higher states, under the assumption that the state approximately preserves unit trace. 

Finally, we note that we perform quantum state tomography experimentally in the single-photon subspace, providing access to the reconstructed stationary state. Note that readout of the higher transmon states when the qubit frequency is set to $\omega_\mathrm{sub}$ is infeasible, as the detuned higher transitions are not protected from dissipation by destructive interference. As shown in Fig.~\ref{fig:fig5}b, the presence of entanglement can also be directly inferred from the structure of the density matrix.

%
%
%
%
\begin{figure*}[t!]
    \centering
    \includegraphics[width=\textwidth]{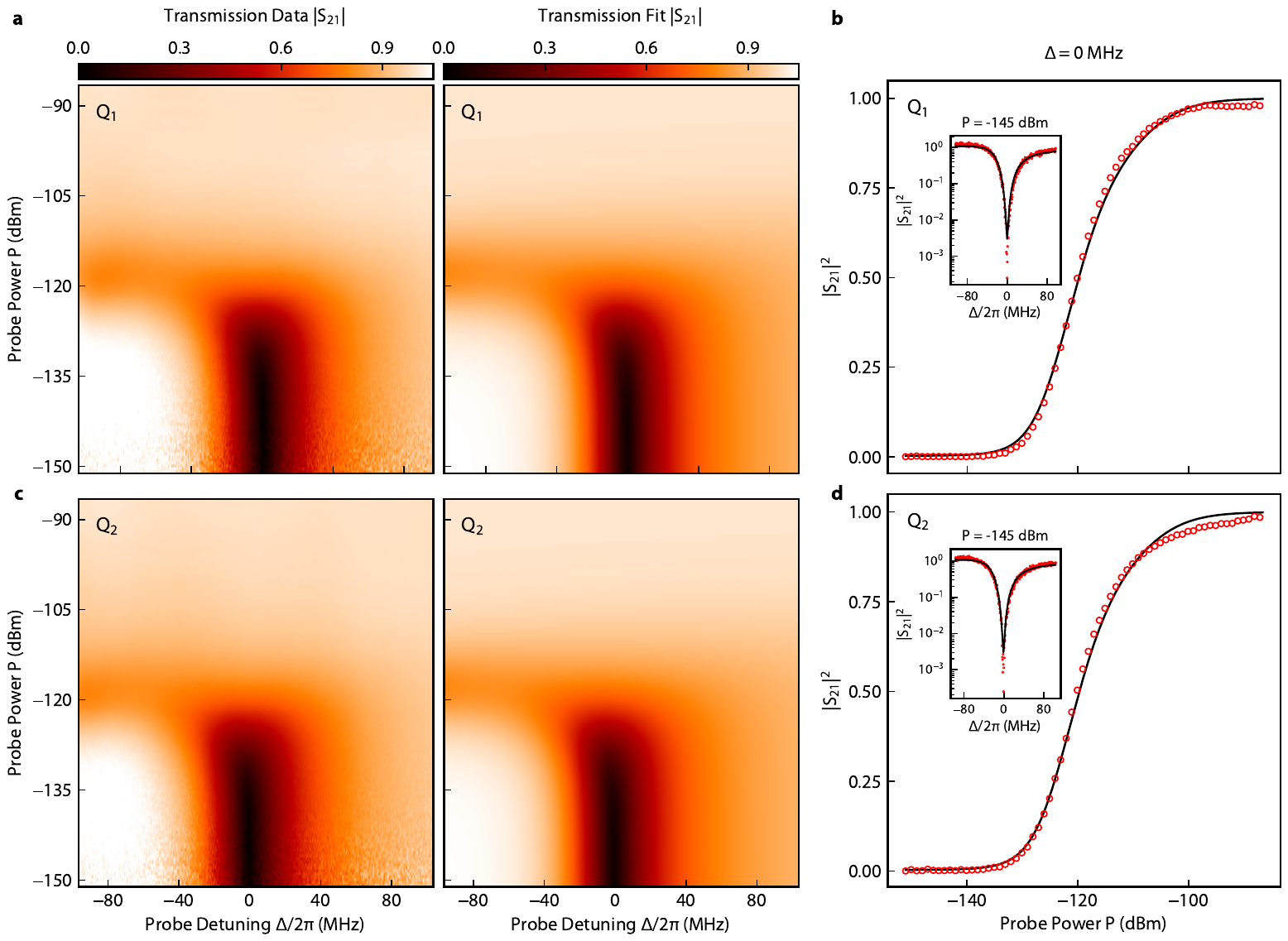}
    
    \caption{\textbf{Individual giant-atom spectroscopy.} \textbf{a, c)} Measured (left) and fitted (right) transmission spectrum $|\mathrm{S_{21}}|$ of a coherent probe incident on Q$_1$ (Q$_2$) through the waveguide as a function of the qubit-probe detuning $\Delta/2\pi$ and the probe power P. The qubits are operated individually at frequencies $\omega_\mathrm{super}/2\pi = \SI{6.208}{\giga\hertz}$. We model the qubits as three-level systems using a master-equation fitting approach. We extract the absolute power delivered to the qubit, the qubit-waveguide coupling rate $\Gamma/2\pi = \SI{59}{\mega\hertz}$, and the anharmonicity $\Lambda/2\pi = \SI{207}{\mega\hertz}$. \textbf{b, d)} Transmittance $|\mathrm{S_{21}}|^2$ as a function of probe power P at zero qubit-probe detuning ($\Delta/2\pi = \SI{0}{\mega\hertz}$). The transmittance begins near zero for low power, but reaches unity as the probe power saturates the qubit response. The measured data is plotted in red, and the simulated fit is plotted in black. The inset shows the Lorentzian frequency response of the qubit at probe power $P = \SI{-145}{\decibel\milli\relax}$. }
    \label{fig:qutrit_power_scan}
\end{figure*}




\section{Spectroscopy of single giant atoms}
\label{subsec:qutrit_power_scan}

In this and the following sections, we characterize the individual and collective giant atom properties using standard waveguide-QED input-output techniques. Particularly, we measure the elastic scattering of a coherent probe in the waveguide when the emitters are set to the superradiant frequency $\omega_\mathrm{super}/2\pi = \SI{6.208}{\giga\hertz}$.
At this frequency, constructive interference between the coupling points enhances the waveguide-induced decay rate of each individual giant atom, yielding $\Gamma= \Gamma_{\mathrm{Q}_i}(\omega_\mathrm{super})= 4\gamma$, as established by Eq.~\eqref{eq:giant_atom_dissipation-wo-non-radiative}. Consequently, for the purpose of analyzing the single-emitter scattering response, each giant atom can be equivalently described as an effective small atom with an enhanced decay rate $\Gamma$ into the waveguide. For low probe powers with average photon numbers $\ll 1$, the qubit acts as a mirror to single photons in the waveguide~\cite{Hoi2011, Astafiev2010, Hoi2013}.

The master equation for a driven emitter-waveguide system is given by~\cite{Mirhosseini2019}
\begin{equation}
    \frac{d\hat{\rho}_{\text{GA}}}{dt}  = -i\big[\hat{H}_{\text{GA}},\hat\rho_{\text{GA}}\big] + \frac{\Gamma + \gamma_\mathrm{nr}}{2} D\big[\hat a \big]\hat\rho_{\text{GA}} + \frac{\gamma_{\phi}}{4}D\big[2\hat a^\dag \hat a\big]\hat \rho_{\text{GA}},
    \label{eq-SM:single-GA}
\end{equation}
where $\hat a^\dag$ and $\hat a$ are the raising and lowering operators, and $\hat \rho_{\text{GA}}$ is the density matrix describing the single giant atom. We recall that $\gamma_{\phi}$ is the pure dephasing rate, and $\gamma_\mathrm{nr}$ is the rate of non-radiative decay to modes other than those in the waveguide.
%
%
We model the quantum emitter as a three-level system using the driven Hamiltonian (in the rotating frame)
\begin{equation}
    \hat H_{\text{GA}} = \Delta \hat a^\dag\hat a - \frac{\Lambda}{2}\hat a^\dag \hat a^\dag \hat a \hat a  -\frac{i}{2}(\Omega \hat a^\dag -\text{H.c.}  ) 
\end{equation}
where $\Delta \equiv \omega - \omega_{\mathrm{d}}$ is the emitter-drive detuning, $\Lambda$ is the qubit anharmonicity, and $\Omega = 2\sqrt{2\gamma P/\omega_\mathrm{d}}$ is the drive strength of the probe with power $P$. 
We note that this master equation is the single giant atom counterpart of Eq.~\eqref{eq-SM:Full-masterEq}; in this case, collective terms are absent because only one emitter is considered.
As discussed in Sec.~\ref{sec-SM:AE-transmon} and Sec.~\ref{sec-SM:conditins-entanglement}, the two-level approximation is not quantitatively sufficient for the present experiment. This is because the waveguide-induced decay rate is appreciable compared to the transmon anharmonicity, $\Gamma\approx\Lambda/3$, and the anharmonicity is not sufficiently large compared to the applied drive strength, $\Lambda\gtrsim\Omega$.

Using a rightward-propagating probe input, we model the radiation emitted from the giant atom as the output field from the right end of the waveguide via the input-output relations~\cite{GardinerQuantumNoise2004}
\begin{equation}
    \hat a_\mathrm{R}(t) = \hat a_\mathrm{R}^\textrm{in}(t) + \sqrt{\frac{\Gamma}{2}}\hat a(t),
\end{equation}
where $\hat a_{\mathrm R}^{\mathrm{in}}(t)$ and $\hat a_{\mathrm R}^{\mathrm{out}}(t)$ are the right-propagating input and output fields, respectively, and $\hat a(t)$ is obtained from the corresponding Heisenberg equation associated with Eq.~\eqref{eq-SM:single-GA}.
From master-equation simulations, we calculate the transmission amplitude $S_{21} = \langle\hat a_\mathrm{R}\rangle/\langle\hat a_\mathrm{R}^\textrm{in}\rangle$~\cite{Mirhosseini2019}. Transmission measurements as a function of probe power $P$ and detuning $\Delta$ are shown in Fig.~\ref{fig:qutrit_power_scan}. We perform a two-dimensional fitting routine using master equations, extracting $\Gamma/2\pi \approx \SI{59}{\mega\hertz}$ and $\Lambda/2\pi \approx \SI{207}{\mega\hertz}$ for each individual giant atom. The non-radiative decay $\gamma_\mathrm{nr}$ and pure dephasing $\gamma_{\phi}$ are negligible compared to $\Gamma$, and thus, they cannot be extracted from this measurement. We also calibrate the absolute power of microwave tones incident on the qubits.

\begin{figure*}[b!]
    \centering
    \includegraphics[width=\textwidth]{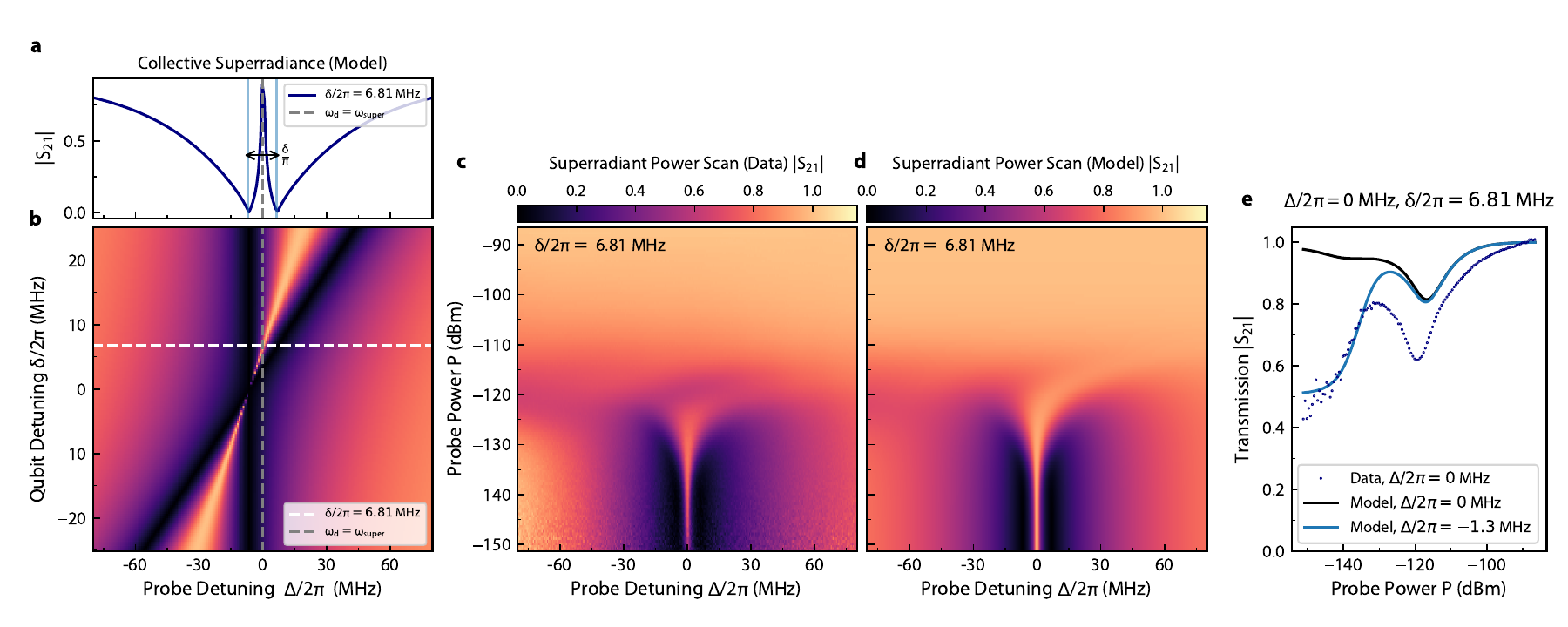}
    \caption{\textbf{Collective superradiance spectroscopy. a,b)} Master equation simulation of the superradiant giant-atom crossing, mirroring the transmission data presented in Fig.~\ref{fig:fig4}. The full crossing near the superradiant frequency is shown in panel \textbf{b}, while the 1D linecut in panel~\textbf{a} focuses on the qubit-drive detuning $\delta/2\pi = 6.81$ MHz that was chosen as an initial operating point for the driven-dissipative entanglement protocol.
    \textbf{c,d)} Power-sweep transmission measurements (\textbf{c}) and master-equation simulation (\textbf{d}) of the two detuned giant atoms centered around $\omega_\mathrm{super}/2\pi = \SI{6.208}{\giga\hertz}$.
    \textbf{e)} Transmission measurements of the two detuned giant atoms as a function of power under a resonant probe $\Delta/2\pi = \SI{0}{\mega\hertz}$. The data and the master-equation model agree qualitatively, though the transmission signal is weaker at lower power. Due to finite frequency resolution in the measurement, the power sweep data appears to be slightly detuned from $\Delta/2\pi = \SI{0}{\mega\hertz}$ upon comparison to the model at small detunings. We also suspect non-Markovian $1/f$ flux noise causes discrepancy between the data and the model at low power, which results in qubit-frequency instability that is averaged over during the continuous-wave measurement. For reference, the entanglement experiment is performed using an input power of approximately $P = \SI{-125}{\decibel\milli\relax}$, where the transmission magnitude dips slightly in the power sweep. The dip in transmission indicates that the probe is pumping population into the dark, protected state. 
    }
    \label{fig:super_sims}
\end{figure*} 

\section{Collective superradiance spectroscopy}
\label{subsec:collective_spec}

Following the same approach as in the previous section, we characterize the two-giant-atom system using transmission spectroscopy near the superradiant frequency, $\omega_\mathrm{super}/2\pi = \SI{6.208}{\giga\hertz}$.
At this operating point, each giant atom can be described as an effective point-like emitter with enhanced waveguide decay rate $\Gamma_i=\Gamma_{\mathrm{Q}i}(\omega_\mathrm{super})=4\gamma$. 
In the following model, we therefore treat the system as two approximately resonant emitters coupled to the waveguide, with $\omega_1\approx \omega_2 \approx \omega_\mathrm{super}$.

The present system is exactly described as in Sec.~\ref{sec-SM:model}, via the master equation presented in Eq.~\eqref{eq-SM:Full-masterEq} [similar to those presented in Refs.~\cite{Lalumiere2013, Gheeraert2020}]. We further impose the triplet dark-state conditions of Eq.~\eqref{eq-SM:dark-state-conditions-driven-triplet}, i.e., $\phi(\omega_{\text{super}})=2\pi m$ and $\theta(\omega_\text{super})=(2m+1)\pi$, with $m\in \mathbb{N}$. Under these conditions, the coherent and dissipative parameters reduce to: $\Gamma_{1/2}\approx4\gamma$, $\Gamma_{\text{coll}}=-2\gamma$, $g=0$, and $\Omega=\Omega_1=-\Omega_2$.
From this description, the input-output relation for the right-propagating field is~\cite{GardinerQuantumNoise2004}%
\begin{equation}
    \hat a_\mathrm{R}(t) =   \hat a_\mathrm{R}^\textrm{in}(t) - \sqrt{\frac{\Gamma}{2}}\hat a_1(t) + \sqrt{\frac{\Gamma}{2}}\hat a_2(t),
\end{equation}
where $\hat a_{\mathrm R}^{\mathrm{in}}(t)$ and $\hat a_{\mathrm R}^{\mathrm{out}}(t)$ are the right-propagating input and output fields, respectively, and $\hat a_i(t)$ is obtained from the corresponding Heisenberg equations associated with Eq.~\eqref{eq-SM:Full-masterEq}.%

Using the single-qubit parameters extracted in the previous section, we numerically simulate the superradiant crossing transmission ($|S_{21}| = |\langle \hat a_\mathrm{R}\rangle/\langle \hat a_\mathrm{in}\rangle |$) measurement presented in Fig.~\ref{fig:fig4}, as shown in Fig.~\ref{fig:super_sims}. We also use the master equation model to study the spectroscopy of the two qubits detuned by $\delta/2\pi = \SI{6.81}{\mega\hertz}$ as a function of probe power $P$ (Fig.~\ref{fig:super_sims}c and d show the data and simulation, respectively). 

In Fig.~\ref{fig:super_sims}e, we focus on the one-dimensional linecut at zero probe detuning ($\Delta/2\pi = \SI{0}{\mega\hertz}$) as a function of power. Comparison to the model at small probe detunings shows that the power-sweep data is also shifted along the frequency axis---an artifact of the limited frequency resolution of the measurement. The data and model agree qualitatively, despite the weak transmission signal at low power likely caused by averaging continuous-wave measurements over slow qubit-frequency fluctuations from non-Markovian $1/f$ flux noise. At high power, power broadening of the interference feature washes out these fluctuations, and eventually the probe saturates the qubits and the signal returns to unity transmission. The dip in transmission near $P= \SI{-125}{\decibel\milli\relax}$---the power used to entangle the qubits in the experiment---indicates the presence of the dark, protected state.

\section{Flux-pulse predistortion}

Fast flux pulses are required to tune the qubit frequencies mid-protocol from $\omega_\mathrm{super}$, where they are strongly coupled to the waveguide, to $\omega_\mathrm{sub}$, where they are decoupled. Flux pulses originating from room-temperature arbitrary waveform generators (AWGs) experience significant distortion along the signal path, resulting in dynamical qubit-frequency shifts that can limit protocol fidelity. To counteract these distortions, we adapt the predistortion calibration routines discussed in Refs.~\cite{sung2020, An2025, Guo2024, Li2024}.
\begin{table}[h!]
\centering
\renewcommand{\arraystretch}{1.5} 
\setlength{\tabcolsep}{3.5pt}       
\begin{tabular}{l | cc | cc | cc | cc | cc | cc}
\hline
\hline
Qubit & $A_{11}$ (mV) & $\tau_{11}$ ($\mu$s) & $A_{21}$ (mV) & $\tau_{21}$ (ns) & $A_3$ (mV) & $\tau_3$ (ns) & $A_4$ (mV) & $\tau_4$ (ns) & $A_5$ (mV) & $\tau_5$ (ns) & $A_6$ (mV) & $\tau_6$ (ns) \\
\hline
Q$_1$ & 4.180 & 32.7 & -11.626 & 452.7 & -9.277 & 44.4 & -4.807 & 41.4 & -2.539 & 22.8 & -2.350 & 46.1 \\
Q$_2$ & 1.119 & 74.6 & -9.544 & 717.7 & -14.565 & 60.0 & -3.483 & 38.4 & -2.412 & 28.6 & -1.717 & 30.2 \\
\hline
\hline
\end{tabular}
\caption{\textbf{Flux-pulse predistortion pole parameters.} Long- and short-time transients represented by a pair of pole amplitude $A_k$ and time constant $\tau_k$, as defined in Eq.~\eqref{eq:poles}. The poles are presented in order of identification, after three to four rounds of corrected voltage-transient measurements.}
\label{tab:predistortion}
\end{table}

The giant atoms experience protection from decoherence into the waveguide within a limited qubit frequency range of approximately \SI{10}{\mega\hertz} centered around $\omega_\mathrm{sub}$, as shown in Fig.~\ref{fig:fig3}b. Consequently, the predistortion routine must be carefully chosen to maintain qubit coherence throughout the calibration sequence. We implement the calibration with the qubit frequencies idling at $\omega_\mathrm{sub}$ to leverage the protection from the waveguide. To minimize sensitivity to first-order flux noise, we design the qubit-frequency tunability to align $\omega_\mathrm{sub}$ with the lower flux sweet spot at $\Phi_\mathrm{ext} = 0.5\Phi_0$.

We operate the qubit as a sensor to characterize the system's step response. Using two $Y_{\pi/2}$ pulses, we capture the dynamic qubit-frequency transient in response to a square-voltage pulse at the flux-line input. Following Ref.~\cite{An2025}, we model the time-transient signal as the convolution of linear time-invariant (LTI) transfer functions $h_k(t)$. The relationship between the input flux pulse $V_\mathrm{in}(t)$ and the signal reaching the qubit $V_\mathrm{out}(t)$ is given by
\begin{equation}
V_\mathrm{out}(t) = V_\mathrm{in}(t) \ast h_1(t) \ast h_2(t) \ast \dots \ast h_k(t).
\end{equation}
By taking the Laplace transform $V(s) = \int_0^{\infty} V(t)e^{-st}dt$, where $s = \sigma + j\omega$ is the complex frequency, the relationship simplifies to a product of transfer functions $H_k(s)$:
\begin{equation}
V_\mathrm{out}(s) = V_\mathrm{in}(s) \prod_{k} H_k(s).
\end{equation}
The total transfer function $H_\mathrm{total}(s) = V_\mathrm{out}(s)/V_\mathrm{in}(s)$ characterizes the flux-line step response. Solving for $H_\mathrm{total}(s)$ enables the application of the inverse filter at the AWG $[V_\mathrm{AWG}(s) = V_\mathrm{target}(s)/H_\mathrm{total}(s)]$, ensuring the signal reaching the qubit approximates an ideal square pulse in the time domain.

In the time domain, we analytically model the step response of each transfer-function component, $s_k(t) = h_k(t) \ast u(t)$, as a combination of exponential decay and damped oscillation:
\begin{equation}
s_k(t) = u(t) \mleft[ 1 + A_{1k}e^{-t/\tau_{1k}} + A_{2k}e^{-t/\tau_{2k}}\cos\mleft(\omega_{2k} t+ \phi \mright) \mright],
\label{eq:poles}
\end{equation}
where $u(t)$ is the unit-step function of duration $T$. The exponential term $A_{1k}e^{-t/\tau_{1k}}$ corresponds to a simple real pole at $s = -1/\tau_{1k}$ in the Laplace domain. The damped oscillation term represents a pair of complex conjugate poles at $s = -1/\tau_{2k} \pm j\omega_{2k}$. Voltage transients can originate from $RC$ time constants due to the charging and discharging of parasitic capacitances along the coaxial flux lines, the skin effect, and also impedance mismatches and reflections along the flux line.

To measure the flux transient, we operate the qubit as a sensor: perturbations in flux voltage results in qubit frequency shifts. We apply a square flux pulse of duration $T$ and, after a variable delay $t_d$, implement a Ramsey-type sequence of two $Y_{\pi/2}$ pulses (\SI{30}{\nano\second} width each) separated by \SI{10}{\nano\second}. We measure the qubit excited-state population as a function of the AWG DC offset. The dynamical frequency shift is captured by comparing the population signals for an input voltage pulse of amplitudes \SI{0}{\volt} and \SI{1}{\volt}; the voltage shift between the population signals given $t_d$ represents $V_\mathrm{out}(t_d)$.

\begin{figure*}[t!]
    \centering
    \includegraphics[width=\textwidth]{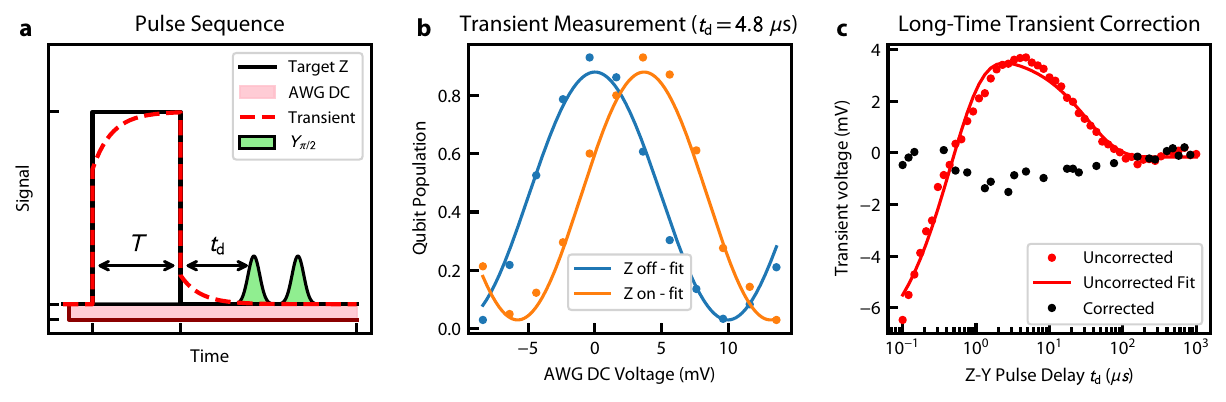}
    \caption{\textbf{Fast-flux-pulse transient calibration. a)} Predistortion calibration sequence using two $Y_{\pi/2}$ pulses. To capture the voltage transient, we apply a square pulse of duration $T = \SI{100}{\micro\second}$ to the flux line of the qubit using an arbitrary waveform generator (AWG). The pulse that arrives at the qubit resembles the red dashed line. Following a time delay of $t_\mathrm{d}$ after the Z pulse, we implement two $Y_{\pi/2}$ pulses spaced by \SI{10}{\nano\second} to detect qubit-frequency shifts in response to the Z pulse. For each delay $t_\mathrm{d}$, we sweep the DC bias of the AWG to map the qubit frequency shift to voltage. 
    \textbf{b)} Example transient measurement for delay $t_\mathrm{d} = \SI{4.8}{\micro\second}$. We measure the Q$_2$ excited-state population as a function of DC voltage when the Z pulse is off and then when it is on. Detuning the qubit from the drive using the AWG DC bias results in a cosine-like interference pattern in excited-state population. The voltage shift between the on and off cosine fits gives the voltage transient at that delay.
    \textbf{c)} Example long-time transient correction. We measure the long-time voltage transients for $t_\mathrm{d} = \SI{100}{\nano\second}$ to \SI{1}{\milli\second} (red dots). Fitting to the model in Eq.~\eqref{eq:Vout_model}, we identify a real and complex pair of poles. We construct the transfer function, predistort the square pulse accordingly, and repeat the voltage transient measurement (black dots), showing the suppression of the transient. 
    }
    \label{fig:predistortion}
\end{figure*}

We first identify long-time transients by repeating these measurements for delay times $t_d = \SI{100}{\nano\second}$ to \SI{1}{\milli\second}. We fit the resulting population signal shifts to the difference of two step responses:
\begin{equation}
V_\mathrm{out}(t_d) = s_k(t_d + T) - s_k(t_d),
\label{eq:Vout_model}
\end{equation}
extracting the pole parameters in Eq.~\eqref{eq:poles}. The pulse sequence and example calibration results are shown in Fig.~\ref{fig:predistortion}. We iteratively repeat this process, correcting the transients as they are revealed, to identify all long-time transients. After correcting long-time transients with a digital filter, we identify short-time transients ($t_d = \qtyrange{1}{100}{\nano\second}$). This multi-stage calibration ensures the qubit frequency remains within the \SI{10}{\mega\hertz} coherence protection window, maintaining high protocol fidelity.

To further suppress long-term transients and ensure signal stability across repeated measurement shots, we implement a net-zero flux-pulse scheme (see Fig.~\ref{fig:fig5}a). Immediately following the readout pulse, we apply an additional flux pulse of equal duration but opposite amplitude to the flux pulse. This ensures that the time-integral of the flux bias over a single experimental cycle is zero. By neutralizing the net flux, we mitigate long-term drifts in the qubit frequency that could otherwise persist despite the predistortion calibration.

\section{Readout and Purcell-filter design}

In dispersive readout, the two main parameters that determine readout performance are the resonator decay rate $\kappa$ into the feedline and the dispersive shift $\chi$ between the resonator and qubit. The optimal signal‑to‑noise ratio (SNR) in the long‑time limit is achieved for $|2\chi|/\kappa$ = 1 \cite{blais2021circuit}. However, good readout SNR has also been demonstrated for lower ratios \cite{walter2017rapid}, providing a margin for targeting error. Additionally, readout speed scales with $\chi$ and $\kappa$. To achieve a readout speed much faster than qubit decoherence, we target a minimum $\kappa/2\pi\gtrsim1.6~$MHz and $|2\chi|/\kappa=1\pm0.5$.

\begin{figure*}[t!]
    \centering
    \includegraphics[width=0.9\textwidth]{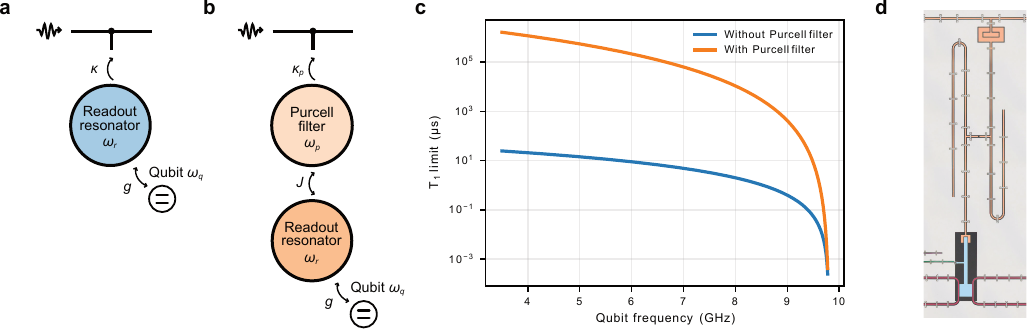}
    \caption{\textbf{Purcell filter scheme.} \textbf{a)} Qubit mode coupled to unfiltered readout resonator mode. \textbf{b)} Qubit mode coupled to readout resonator mode coupled to a Purcell filter mode. \textbf{c)} Comparison of the $T_1$ limit seen by the qubit for a) and b) respectively, for the same effective $\kappa/2\pi=|2\chi|/2\pi=\SI{1.6}{\mega\hertz}$. \textbf{d)} Layout of the readout resonator-Purcell system on the device. 
    }
    \label{fig:Purcell_scheme}
\end{figure*}

A resonator also introduces an undesired relaxation channel to the qubit, through a process known as the Purcell effect \cite{PurcellResonanceAbsorption1946, haroche2006exploring}. The Purcell rate increases with $\chi$ and $\kappa$, providing a natural upper bound to these two parameters before the readout resonator limits the qubit $T_1$ time. Intuitively, the transmission of a qubit excitation through the resonator and into the readout waveguide (feedline) is largest near the resonance frequency and suppressed according to a Lorentzian lineshape further away. For a qubit dispersively coupled to a single readout resonator (see Fig.~\Ref{fig:Purcell_scheme}a), the inherited $T_1$ limit is given by \cite{blais2021circuit, sete2015quantum}
\begin{equation}
    1/T_{1,\text{np}} = \kappa\frac{g^2}{\Delta_{qr}^2},
    \label{app_read:T1_no_purcell}
\end{equation}
with $g$ the coupling strength between the qubit and readout resonator, and $\Delta_{qr}=(\omega_q-\omega_r)$, where $\omega_q$ and $\omega_r$ are the frequencies of the qubit and resonator, respectively. Note that $\chi$ and $g$ are related by the expression \cite{blais2021circuit}
\begin{equation}
\chi=-\frac{g^2\Lambda}{\Delta_{qr}(\Delta_{qr}-\Lambda)},
\label{Eq:chi}
\end{equation}
with $\Lambda$ the anharmonicity of the qubit. 

In this experiment, the qubits are designed to have a relatively large $g/2\pi= \SI{380}{\mega\hertz}$ and operate at a minimum of $\Delta_{qr}/2\pi= \SI{3.6}{\giga\hertz}$. With $\kappa/2\pi=\SI{1.6}{\mega\hertz}$, we find a limit of $T_{1,p}=\SI{8.9}{\micro\second}$, which is prohibitive. In Fig.~\ref{fig:Purcell_scheme}c, we plot the $T_1$ limit versus Qubit frequency and find it to be low across the entire spectrum. To achieve our target $T_{1,p}> \SI{1}{\milli\second}$, while maintaining the same effective $\kappa$, we employ Purcell filters~\cite{reed2010fast,jeffrey2014fast,bronn2015broadband,sete2015quantum, walter2017rapid, heinsoo2018rapid, swiadek2024enhancing}. This filter is an additional linear resonator coupled between the readout resonator and the feedline, as shown in Fig.~\ref{fig:Purcell_scheme}b. Intuitively, a qubit excitation would need to propagate through both resonators in series to decay into the feedline, resulting in the transmission at the qubit frequency being strongly suppressed. This results in the modification of the Purcell decay rate seen by the qubit \cite{sete2015quantum}:
\begin{equation}
    1/T_{1,p} = \kappa_p\frac{J^2}{\Delta_{qp}^2+(\kappa_p/2)^2}\frac{g^2}{\Delta_{qr}^2} \approx\kappa_p\frac{J^2}{\Delta_{qp}^2}\frac{g^2}{\Delta_{qr}^2},
    \label{app_read:T1_purcell}
\end{equation}
where $J$ is the coupling strength between the Purcell filter and the readout resonator, $\Delta_{qp}=(\omega_q-\omega_p)$, $\omega_p$ is the frequency of the Purcell filter, $\kappa_p$ is the decay rate of the Purcell filter into the feedline, and we assume $\kappa_p\ll\Delta_{qp}$. The main contributing factor to the enhancement of $T_{1,p}$ is the additional $\Delta_{qp}^{-2}$ factor, as the filter provides a second Lorentzian suppression. For typical ranges of $\kappa_p$ and $J$, the $T_1$ limit typically gains two to five orders of magnitude, while maintaining the same effective $\kappa_{\text{eff}}$ which governs the readout dynamics (see Fig.~\ref{fig:Purcell_scheme}c).

To find the relevant dynamics of a system of two coupled resonators, we formulate the equations of motion of the coherent fields $\alpha$ and $\beta$ of the readout resonator and the Purcell filter, respectively~\cite{gardiner1985input, sete2015quantum, swiadek2024enhancing}: 
\begin{equation}
\begin{bmatrix}
\dot{\alpha}^{g/e} \\
\dot{\beta}^{g/e}
\end{bmatrix}
= 
-i
\begin{bmatrix}
\omega_r^{g/e} & J \\
J & \omega_p - i \kappa_p / 2
\end{bmatrix}
\begin{bmatrix}
\alpha^{g/e} \\
\beta^{g/e}
\end{bmatrix}.
\label{eq:eq_of_motion}
\end{equation} 
Quantities with the superscript $g/e$ are dependent on the qubit being in $g$ or in $e$. If $\kappa_p\gg J$, the Purcell filter can be adiabatically eliminated and we can solve for the quasisteady dynamics by setting $\dot{\beta}^{g/e}=0$. If the mode amplitude is of the form $\alpha(t)\sim e^{\lambda_\alpha t}$, with $\lambda_\alpha=-i\omega-\kappa/2$, then we find $\kappa_{\text{eff}}=-2\Re({\lambda_\alpha})$:
\begin{equation}
    \kappa_{\text{eff}}^{g/e}=\frac{4J^2}{\kappa_p}\frac{1}{1+[2(\omega_r^{g/e}-\omega_p)/\kappa_p]^2}.
    \label{app_read:keff_approx}
\end{equation}
%

To achieve higher levels of $T_1$ protection, however, we often choose $\kappa_p\approx J$, in which case Eq.~\eqref{app_read:keff_approx} is no longer valid. Instead, the readout resonator and Purcell filter hybridize, and the readout dynamics are governed by the normal modes of the system~\cite{heinsoo2018rapid,swiadek2024enhancing}. The frequency and linewidth of the normal modes are again recovered from the real and imaginary parts of the eigenvalues of Eq.~\eqref{eq:eq_of_motion}:
\begin{align}
\kappa^{g/e}_{1,2}
&=
\frac{\kappa_p}{2}
\mp
\mathrm{Im}
\sqrt{
\mleft(
\Delta_{rp}^{g/e}
+
\frac{i \kappa_p}{2}
\mright)^2
+
4J^2
},
\label{eq:keff}
\\[6pt]
\omega^{g/e}_{1,2}
&=
\frac{\omega_r^{g/e} + \omega_p}{2}
\pm
\frac{1}{2}
\mathrm{Re}
\sqrt{
\mleft(
\Delta_{rp}^{g/e}
+
\frac{i \kappa_p}{2}
\mright)^2
+
4J^2
},
\label{eq:weff}
\\[6pt]
2\chi_{1,2}
&=
\omega_{1,2}^e
-
\omega_{1,2}^g .
\label{eq:chieff}
\end{align}
with $\Delta_{rp}^{g/e}=\omega_r^{g/e}-\omega_p$ the qubit-state dependent detuning between the readout resonator and Purcell filter. Different combinations of bare mode parameters can result in the same normal mode linewidth, while leading to varying degrees of Purcell protection. However, for our target value of $\kappa_1/2\pi= \SI{1.6}{\mega\hertz}$, any reasonable choice of bare mode parameters will sufficiently suppress Purcell decay. As an example, $\kappa_p/2\pi= \SI{50}{\mega\hertz}$, $J/2\pi= \SI{4.4}{\mega\hertz}$, and $\Delta_{rp}/2\pi= \SI{0}{\mega\hertz}$ results in $\kappa_1/2\pi= \SI{1.6}{\mega\hertz}$ and a Purcell limit of $T_{1,{\text{p}}}= \SI{190}{\milli\second}$, which is five orders of magnitude larger than without a Purcell filter. These are the bare mode values used in Fig.~\ref{fig:Purcell_scheme}c. Note, that, while these normal modes are superpositions of both the Purcell filter and readout resonator, one of them is typically more readout-like and the other Purcell-like, especially if they are not perfectly hybridized. We will refer to the readout-like mode as mode 1.

When designing readout resonators Purcell filters, it is important to design for robustness to expected parameter mistargeting. In practice, achieving $\Delta_{rp}/2\pi= \SI{0}{\mega\hertz}$ is challenging. A finite detuning reduces $\kappa_1$, so we intentionally overshoot our target $\kappa_1$. It is also hard to realize the targeted $\kappa_p$ as it is affected by reflections on the feedline arising from impedance mismatches, potentially even off the chip. These reflections can enhance or reduce $\kappa_p$ through interference. To prevent $\kappa_p$ from being too low, we target a large $\kappa_p= \SI{50}{\mega\hertz}$. We design for $\pm \SI{50}{\mega\hertz}$ variations on $\Delta_{rp}$ and $\times2$ variations in $\kappa_p$.

We apply a combination of microwave circuit simulation and finite-element-model (FEM) electro-magnetic simulations to find the correct coplanar waveguide lengths and capacitor sizes to realize our target parameters. See Fig.~\ref{fig:Purcell_scheme}d for the device layout of the readout circuitry. To extract the parameters of a design, we simulate the $S_{21}$ transmission through our feedline to which the readout-Purcell system is coupled and fit a model derived from the equations of motion for this particular circuit topology~\cite{heinsoo2018rapid,swiadek2024enhancing}: 
\begin{equation}
\mleft| S_{\text{2,1}}(\omega) \mright| = \mleft[ A + k(\omega - \omega_r) \mright] \times 
\mleft| 
\cos(\phi) - e^{i\phi} \frac{ \kappa_p(-2i\Delta_r) }{ 4J^2 + (\kappa_p - 2i\Delta_p)(-2i\Delta_r) } 
\mright|.
\label{eq:s21_model}
\end{equation}
Here, A is the amplitude, $k$ a linear slope in the spectrum centered at $\omega_r$, and $\phi$ a phase rotation, all three of which are typically caused by reflections on the feedline. The term $\Delta_p=\omega-\omega_p$ is the detuning between the drive frequency $\omega$ and the Purcell filter, and $\Delta_r=\omega-\omega_r$ is the detuning between the drive frequency and the readout resonator. The extracted bare mode parameters can be used to calculate the normal mode parameters using Eqs.~\eqref{eq:keff}, \eqref{eq:weff}, and \eqref{eq:chieff}. Alternatively, one can also simulate the eigenmodes of the system using circuit or FEM solvers and invert these expressions to retrieve the bare mode parameters. 

We target $g$ by simulating the capacitance matrix between the qubit and resonator and use the expression~\cite{blais2021circuit}
\begin{equation}
g \simeq \frac{C_g}{2\sqrt{C_q C_r}}\sqrt{\omega_q\omega_r},
\end{equation}
where $C_g$ is the coupling capacitance between the resonator and the qubit, $C_q$ is the mode capacitance of the qubit, and $C_r$ is the mode capacitance of our resonator. For a quarter-wave resonator it is given as $C_r=\pi/(4\omega_rZ_0)$~\cite{pozar2021microwave}, with $Z_0$ the characteristic impedance of the transmission line. We target $\chi= \SI{-1.6}{\mega\hertz}$.

\begin{figure*}
    \centering
    \includegraphics[width=0.9\textwidth]{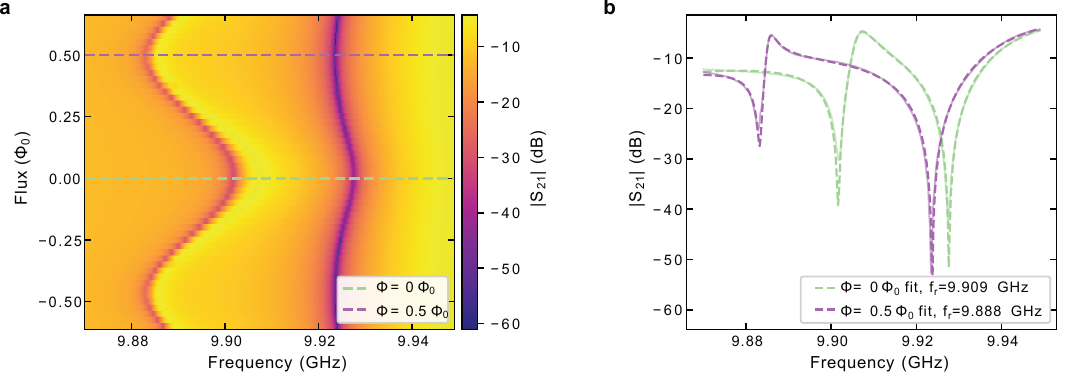}
    \caption{\textbf{Resonator fit.} \textbf{a)} Resonator spectroscopy versus qubit flux shows two tunable resonator normal modes. The thinner feature with more tuning can be attributed to the readout-like mode. \textbf{b)} Plot of the data slice at 0~$\Phi_0$ (green) and 0.5~$\Phi_0$ (magenta) with fits overlayed in dashed lines. The fit uses the model Eq.~\eqref{eq:s21_model} for hybridized readout resonator-Purcell systems and shows good agreement with the data.
    }
    \label{fig:resonator_fit}
\end{figure*}

In Fig.~\ref{fig:resonator_fit}a, we plot the resonator spectrum while sweeping the flux of the qubit. We see two modes, both shifting with flux. This is because both normal modes participate in the readout resonator, inheriting $\chi_{1,2}$. We can attribute the thinner mode with larger $\chi$ to the readout-like mode. Figure \ref{fig:resonator_fit}b shows the line cuts at 0~$\Phi_0$ and 0.5~$\Phi_0$, with a fit of Eq.~\eqref{eq:s21_model} overlayed, demonstrating good agreement. We extract the fitted parameters at the upper and lower sweetspots and use them together with the qubit frequencies at those points to infer $\chi_{1,2}$. All extracted parameters are listed in Table~\ref{tab:device_readout_summary}.

\begin{table*}[h!]
\centering
\footnotesize
\resizebox{\textwidth}{!}{%
\begin{tabular}{lcccccccccccccccc}
\hline
\hline

 & $\kappa_p$ & $J$ & $\omega_p$ & $\omega_r$ & $|\Delta_{rp}|$
 & $\kappa_1$ & $\kappa_2$ & $\omega_1$ & $\omega_2$
 & $2\chi$ & $2\chi_1$ & $2\chi_2$
 & $|2\chi_1|/\kappa_1$
 & $T_{1,\text{p}}$
 & $T_{1,\text{np}}$ \\

 & (MHz) & (MHz) & (GHz) & (GHz) & (MHz)
 & (MHz) & (MHz) & (GHz) & (GHz)
 & (MHz) & (MHz) & (MHz)
 &  & (ms) & ($\mu$s) \\

\hline

\\[-0.9em]

r1 target
& 50
& 22
& 
& 9.800
& $\lesssim 50$
& $\mathbf{\gtrsim 1.6}$
& 
& 
& 
& 
& 
& 
& $\mathbf{1 \pm 0.5}$
& $\mathbf{> 1.0}$
& \\

r1 upper
& 12.6
& 15.8
& 9.841
& 9.809
& 32
& $\mathbf{1.9}$
& 10.7
& 9.803
& 9.847
& -5.4
& -4.7
& -0.8
& $\mathbf{2.5}$
& $\mathbf{49.2}$
& 7.1 \\

r1 lower
& 11.10
& 22.19
& 9.840
& 9.794
& 46
& $\mathbf{1.5}$
& 9.6
& 9.785
& 9.849
& -2.7
& -2.3
& -0.4
& $\mathbf{1.5}$
& $\mathbf{117.7}$
& 17.7 \\

\\[-0.2em]

r2 target
& 50
& 13
& 
& 9.900
& $\lesssim 50$
& $\mathbf{\gtrsim 1.6}$
& 
& 
& 
& 
& 
& 
& $\mathbf{1 \pm 0.5}$
& $\mathbf{> 1.0}$
& \\

r2 upper
& 49.15
& 11.9
& 9.932
& 9.909
& 23
& $\mathbf{5.3}$
& 43.8
& 9.906
& 9.934
& -7.6
& -6.7
& -0.8
& $\mathbf{1.26}$
& $\mathbf{17.1}$
& 1.7 \\

r2 lower
& 50.4
& 13.4
& 9.932
& 9.888
& 44
& $\mathbf{3.21}$
& 47.19
& 9.885
& 9.934
& -3.8
& -3.5
& -0.23
& $\mathbf{1.10}$
& $\mathbf{52.3}$
& 5.5 \\

\hline
\hline
\end{tabular}%
}

\caption{
\textbf{Target and extracted parameters for resonator devices $r1$ and $r2$.}
Frequency-like quantities are reported as angular frequencies divided by $2\pi$. Bare mode parameters are extracted from the fits in Fig.~\ref{fig:resonator_fit}, and converted to normal mode parameters using Eqs.~\eqref{eq:keff}, \eqref{eq:weff}, and \eqref{eq:chieff}. Target parameters (in bold) were achieved well. Good readout is possible across the entire readout spectrum while providing sufficient $T_1$ protection. We also quote the equivalent $T_{1,\text{np}}$ limit if $\kappa_1$ and $2\chi_1$ were realized without a Purcell filter [Eq.~\eqref{app_read:T1_no_purcell}].
}
\label{tab:device_readout_summary}
\end{table*}

\noindent The main objectives of our design were $\kappa_1\gtrsim \SI{1.6}{\mega\hertz}$ and $|2\chi_1|/\kappa_1=1\pm0.5$, as well as $T_1> \SI{1}{\milli\second}$. These targets have been achieved well, despite underlying variations in $\kappa_p$ and $\Delta_{rp}$. Since the extracted $T_1$ limits are very large, future designs could afford the resonators to be placed much closer to the qubits. Alternatively, one could consider using one broadband Purcell filter coupled to both qubits, with a $\kappa_p/2\pi\gtrsim \SI{300}{\mega\hertz}$ being potentially easier to target. 

%% file: AVV_Refs.bib
@article{AgustiNonMarkovianThermal2026,
  title = {Non-{{Markovian}} Thermal Reservoirs for Autonomous Entanglement Distribution},
  author = {Agust{\'i}, Joan and Schneider, Christian M. F. and Fedorov, Kirill G. and Filipp, Stefan and Rabl, Peter},
  year = 2026,
  month = apr,
  journal = {Quantum},
  volume = {10},
  pages = {2066},
  issn = {2521-327X},
  doi = {10.22331/q-2026-04-15-2066},
  urldate = {2026-05-13},
  langid = {english}
}

@article{BraskImprovedQuantum2015,
  title = {Improved {{Quantum Magnetometry}} beyond the {{Standard Quantum Limit}}},
  author = {Brask, J. B. and Chaves, R. and Ko{\l}ody{\'n}ski, J.},
  year = 2015,
  month = jul,
  journal = {Physical Review X},
  volume = {5},
  number = {3},
  pages = {031010},
  issn = {2160-3308},
  doi = {10.1103/PhysRevX.5.031010},
  urldate = {2023-09-29},
  langid = {english}
}

@book{BreuerTheoryOpen2007,
  title = {The {{Theory}} of {{Open Quantum Systems}}},
  author = {Breuer, Heinz Peter and Petruccione, Francesco},
  year = 2007,
  month = jan,
  journal = {The Theory of Open Quantum Systems},
  publisher = {Oxford University Press},
  address = {Oxford},
  doi = {10.1093/acprof:oso/9780199213900.001.0001},
  isbn = {978-0-19-921390-0}
}

@article{BrownUnravelingMetastable2024,
  title = {Unraveling Metastable {{Markovian}} Open Quantum Systems},
  author = {Brown, Calum A. and Macieszczak, Katarzyna and Jack, Robert L.},
  year = 2024,
  month = feb,
  journal = {Physical Review A},
  volume = {109},
  number = {2},
  pages = {022244},
  issn = {2469-9926, 2469-9934},
  doi = {10.1103/PhysRevA.109.022244},
  urldate = {2025-01-23},
  langid = {english}
}

@article{CarmichaelPhotoelectronWaiting1989,
  title = {Photoelectron Waiting Times and Atomic State Reduction in Resonance Fluorescence},
  author = {Carmichael, H. J. and Singh, Surendra and Vyas, Reeta and Rice, P. R.},
  year = 1989,
  month = feb,
  journal = {Physical Review A},
  volume = {39},
  number = {3},
  pages = {1200--1218},
  issn = {0556-2791},
  doi = {10.1103/PhysRevA.39.1200},
  urldate = {2024-09-01},
  copyright = {http://link.aps.org/licenses/aps-default-license},
  langid = {english}
}

@book{Cohen-TannoudjiAtomPhotonInteractions1998,
  title = {Atom-{{Photon Interactions}}},
  author = {Cohen-Tannoudji, Claude and Dupont-Roc, Jacques and Grynberg, Gilbert},
  year = 1998,
  month = apr,
  journal = {Atom---Photon Interactions},
  publisher = {Wiley},
  address = {New York, USA},
  doi = {10.1002/9783527617197},
  isbn = {978-0-471-29336-1}
}

@article{CombesSLHFramework2017,
  title = {The {{SLH}} Framework for Modeling Quantum Input-Output Networks},
  author = {Combes, Joshua and Kerckhoff, Joseph and Sarovar, Mohan},
  year = 2017,
  month = may,
  journal = {Advances in Physics: X},
  volume = {2},
  number = {3},
  pages = {784--888},
  issn = {2374-6149},
  doi = {10.1080/23746149.2017.1343097},
  urldate = {2025-09-22},
  langid = {english}
}

@article{DiehlQuantumStates2008,
  title = {Quantum States and Phases in Driven Open Quantum Systems with Cold Atoms},
  author = {Diehl, S. and Micheli, A. and Kantian, A. and Kraus, B. and B{\"u}chler, H. P. and Zoller, P.},
  year = 2008,
  month = nov,
  journal = {Nature Physics},
  volume = {4},
  number = {11},
  pages = {878--883},
  issn = {1745-2473, 1745-2481},
  doi = {10.1038/nphys1073},
  urldate = {2022-06-22},
  langid = {english}
}

@article{EvansGeneratorsPositive1979,
  title = {The Generators of Positive Semigroups},
  author = {Evans, David E and {Hanche-Olsen}, Harald},
  year = 1979,
  month = may,
  journal = {Journal of Functional Analysis},
  volume = {32},
  number = {2},
  pages = {207--212},
  issn = {00221236},
  doi = {10.1016/0022-1236(79)90054-5},
  urldate = {2024-09-02},
  copyright = {https://www.elsevier.com/tdm/userlicense/1.0/},
  langid = {english}
}

@article{EvansIrreducibleQuantum1977,
  title = {Irreducible Quantum Dynamical Semigroups},
  author = {Evans, David E.},
  year = 1977,
  month = oct,
  journal = {Communications in Mathematical Physics},
  volume = {54},
  number = {3},
  pages = {293--297},
  issn = {0010-3616, 1432-0916},
  doi = {10.1007/BF01614091},
  urldate = {2024-09-02},
  copyright = {http://www.springer.com/tdm},
  langid = {english}
}

@article{EvansPhotonmediatedInteractions2018,
  title = {Photon-Mediated Interactions between Quantum Emitters in a Diamond Nanocavity},
  author = {Evans, R. E. and Bhaskar, M. K. and Sukachev, D. D. and Nguyen, C. T. and Sipahigil, A. and Burek, M. J. and Machielse, B. and Zhang, G. H. and Zibrov, A. S. and Bielejec, E. and Park, H. and Lon{\v c}ar, M. and Lukin, M. D.},
  year = 2018,
  month = nov,
  journal = {Science},
  volume = {362},
  number = {6415},
  pages = {662--665},
  issn = {0036-8075, 1095-9203},
  doi = {10.1126/science.aau4691},
  urldate = {2023-02-21},
  langid = {english}
}

@book{GarciaRipollQuantumInformation2022,
  title = {Quantum {{Information}} and {{Quantum Optics}} with {{Superconducting Circuits}}},
  author = {Garc{\'i}a Ripoll, Juan Jos{\'e}},
  year = 2022,
  month = jul,
  edition = {1},
  publisher = {Cambridge University Press},
  address = {Cambridge},
  doi = {10.1017/9781316779460},
  urldate = {2023-01-15},
  isbn = {978-1-316-77946-0 978-1-107-17291-3},
  langid = {english}
}

@book{GardinerQuantumNoise2004,
  title = {Quantum Noise: A Handbook of {{Markovian}} and Non-{{Markovian}} Quantum Stochastic Methods with Applications to Quantum Optics},
  shorttitle = {Quantum Noise},
  author = {Gardiner, C. W. and Zoller, P.},
  year = 2004,
  series = {Springer Series in Synergetics},
  edition = {3rd ed},
  publisher = {Springer},
  address = {Berlin ; New York},
  isbn = {978-3-540-22301-6},
  lccn = {QC446.2 .G37 2004}
}

@article{GiovannettiAdvancesQuantum2011,
  title = {Advances in Quantum Metrology},
  author = {Giovannetti, Vittorio and Lloyd, Seth and Maccone, Lorenzo},
  year = 2011,
  month = apr,
  journal = {Nature Photonics},
  volume = {5},
  number = {4},
  pages = {222--229},
  issn = {1749-4885, 1749-4893},
  doi = {10.1038/nphoton.2011.35},
  urldate = {2024-10-07},
  copyright = {http://www.springer.com/tdm},
  langid = {english}
}

@article{GiovannettiQuantumMetrology2006,
  title = {Quantum {{Metrology}}},
  author = {Giovannetti, Vittorio and Lloyd, Seth and Maccone, Lorenzo},
  year = 2006,
  month = jan,
  journal = {Physical Review Letters},
  volume = {96},
  number = {1},
  pages = {010401},
  issn = {0031-9007, 1079-7114},
  doi = {10.1103/PhysRevLett.96.010401},
  urldate = {2026-03-26},
  copyright = {http://link.aps.org/licenses/aps-default-license},
  langid = {english}
}

@article{Gonzalez-BallesteroTutorialProjector2024,
  title = {Tutorial: Projector Approach to Master Equations for Open Quantum Systems},
  shorttitle = {Tutorial},
  author = {{Gonzalez-Ballestero}, C.},
  year = 2024,
  month = aug,
  journal = {Quantum},
  volume = {8},
  pages = {1454},
  issn = {2521-327X},
  doi = {10.22331/q-2024-08-29-1454},
  urldate = {2024-09-02},
  langid = {english}
}

@article{Gonzalez-TudelaEntanglementTwo2011,
  title = {Entanglement of {{Two Qubits Mediated}} by {{One-Dimensional Plasmonic Waveguides}}},
  author = {{Gonzalez-Tudela}, A. and {Martin-Cano}, D. and Moreno, E. and {Martin-Moreno}, L. and Tejedor, C. and {Garcia-Vidal}, F. J.},
  year = 2011,
  month = jan,
  journal = {Physical Review Letters},
  volume = {106},
  number = {2},
  pages = {020501},
  issn = {0031-9007, 1079-7114},
  doi = {10.1103/PhysRevLett.106.020501},
  urldate = {2022-07-07},
  langid = {english}
}

@article{Gonzalez-TudelaLightMatter2024,
  title = {Light--Matter Interactions in Quantum Nanophotonic Devices},
  author = {{Gonz{\'a}lez-Tudela}, Alejandro and Reiserer, Andreas and {Garc{\'i}a-Ripoll}, Juan Jos{\'e} and {Garc{\'i}a-Vidal}, Francisco J.},
  year = 2024,
  month = jan,
  journal = {Nature Reviews Physics},
  volume = {6},
  number = {3},
  pages = {166--179},
  issn = {2522-5820},
  doi = {10.1038/s42254-023-00681-1},
  urldate = {2024-09-01},
  langid = {english}
}

@article{GrossSuperradianceEssay1982,
  title = {Superradiance: {{An}} Essay on the Theory of Collective Spontaneous Emission},
  shorttitle = {Superradiance},
  author = {Gross, M. and Haroche, S.},
  year = 1982,
  month = dec,
  journal = {Physics Reports},
  volume = {93},
  number = {5},
  pages = {301--396},
  issn = {0370-1573},
  doi = {10.1016/0370-1573(82)90102-8},
  urldate = {2022-12-02},
  langid = {english}
}

@article{HorodeckiQuantumEntanglement2009,
  title = {Quantum Entanglement},
  author = {Horodecki, Ryszard and Horodecki, Pawe{\l} and Horodecki, Micha{\l} and Horodecki, Karol},
  year = 2009,
  month = jun,
  journal = {Reviews of Modern Physics},
  volume = {81},
  number = {2},
  pages = {865--942},
  issn = {0034-6861},
  doi = {10.1103/RevModPhys.81.865}
}

@article{JozsaFidelityMixed1994,
  title = {Fidelity for {{Mixed Quantum States}}},
  author = {Jozsa, Richard},
  year = 1994,
  month = dec,
  journal = {Journal of Modern Optics},
  volume = {41},
  number = {12},
  pages = {2315--2323},
  issn = {0950-0340, 1362-3044},
  doi = {10.1080/09500349414552171},
  urldate = {2024-12-16},
  langid = {english}
}

@article{KastoryanoDissipativePreparation2011,
  title = {Dissipative {{Preparation}} of {{Entanglement}} in {{Optical Cavities}}},
  author = {Kastoryano, M. J. and Reiter, F. and S{\o}rensen, A. S.},
  year = 2011,
  month = feb,
  journal = {Physical Review Letters},
  volume = {106},
  number = {9},
  pages = {090502},
  issn = {0031-9007, 1079-7114},
  doi = {10.1103/PhysRevLett.106.090502},
  urldate = {2022-06-07},
  langid = {english}
}

@article{KesslerDissipativePhase2012,
  title = {Dissipative Phase Transition in a Central Spin System},
  author = {Kessler, E M and Giedke, G and Imamoglu, A and Yelin, S F and Lukin, M D and Cirac, J I},
  year = 2012,
  month = jul,
  journal = {Physical Review A},
  volume = {86},
  number = {1},
  pages = {012116},
  issn = {1050-2947},
  doi = {10.1103/PhysRevA.86.012116}
}

@article{KrausPreparationEntangled2008,
  title = {Preparation of Entangled States by Quantum {{Markov}} Processes},
  author = {Kraus, B. and B{\"u}chler, H. P. and Diehl, S. and Kantian, A. and Micheli, A. and Zoller, P.},
  year = 2008,
  month = oct,
  journal = {Physical Review A},
  volume = {78},
  number = {4},
  pages = {042307},
  issn = {1050-2947, 1094-1622},
  doi = {10.1103/PhysRevA.78.042307},
  urldate = {2022-06-22},
  langid = {english}
}

@article{LingenfelterExactResults2024,
  title = {Exact {{Results}} for a {{Boundary-Driven Double Spin Chain}} and {{Resource-Efficient Remote Entanglement Stabilization}}},
  author = {Lingenfelter, Andrew and Yao, Mingxing and Pocklington, Andrew and Wang, Yu-Xin and Irfan, Abdullah and Pfaff, Wolfgang and Clerk, Aashish A.},
  year = 2024,
  month = may,
  journal = {Physical Review X},
  volume = {14},
  number = {2},
  pages = {021028},
  issn = {2160-3308},
  doi = {10.1103/PhysRevX.14.021028},
  urldate = {2026-05-20},
  langid = {english}
}

@article{MacieszczakTheoryClassical2021,
  title = {Theory of Classical Metastability in Open Quantum Systems},
  author = {Macieszczak, Katarzyna and Rose, Dominic C and Lesanovsky, Igor and Garrahan, Juan P},
  year = 2021,
  month = jul,
  journal = {Physical Review Research},
  volume = {3},
  number = {3},
  pages = {033047},
  publisher = {American Physical Society},
  issn = {2643-1564},
  doi = {10.1103/PhysRevResearch.3.033047}
}

@article{MacieszczakTheoryMetastability2016,
  title = {Towards a Theory of Metastability in Open Quantum Dynamics},
  author = {Macieszczak, Katarzyna and Gu{\c t}{\u a}, M{\u a}d{\u a}lin and Lesanovsky, Igor and Garrahan, Juan P},
  year = 2016,
  month = jun,
  journal = {Physical Review Letters},
  volume = {116},
  number = {24},
  pages = {240404},
  issn = {0031-9007},
  doi = {10.1103/PhysRevLett.116.240404}
}

@article{MingantiQuantumExceptional2019,
  title = {Quantum Exceptional Points of Non-{{Hermitian Hamiltonians}} and {{Liouvillians}}: {{The}} Effects of Quantum Jumps},
  author = {Minganti, Fabrizio and Miranowicz, Adam and Chhajlany, Ravindra W and Nori, Franco},
  year = 2019,
  month = dec,
  journal = {Physical Review A},
  volume = {100},
  number = {6},
  pages = {062131},
  publisher = {American Physical Society},
  issn = {2469-9926},
  doi = {10.1103/PhysRevA.100.062131}
}

@article{MiriExceptionalPoints2019,
  title = {Exceptional Points in Optics and Photonics},
  author = {Miri, Mohammad-Ali and Al{\`u}, Andrea},
  year = 2019,
  month = jan,
  journal = {Science (New York, N.Y.)},
  volume = {363},
  number = {6422},
  pages = {eaar7709},
  issn = {0036-8075, 1095-9203},
  doi = {10.1126/science.aar7709},
  urldate = {2022-07-06},
  langid = {english}
}

@article{MolmerMonteCarlo1993,
  title = {Monte {{Carlo}} Wave-Function Method in Quantum Optics},
  author = {M{\o}lmer, Klaus and Castin, Yvan and Dalibard, Jean},
  year = 1993,
  month = mar,
  journal = {Journal of the Optical Society of America B},
  volume = {10},
  number = {3},
  pages = {524},
  issn = {0740-3224, 1520-8540},
  doi = {10.1364/JOSAB.10.000524},
  urldate = {2024-09-01},
  copyright = {https://doi.org/10.1364/OA\_License\_v1\#VOR},
  langid = {english}
}

@article{MolmerMonteCarlo1996,
  title = {Monte {{Carlo}} Wavefunctions in Quantum Optics},
  author = {M{\o}lmer, Klaus and Castin, Yvan},
  year = 1996,
  month = feb,
  journal = {Quantum and Semiclassical Optics: Journal of the European Optical Society Part B},
  volume = {8},
  number = {1},
  pages = {49--72},
  issn = {1355-5111, 1361-6625},
  doi = {10.1088/1355-5111/8/1/007},
  urldate = {2024-09-01}
}

@article{MullerExceptionalPoints2008,
  title = {Exceptional Points in Open Quantum Systems},
  author = {M{\"u}ller, Markus and Rotter, Ingrid},
  year = 2008,
  month = jun,
  journal = {Journal of Physics A: Mathematical and Theoretical},
  volume = {41},
  number = {24},
  pages = {244018},
  issn = {1751-8113, 1751-8121},
  doi = {10.1088/1751-8113/41/24/244018},
  urldate = {2025-01-21}
}

@article{NakajimaQuantumTheory1958,
  title = {On {{Quantum Theory}} of {{Transport Phenomena}}: {{Steady Diffusion}}},
  shorttitle = {On {{Quantum Theory}} of {{Transport Phenomena}}},
  author = {Nakajima, Sadao},
  year = 1958,
  month = dec,
  journal = {Progress of Theoretical Physics},
  volume = {20},
  number = {6},
  pages = {948--959},
  issn = {0033-068X},
  doi = {10.1143/PTP.20.948},
  urldate = {2024-09-02},
  langid = {english}
}

@book{NielsenQuantumComputation2012,
  title = {Quantum {{Computation}} and {{Quantum Information}}: 10th {{Anniversary Edition}}},
  shorttitle = {Quantum {{Computation}} and {{Quantum Information}}},
  author = {Nielsen, Michael A. and Chuang, Isaac L.},
  year = 2012,
  month = jun,
  edition = {1},
  publisher = {Cambridge University Press},
  doi = {10.1017/CBO9780511976667},
  urldate = {2023-02-16},
  isbn = {978-1-107-00217-3 978-0-511-97666-7}
}

@article{OzdemirParityTime2019,
  title = {Parity--Time Symmetry and Exceptional Points in Photonics},
  author = {{\"O}zdemir, {\c S}. K. and Rotter, S. and Nori, F. and Yang, L.},
  year = 2019,
  month = aug,
  journal = {Nature Materials},
  volume = {18},
  number = {8},
  pages = {783--798},
  issn = {1476-1122, 1476-4660},
  doi = {10.1038/s41563-019-0304-9},
  urldate = {2024-09-23},
  langid = {english}
}

@article{PezzeQuantumMetrology2018,
  title = {Quantum Metrology with Nonclassical States of Atomic Ensembles},
  author = {Pezz{\`e}, Luca and Smerzi, Augusto and Oberthaler, Markus K. and Schmied, Roman and Treutlein, Philipp},
  year = 2018,
  month = sep,
  journal = {Reviews of Modern Physics},
  volume = {90},
  number = {3},
  pages = {035005},
  issn = {0034-6861, 1539-0756},
  doi = {10.1103/RevModPhys.90.035005},
  urldate = {2023-09-29},
  langid = {english}
}

@article{PlenioCavitylossinducedGeneration1999,
  title = {Cavity-Loss-Induced Generation of Entangled Atoms},
  author = {Plenio, M. B. and Huelga, S. F. and Beige, A. and Knight, P. L.},
  year = 1999,
  month = mar,
  journal = {Physical Review A},
  volume = {59},
  number = {3},
  pages = {2468--2475},
  issn = {1050-2947, 1094-1622},
  doi = {10.1103/PhysRevA.59.2468},
  urldate = {2022-06-22},
  langid = {english}
}

@article{PlenioIntroductionEntanglement2007,
  title = {An Introduction to Entanglement Measures},
  author = {Plenio, M.B. and Virmani, S.},
  year = 2007,
  month = jan,
  journal = {Quantum Information and Computation},
  volume = {7},
  number = {1\&2},
  pages = {1--51},
  issn = {15337146, 15337146},
  doi = {10.26421/QIC7.1-2-1},
  urldate = {2022-07-08}
}

@article{PlenioLogarithmicNegativity2005,
  title = {Logarithmic {{Negativity}}: {{A Full Entanglement Monotone That}} Is Not {{Convex}}},
  shorttitle = {Logarithmic {{Negativity}}},
  author = {Plenio, M. B.},
  year = 2005,
  month = aug,
  journal = {Physical Review Letters},
  volume = {95},
  number = {9},
  pages = {090503},
  issn = {0031-9007, 1079-7114},
  doi = {10.1103/PhysRevLett.95.090503},
  urldate = {2024-12-13},
  copyright = {http://link.aps.org/licenses/aps-default-license},
  langid = {english}
}

@article{PlenioQuantumjumpApproach1998,
  title = {The Quantum-Jump Approach to Dissipative Dynamics in Quantum Optics},
  author = {Plenio, M B and Knight, P L},
  year = 1998,
  month = jan,
  journal = {Reviews of Modern Physics},
  volume = {70},
  number = {1},
  pages = {101--144},
  issn = {0034-6861},
  doi = {10.1103/RevModPhys.70.101}
}

@article{PoyatosQuantumReservoir1996,
  title = {Quantum {{Reservoir Engineering}} with {{Laser Cooled Trapped Ions}}},
  author = {Poyatos, J. F. and Cirac, J. I. and Zoller, P.},
  year = 1996,
  month = dec,
  journal = {Physical Review Letters},
  volume = {77},
  number = {23},
  pages = {4728--4731},
  issn = {0031-9007, 1079-7114},
  doi = {10.1103/PhysRevLett.77.4728},
  urldate = {2023-09-29},
  langid = {english}
}

@article{PurcellResonanceAbsorption1946,
  title = {Resonance {{Absorption}} by {{Nuclear Magnetic Moments}} in a {{Solid}}},
  author = {Purcell, E. M. and Torrey, H. C. and Pound, R. V.},
  year = 1946,
  month = jan,
  journal = {Physical Review},
  volume = {69},
  number = {1-2},
  pages = {37--38},
  issn = {0031-899X},
  doi = {10.1103/PhysRev.69.37},
  urldate = {2022-12-02},
  langid = {english}
}

@article{ReiterEffectiveOperator2012,
  title = {Effective Operator Formalism for Open Quantum Systems},
  author = {Reiter, Florentin and S{\o}rensen, Anders S},
  year = 2012,
  month = mar,
  journal = {Physical Review A},
  volume = {85},
  number = {3},
  pages = {032111},
  issn = {1050-2947},
  doi = {10.1103/PhysRevA.85.032111}
}

@book{RivasOpenQuantum2012,
  title = {Open {{Quantum Systems}}},
  author = {Rivas, {\'A}ngel and Huelga, Susana F.},
  year = 2012,
  series = {{{SpringerBriefs}} in {{Physics}}},
  publisher = {Springer Berlin Heidelberg},
  address = {Berlin, Heidelberg},
  doi = {10.1007/978-3-642-23354-8},
  urldate = {2022-04-04},
  isbn = {978-3-642-23353-1 978-3-642-23354-8},
  langid = {english}
}

@book{SakuraiModernQuantum2017,
  title = {Modern {{Quantum Mechanics}}},
  shorttitle = {Modern {{Quantum Mechanics}}},
  author = {Sakurai, J. J. and Napolitano, Jim},
  year = 2017,
  month = sep,
  edition = {2},
  publisher = {Cambridge University Press},
  doi = {10.1017/9781108499996},
  urldate = {2022-04-04},
  isbn = {978-1-108-49999-6}
}

@article{TiranovCollectiveSuper2023,
  title = {Collective Super- and Subradiant Dynamics between Distant Optical Quantum Emitters},
  author = {Tiranov, Alexey and Angelopoulou, Vasiliki and {van Diepen}, Cornelis Jacobus and Schrinski, Bj{\"o}rn and Sandberg, Oliver August Dall'Alba and Wang, Ying and Midolo, Leonardo and Scholz, Sven and Wieck, Andreas Dirk and Ludwig, Arne and S{\o}rensen, Anders S{\o}ndberg and Lodahl, Peter},
  year = 2023,
  month = jan,
  journal = {Science},
  volume = {379},
  number = {6630},
  pages = {389--393},
  issn = {0036-8075, 1095-9203},
  doi = {10.1126/science.ade9324},
  urldate = {2023-01-31},
  langid = {english}
}

@article{VidalComputableMeasure2002,
  title = {Computable Measure of Entanglement},
  author = {Vidal, G. and Werner, R. F.},
  year = 2002,
  month = feb,
  journal = {Physical Review A},
  volume = {65},
  number = {3},
  pages = {032314},
  issn = {1050-2947, 1094-1622},
  doi = {10.1103/PhysRevA.65.032314},
  urldate = {2024-10-02},
  copyright = {http://link.aps.org/licenses/aps-default-license},
  langid = {english}
}

@article{Vivas-VianaDissipativeStabilization2024,
  title = {Dissipative Stabilization of Maximal Entanglement between Nonidentical Emitters via Two-Photon Excitation},
  author = {{Vivas-Via{\~n}a}, Alejandro and {Mart{\'i}n-Cano}, Diego and Mu{\~n}oz, Carlos S{\'a}nchez},
  year = 2024,
  month = oct,
  journal = {Physical Review Research},
  volume = {6},
  number = {4},
  pages = {043051},
  issn = {2643-1564},
  doi = {10.1103/PhysRevResearch.6.043051},
  urldate = {2025-01-04},
  langid = {english}
}

@article{Vivas-VianaFrequencyResolvedPurcell2024,
  title = {Frequency-{{Resolved Purcell Effect}} for the {{Dissipative Generation}} of {{Steady-State Entanglement}}},
  author = {{Vivas-Via{\~n}a}, Alejandro and {Mart{\'i}n-Cano}, Diego and Mu{\~n}oz, Carlos S{\'a}nchez},
  year = 2024,
  month = oct,
  journal = {Physical Review Letters},
  volume = {133},
  number = {17},
  pages = {173601},
  issn = {0031-9007, 1079-7114},
  doi = {10.1103/PhysRevLett.133.173601},
  urldate = {2025-01-04},
  langid = {english}
}

@misc{Vivas-VianaNonclassicalDrivenDissipative2025,
  title = {Nonclassical {{Driven-Dissipative Dynamics}} in {{Collective Quantum Optics}}, {{Ph}}.{{D}}. {{Thesis}}, {{Universidad Aut\'onoma}} de {{Madrid}}},
  author = {{Vivas-Via{\~n}a}, Alejandro},
  year = 2025,
  number = {arXiv:2509.10672 [quant-ph]},
  eprint = {2509.10672},
  primaryclass = {quant-ph},
  publisher = {arXiv},
  urldate = {2026-05-02},
  archiveprefix = {arXiv},
  copyright = {Creative Commons Attribution 4.0 International}
}

@article{Vivas-VianaUnconventionalMechanism2022,
  title = {Unconventional Mechanism of Virtual-State Population through Dissipation},
  author = {{Vivas-Via{\~n}a}, Alejandro and {Gonz{\'a}lez-Tudela}, Alejandro and Mu{\~n}oz, Carlos S{\'a}nchez},
  year = 2022,
  month = jul,
  journal = {Physical Review A},
  volume = {106},
  number = {1},
  pages = {012217},
  issn = {2469-9926, 2469-9934},
  doi = {10.1103/PhysRevA.106.012217},
  urldate = {2022-11-25},
  langid = {english}
}

@book{WallsQuantumOptics2008,
  title = {Quantum {{Optics}}},
  editor = {Walls, D.F. and Milburn, Gerard J.},
  year = 2008,
  publisher = {Springer Berlin Heidelberg},
  address = {Berlin, Heidelberg},
  doi = {10.1007/978-3-540-28574-8},
  urldate = {2023-02-16},
  isbn = {978-3-540-28573-1},
  langid = {english}
}

@article{WoottersEntanglementFormation1998,
  title = {Entanglement of Formation of an Arbitrary State of Two Qubits},
  author = {Wootters, William K},
  year = 1998,
  month = mar,
  journal = {Physical Review Letters},
  volume = {80},
  number = {10},
  pages = {2245--2248},
  issn = {0031-9007},
  doi = {10.1103/PhysRevLett.80.2245}
}

@article{WoottersEntanglementFormation2001,
  title = {Entanglement of Formation and Concurrence},
  author = {Wootters, William K},
  year = 2001,
  month = jul,
  journal = {Quantum Information and Computation},
  volume = {1},
  number = {1},
  pages = {27--44},
  issn = {15337146},
  doi = {10.26421/QIC1.1-3}
}

@article{YangEntanglementPhotonic2025,
  title = {Entanglement of Photonic Modes from a Continuously Driven Two-Level System},
  author = {Yang, Jiaying and Strandberg, Ingrid and {Vivas-Via{\~n}a}, Alejandro and Gaikwad, Akshay and {Castillo-Moreno}, Claudia and Kockum, Anton Frisk and Ullah, Muhammad Asad and Mu{\~n}oz, Carlos S{\'a}nchez and Eriksson, Axel Martin and Gasparinetti, Simone},
  year = 2025,
  month = apr,
  journal = {npj Quantum Information},
  volume = {11},
  number = {1},
  pages = {69},
  issn = {2056-6387},
  doi = {10.1038/s41534-025-00995-1},
  urldate = {2025-05-02},
  langid = {english}
}

@article{ZwanzigEnsembleMethod1960,
  title = {Ensemble {{Method}} in the {{Theory}} of {{Irreversibility}}},
  author = {Zwanzig, Robert},
  year = 1960,
  month = nov,
  journal = {The Journal of Chemical Physics},
  volume = {33},
  number = {5},
  pages = {1338--1341},
  issn = {0021-9606, 1089-7690},
  doi = {10.1063/1.1731409},
  urldate = {2024-09-02},
  langid = {english}
}


%% file: main.bib
@PREAMBLE{
 "\providecommand{\noopsort}[1]{}" 
 # "\providecommand{\singleletter}[1]{#1}%" 
}

@Article{Mlynek2014,
author={Mlynek, J. A.
and Abdumalikov, A. A.
and Eichler, C.
and Wallraff, A.},
title={Observation of Dicke superradiance for two artificial atoms in a cavity with high decay rate},
journal={Nature Communications},
year={2014},
month={Nov},
day={04},
volume={5},
number={1},
pages={5186},
issn={2041-1723},
doi={10.1038/ncomms6186},
url={https://doi.org/10.1038/ncomms6186}
}

@misc{Andresjuanes2025,
      title={Entangling remote qubits through a two-mode squeezed reservoir}, 
      author={A. Andrés-Juanes and J. Agustí and R. Sett and E. S. Redchenko and L. Kapoor and S. Hawaldar and P. Rabl and J. M. Fink},
      year={2025},
      eprint={2510.07139},
      archivePrefix={arXiv},
      primaryClass={quant-ph},
      url={https://arxiv.org/abs/2510.07139}, 
}

@misc{guo2026,
      title={{Entangling Superconducting Qubits via Energy-Selective Local Reservoirs}}, 
      author={Qihao Guo and Botao Du and Ruichao Ma},
      year={2026},
      eprint={2605.12429},
      archivePrefix={arXiv}
}

@phdthesis{Almanakly2025thesis,
  author  = {Almanakly, Aziza},
  title   = {Quantum Networking using Waveguide Quantum Electrodynamics},
  school  = {Massachusetts Institute of Technology},
  year    = {2025},
  address = {Cambridge, MA},
  month   = {September},
  url     = {https://hdl.handle.net/1721.1/164572}}

@article{Almanakly2026,
  title = {{Probing Sensitivity Near a Quantum Exceptional Point Using Waveguide Quantum Electrodynamics}},
  author = {Almanakly, Aziza and Assouly, R\'eouven and Kang, Harry Hanlim and Gingras, Michael and Niedzielski, Bethany M. and Stickler, Hannah and Schwartz, Mollie E. and Serniak, Kyle and Hays, Max and Grover, Jeffrey A. and Oliver, William D.},
  journal = {Physical Review Letters},
  volume = {136},
  issue = {18},
  pages = {183601},
  numpages = {7},
  year = {2026},
  month = {May},
  publisher = {American Physical Society},
  doi = {10.1103/gd4s-fgwt},
  url = {https://link.aps.org/doi/10.1103/gd4s-fgwt}
}

@misc{An2025,
      title={{ZZ-Free Two-Transmon CZ Gate Mediated by a Fluxonium Coupler}}, 
      author={Junyoung An and Helin Zhang and Qi Ding and Leon Ding and Youngkyu Sung and Roni Winik and Junghyun Kim and Ilan T. Rosen and Kate Azar and Renee DePencier Piñero and Jeffrey M. Gertler and Michael Gingras and Bethany M. Niedzielski and Hannah Stickler and Mollie E. Schwartz and Joel I-j. Wang and Terry P. Orlando and Simon Gustavsson and Max Hays and Jeffrey A. Grover and Kyle Serniak and William D. Oliver},
      year={2025},
      eprint={2511.02115},
      archivePrefix={arXiv}
}

@article{Chen2025,
  title = {Hardware-efficient stabilization of entanglement via engineered dissipation in superconducting circuits},
  author = {Chen, Changling and Tang, Kai and Zhou, Yuxuan and Yi, KangYuan and Zhang, Xuan and Zhang, Xu and Guo, Haosheng and Liu, Song and Chen, Yuanzhen and Yan, Tongxing and Yu, Dapeng},
  journal = {Physical Review Research},
  volume = {7},
  issue = {2},
  pages = {L022018},
  numpages = {6},
  year = {2025},
  month = {Apr},
  publisher = {American Physical Society},
  doi = {10.1103/PhysRevResearch.7.L022018},
  url = {https://link.aps.org/doi/10.1103/PhysRevResearch.7.L022018}
}

@article{Li2024,
  title = {{Realization of High-Fidelity CZ Gate Based on a Double-Transmon Coupler}},
  author = {Li, Rui and Kubo, Kentaro and Ho, Yinghao and Yan, Zhiguang and Nakamura, Yasunobu and Goto, Hayato},
  journal = {Physical Review X},
  volume = {14},
  issue = {4},
  pages = {041050},
  numpages = {30},
  year = {2024},
  month = {Nov},
  publisher = {American Physical Society},
  doi = {10.1103/PhysRevX.14.041050},
  url = {https://link.aps.org/doi/10.1103/PhysRevX.14.041050}
}

@article{Guo2024,
  title = {Universal scalable characterization and correction of pulse distortions in controlled quantum systems},
  author = {Guo, Liang-Liang and Duan, Peng and Zhang, Sheng and Yang, Xin-Xin and Zhang, Chi and Du, Lei and Zhang, Hai-Feng and Tao, Hao-Ran and Wang, Tian-Le and Jia, Zhi-Long and Chen, Zhao-Yun and Guo, Guo-Ping},
  journal = {Physical Review Applied},
  volume = {21},
  issue = {6},
  pages = {064060},
  numpages = {11},
  year = {2024},
  month = {Jun},
  publisher = {American Physical Society},
  doi = {10.1103/PhysRevApplied.21.064060},
  url = {https://link.aps.org/doi/10.1103/PhysRevApplied.21.064060}
}

@Article{Zhang2026,
author={Zhang, Jiajian
and Wang, Lingna
and Hai, Yong-Ju
and Zhang, Jiawei
and Chu, Ji
and Jiang, Ji
and Huang, Wenhui
and Liang, Yongqi
and Qiu, Jiawei
and Sun, Xuandong
and Tao, Ziyu
and Zhang, Libo
and Zhou, Yuxuan
and Chen, Yuanzhen
and Guo, Weijie
and Linpeng, Xiayu
and Liu, Song
and Ren, Wenhui
and Zhong, Youpeng
and Niu, Jingjing
and Yuan, Haidong
and Yu, Dapeng},
title={Distributed multi-parameter quantum metrology with a superconducting quantum network},
journal={Nature Communications},
year={2026},
month={Jan},
day={20},
volume={17},
number={1},
pages={1825},
issn={2041-1723},
doi={10.1038/s41467-026-68535-9},
url={https://doi.org/10.1038/s41467-026-68535-9}
}

@article{Chitambar2019,
  title = {Quantum resource theories},
  author = {Chitambar, Eric and Gour, Gilad},
  journal = {Reviews of Modern Physics},
  volume = {91},
  issue = {2},
  pages = {025001},
  numpages = {48},
  year = {2019},
  month = {Apr},
  publisher = {American Physical Society},
  doi = {10.1103/RevModPhys.91.025001},
  url = {https://link.aps.org/doi/10.1103/RevModPhys.91.025001}
}

@Article{Almanakly2025,
author={Almanakly, Aziza
and Yankelevich, Beatriz
and Hays, Max
and Kannan, Bharath
and Assouly, R{\'e}ouven
and Greene, Alex
and Gingras, Michael
and Niedzielski, Bethany M.
and Stickler, Hannah
and Schwartz, Mollie E.
and Serniak, Kyle
and Wang, Joel {\^I}-j.
and Orlando, Terry P.
and Gustavsson, Simon
and Grover, Jeffrey A.
and Oliver, William D.},
title={Deterministic remote entanglement using a chiral quantum interconnect},
journal={Nature Physics},
year={2025},
month={Mar},
day={21},
volume={21},
pages={825},
issn={1745-2481},
doi={10.1038/s41567-025-02811-1},
url={https://doi.org/10.1038/s41567-025-02811-1}
}

@article{Soro2022,
  title = {Chiral quantum optics with giant atoms},
  author = {Soro, Ariadna and Kockum, Anton Frisk},
  journal = {Physical Review A},
  volume = {105},
  issue = {2},
  pages = {023712},
  numpages = {16},
  year = {2022},
  month = {Feb},
  publisher = {American Physical Society},
  doi = {10.1103/PhysRevA.105.023712},
  url = {https://link.aps.org/doi/10.1103/PhysRevA.105.023712}
}

@misc{ramette2023,
  title={{Fault-Tolerant Connection of Error-Corrected Qubits with Noisy Links}}, 
  author={Joshua Ramette and Josiah Sinclair and Nikolas P. Breuckmann and Vladan Vuletić},
  year={2023},
  eprint={2302.01296},
  archivePrefix = {arXiv}
}

@article{Pita_Vidal_2024,
   title={Strong tunable coupling between two distant superconducting spin qubits},
   ISSN={1745-2481},
   url={http://dx.doi.org/10.1038/s41567-024-02497-x},
   DOI={10.1038/s41567-024-02497-x},
   journal={Nature Physics},
   publisher={Springer Science and Business Media LLC},
   author={Pita-Vidal, Marta and Wesdorp, Jaap J. and Splitthoff, Lukas J. and Bargerbos, Arno and Liu, Yu and Kouwenhoven, Leo P. and Andersen, Christian Kraglund},
   year={2024},
   volume={20},
   pages={1158},
   month=may }

@Article{Hensen2015,
author={Hensen, B.
and Bernien, H.
and Dr{\'e}au, A. E.
and Reiserer, A.
and Kalb, N.
and Blok, M. S.
and Ruitenberg, J.
and Vermeulen, R. F. L.
and Schouten, R. N.
and Abell{\'a}n, C.
and Amaya, W.
and Pruneri, V.
and Mitchell, M. W.
and Markham, M.
and Twitchen, D. J.
and Elkouss, D.
and Wehner, S.
and Taminiau, T. H.
and Hanson, R.},
title={{Loophole-free Bell inequality violation using electron spins separated by 1.3 kilometres}},
journal={Nature},
year={2015},
month={Oct},
day={01},
volume={526},
number={7575},
pages={682-686},
issn={1476-4687},
doi={10.1038/nature15759},
url={https://doi.org/10.1038/nature15759}
}

@Article{Zanner2022,
author={Zanner, Maximilian
and Orell, Tuure
and Schneider, Christian M. F.
and Albert, Romain
and Oleschko, Stefan
and Juan, Mathieu L.
and Silveri, Matti
and Kirchmair, Gerhard},
title={Coherent control of a multi-qubit dark state in waveguide quantum electrodynamics},
journal={Nature Physics},
year={2022},
month={May},
day={01},
volume={18},
number={5},
pages={538-543},
issn={1745-2481},
doi={10.1038/s41567-022-01527-w},
url={https://doi.org/10.1038/s41567-022-01527-w}
}

@Article{Chou2018,
author={Chou, Kevin S.
and Blumoff, Jacob Z.
and Wang, Christopher S.
and Reinhold, Philip C.
and Axline, Christopher J.
and Gao, Yvonne Y.
and Frunzio, L.
and Devoret, M. H.
and Jiang, Liang
and Schoelkopf, R. J.},
title={Deterministic teleportation of a quantum gate between two logical qubits},
journal={Nature},
year={2018},
month={Sep},
day={01},
volume={561},
number={7723},
pages={368-373},
issn={1476-4687},
doi={10.1038/s41586-018-0470-y},
url={https://doi.org/10.1038/s41586-018-0470-y}
}

@article{Pichler2015,
  title = {Quantum optics of chiral spin networks},
  author = {Pichler, Hannes and Ramos, Tom\'as and Daley, Andrew J. and Zoller, Peter},
  journal = {Physical Review A},
  volume = {91},
  issue = {4},
  pages = {042116},
  numpages = {19},
  year = {2015},
  month = {Apr},
  publisher = {American Physical Society},
  doi = {10.1103/PhysRevA.91.042116},
  url = {https://link.aps.org/doi/10.1103/PhysRevA.91.042116}
}

@article{kannan2023,
	author = {Kannan, Bharath and Almanakly, Aziza and Sung, Youngkyu and Di Paolo, Agustin and Rower, David A. and Braum{\"u}ller, Jochen and Melville, Alexander and Niedzielski, Bethany M. and Karamlou, Amir and Serniak, Kyle and Veps{\"a}l{\"a}inen, Antti and Schwartz, Mollie E. and Yoder, Jonilyn L. and Winik, Roni and Wang, Joel I-Jan and Orlando, Terry P. and Gustavsson, Simon and Grover, Jeffrey A. and Oliver, William D.},
	date = {2023/03/01},
	doi = {10.1038/s41567-022-01869-5},
	id = {Kannan2023},
	isbn = {1745-2481},
	journal = {Nature Physics},
	number = {3},
	pages = {394--400},
	title = {On-demand directional microwave photon emission using waveguide quantum electrodynamics},
	url = {https://doi.org/10.1038/s41567-022-01869-5},
	volume = {19},
	year = {2023}}

@article{Gheeraert2020,
  title = {Programmable directional emitter and receiver of itinerant microwave photons in a waveguide},
  author = {Gheeraert, Nicolas and Kono, Shingo and Nakamura, Yasunobu},
  journal = {Phys. Rev. A},
  volume = {102},
  issue = {5},
  pages = {053720},
  numpages = {14},
  year = {2020},
  month = {Nov},
  publisher = {American Physical Society},
  doi = {10.1103/PhysRevA.102.053720},
  url = {https://link.aps.org/doi/10.1103/PhysRevA.102.053720}
}

@article{Cirac1999,
  title = {Distributed quantum computation over noisy channels},
  author = {Cirac, J. I. and Ekert, A. K. and Huelga, S. F. and Macchiavello, C.},
  journal = {Physical Review A},
  volume = {59},
  issue = {6},
  pages = {4249--4254},
  numpages = {0},
  year = {1999},
  month = {Jun},
  publisher = {American Physical Society},
  doi = {10.1103/PhysRevA.59.4249},
  url = {https://link.aps.org/doi/10.1103/PhysRevA.59.4249}
}

@article{Koch2007,
  title = {{Charge-insensitive qubit design derived from the Cooper pair box}},
  author = {Koch, Jens and Yu, Terri M. and Gambetta, Jay and Houck, A. A. and Schuster, D. I. and Majer, J. and Blais, Alexandre and Devoret, M. H. and Girvin, S. M. and Schoelkopf, R. J.},
  journal = {Physical Review A},
  volume = {76},
  issue = {4},
  pages = {042319},
  numpages = {19},
  year = {2007},
  month = {Oct},
  publisher = {American Physical Society},
  doi = {10.1103/PhysRevA.76.042319},
  url = {https://link.aps.org/doi/10.1103/PhysRevA.76.042319}
}

@article {Astafiev2010,
	author = {Astafiev, O. and Zagoskin, A. M. and Abdumalikov, A. A. and Pashkin, Yu. A. and Yamamoto, T. and Inomata, K. and Nakamura, Y. and Tsai, J. S.},
	title = {Resonance Fluorescence of a Single Artificial Atom},
	volume = {327},
	number = {5967},
	pages = {840--843},
	year = {2010},
	doi = {10.1126/science.1181918},
	publisher = {American Association for the Advancement of Science},
	issn = {0036-8075},
	URL = {https://science.sciencemag.org/content/327/5967/840},
	journal = {Science}
}

@article{Jiang2007,
  title = {Distributed quantum computation based on small quantum registers},
  author = {Jiang, Liang and Taylor, Jacob M. and S\o{}rensen, Anders S. and Lukin, Mikhail D.},
  journal = {Physical Review A},
  volume = {76},
  issue = {6},
  pages = {062323},
  numpages = {22},
  year = {2007},
  month = {Dec},
  publisher = {American Physical Society},
  doi = {10.1103/PhysRevA.76.062323},
  url = {https://link.aps.org/doi/10.1103/PhysRevA.76.062323}
}

@article{Hoi2011,
  title = {Demonstration of a Single-Photon Router in the Microwave Regime},
  author = {Hoi, Io-Chun and Wilson, C. M. and Johansson, G\"oran and Palomaki, Tauno and Peropadre, Borja and Delsing, Per},
  journal = {Phys. Rev. Lett.},
  volume = {107},
  issue = {7},
  pages = {073601},
  numpages = {5},
  year = {2011},
  month = {Aug},
  publisher = {American Physical Society},
  doi = {10.1103/PhysRevLett.107.073601},
  url = {https://link.aps.org/doi/10.1103/PhysRevLett.107.073601}
}

@article{Hoi2013,
	doi = {10.1088/1367-2630/15/2/025011},
	url = {https://doi.org/10.1088%2F1367-2630%2F15%2F2%2F025011},
	year = 2013,
	month = {feb},
	publisher = {{IOP} Publishing},
	volume = {15},
	number = {2},
	pages = {025011},
	author = {Io-Chun Hoi and C M Wilson and Göran Johansson and Joel Lindkvist and Borja Peropadre and Tauno Palomaki and Per Delsing},
	title = {Microwave quantum optics with an artificial atom in one-dimensional open space},
	journal = {New Journal of Physics}
}

@article{Lalumiere2013,
  title = {Input-output theory for waveguide {QED} with an ensemble of inhomogeneous atoms},
  author = {Lalumi\`ere, Kevin and Sanders, Barry C. and van Loo, A. F. and Fedorov, A. and Wallraff, A. and Blais, A.},
  journal = {Physical Review A},
  volume = {88},
  issue = {4},
  pages = {043806},
  numpages = {15},
  year = {2013},
  month = {Oct},
  publisher = {American Physical Society},
  doi = {10.1103/PhysRevA.88.043806},
  url = {https://link.aps.org/doi/10.1103/PhysRevA.88.043806}
}

@article{FriskKockum2014,
  title = {{Designing frequency-dependent relaxation rates and Lamb shifts for a giant artificial atom}},
  author = {Frisk Kockum, Anton and Delsing, Per and Johansson, G\"oran},
  journal = {Physical Review A},
  volume = {90},
  issue = {1},
  pages = {013837},
  numpages = {13},
  year = {2014},
  month = {Jul},
  publisher = {American Physical Society},
  doi = {10.1103/PhysRevA.90.013837},
  url = {https://link.aps.org/doi/10.1103/PhysRevA.90.013837}
}

@Article{Mirhosseini2019,
author={Mirhosseini, Mohammad
and Kim, Eunjong
and Zhang, Xueyue
and Sipahigil, Alp
and Dieterle, Paul B.
and Keller, Andrew J.
and Asenjo-Garcia, Ana
and Chang, Darrick E.
and Painter, Oskar},
title={Cavity quantum electrodynamics with atom-like mirrors},
journal={Nature},
year={2019},
volume={569},
number={7758},
pages={692-697},
issn={1476-4687},
doi={10.1038/s41586-019-1196-1},
url={https://doi.org/10.1038/s41586-019-1196-1}
}

@article{Dicke1954,
  title = {{Coherence in Spontaneous Radiation Processes}},
  author = {Dicke, R. H.},
  journal = {Physical Review},
  volume = {93},
  issue = {1},
  pages = {99--110},
  numpages = {0},
  year = {1954},
  month = {Jan},
  publisher = {American Physical Society},
  doi = {10.1103/PhysRev.93.99},
  url = {https://link.aps.org/doi/10.1103/PhysRev.93.99}
}

@article {VanLoo2013,
	author = {van Loo, Arjan F. and Fedorov, Arkady and Lalumi{\`e}re, Kevin and Sanders, Barry C. and Blais, Alexandre and Wallraff, Andreas},
	title = {Photon-Mediated Interactions Between Distant Artificial Atoms},
	volume = {342},
	number = {6165},
	pages = {1494--1496},
	year = {2013},
	doi = {10.1126/science.1244324},
	publisher = {American Association for the Advancement of Science},
	issn = {0036-8075},
	URL = {https://science.sciencemag.org/content/342/6165/1494},
	journal = {Science}
}

@misc{Irfan2025,
  title={Autonomous stabilization of remote entanglement in a cascaded quantum network},
  author={Irfan, A and Singirikonda, K and Yao, M and Lingenfelter, A and Mollenhauer, M and Cao, X and Clerk, AA and Pfaff, W},
  eprint={2509.11872},
  archivePrefix = {arXiv},
  year = {2025}
}

@Article{Brown2022,
author={Brown, T.
and Doucet, E.
and Rist{\`e}, D.
and Ribeill, G.
and Cicak, K.
and Aumentado, J.
and Simmonds, R.
and Govia, L.
and Kamal, A.
and Ranzani, L.},
title={Trade off-free entanglement stabilization in a superconducting qutrit-qubit system},
journal={Nature Communications},
year={2022},
month={Jul},
day={09},
volume={13},
number={1},
pages={3994},
issn={2041-1723},
doi={10.1038/s41467-022-31638-0},
url={https://doi.org/10.1038/s41467-022-31638-0}
}

@Article{Lin2013,
author={Lin, Y.
and Gaebler, J. P.
and Reiter, F.
and Tan, T. R.
and Bowler, R.
and S{\o}rensen, A. S.
and Leibfried, D.
and Wineland, D. J.},
title={Dissipative production of a maximally entangled steady state of two quantum bits},
journal={Nature},
year={2013},
month={Dec},
day={01},
volume={504},
number={7480},
pages={415-418},
issn={1476-4687},
doi={10.1038/nature12801},
url={https://doi.org/10.1038/nature12801}
}

@article{Chiaro2025,
  title = {{Active Leakage Cancellation in Single Qubit Gates}},
  author = {Chiaro, Ben and Zhang, Yaxing},
  journal = {Physical Review Letters},
  volume = {135},
  issue = {13},
  pages = {130601},
  numpages = {8},
  year = {2025},
  month = {Sep},
  publisher = {American Physical Society},
  doi = {10.1103/4kz9-w97h},
  url = {https://link.aps.org/doi/10.1103/4kz9-w97h}
}

@Article{Yan2016,
author={Yan, Fei
and Gustavsson, Simon
and Kamal, Archana
and Birenbaum, Jeffrey
and Sears, Adam P.
and Hover, David
and Gudmundsen, Ted J.
and Rosenberg, Danna
and Samach, Gabriel
and Weber, S.
and Yoder, Jonilyn L.
and Orlando, Terry P.
and Clarke, John
and Kerman, Andrew J.
and Oliver, William D.},
title={The flux qubit revisited to enhance coherence and reproducibility},
journal={Nature Communications},
year={2016},
month={Nov},
day={03},
volume={7},
number={1},
pages={12964},
issn={2041-1723},
doi={10.1038/ncomms12964},
url={https://doi.org/10.1038/ncomms12964}
}

@misc{haug2025,
      title={{Lattice surgery with Bell measurements: Modular fault-tolerant quantum computation at low entanglement cost}}, 
      author={Trond Hjerpekjøn Haug and Timo Hillmann and Anton Frisk Kockum and Raphaël {Van Laer}},
      year={2025},
      eprint={2510.13541},
      archivePrefix={arXiv}
}

@article{Shah2024,
  title = {Stabilizing Remote Entanglement via Waveguide Dissipation},
  author = {Shah, Parth S. and Yang, Frank and Joshi, Chaitali and Mirhosseini, Mohammad},
  journal = {PRX Quantum},
  volume = {5},
  issue = {3},
  pages = {030346},
  numpages = {23},
  year = {2024},
  month = {Sep},
  publisher = {American Physical Society},
  doi = {10.1103/PRXQuantum.5.030346},
  url = {https://link.aps.org/doi/10.1103/PRXQuantum.5.030346}
}

@Article{Shankar2013,
author={Shankar, S.
and Hatridge, M.
and Leghtas, Z.
and Sliwa, K. M.
and Narla, A.
and Vool, U.
and Girvin, S. M.
and Frunzio, L.
and Mirrahimi, M.
and Devoret, M. H.},
title={Autonomously stabilized entanglement between two superconducting quantum bits},
journal={Nature},
year={2013},
month={Dec},
day={01},
volume={504},
number={7480},
pages={419-422},
issn={1476-4687},
doi={10.1038/nature12802},
url={https://doi.org/10.1038/nature12802}
}

@Article{Storz2023,
author={Storz, Simon
and Sch{\"a}r, Josua
and Kulikov, Anatoly
and Magnard, Paul
and Kurpiers, Philipp
and L{\"u}tolf, Janis
and Walter, Theo
and Copetudo, Adrian
and Reuer, Kevin
and Akin, Abdulkadir
and Besse, Jean-Claude
and Gabureac, Mihai
and Norris, Graham J.
and Rosario, Andr{\'e}s
and Martin, Ferran
and Martinez, Jos{\'e}
and Amaya, Waldimar
and Mitchell, Morgan W.
and Abellan, Carlos
and Bancal, Jean-Daniel
and Sangouard, Nicolas
and Royer, Baptiste
and Blais, Alexandre
and Wallraff, Andreas},
title={{Loophole-free Bell inequality violation with superconducting circuits}},
journal={Nature},
year={2023},
month={May},
day={01},
volume={617},
number={7960},
pages={265-270},
issn={1476-4687},
doi={10.1038/s41586-023-05885-0},
url={https://doi.org/10.1038/s41586-023-05885-0}
}

@Article{Kurpiers2018,
author={Kurpiers, P.
and Magnard, P.
and Walter, T.
and Royer, B.
and Pechal, M.
and Heinsoo, J.
and Salath{\'e}, Y.
and Akin, A.
and Storz, S.
and Besse, J.-C.
and Gasparinetti, S.
and Blais, A.
and Wallraff, A.},
title={Deterministic quantum state transfer and remote entanglement using microwave photons},
journal={Nature},
year={2018},
volume={558},
number={7709},
pages={264-267},
issn={1476-4687},
doi={10.1038/s41586-018-0195-y},
url={https://doi.org/10.1038/s41586-018-0195-y}
}

@article{Kimble2008,
author={Kimble, H. J.},
title={The quantum internet},
journal={Nature},
year={2008},
volume={453},
number={7198},
pages={1023-1030},
issn={1476-4687},
doi={10.1038/nature07127},
url={https://doi.org/10.1038/nature07127}
}

@Article{Kannan2020,
author={Kannan, Bharath
and Ruckriegel, Max J.
and Campbell, Daniel L.
and Frisk Kockum, Anton
and Braum{\"u}ller, Jochen
and Kim, David K.
and Kjaergaard, Morten
and Krantz, Philip
and Melville, Alexander
and Niedzielski, Bethany M.
and Veps{\"a}l{\"a}inen, Antti
and Winik, Roni
and Yoder, Jonilyn L.
and Nori, Franco
and Orlando, Terry P.
and Gustavsson, Simon
and Oliver, William D.},
title={Waveguide quantum electrodynamics with superconducting artificial giant atoms},
journal={Nature},
year={2020},
month={Jul},
day={01},
volume={583},
number={7818},
pages={775-779},
issn={1476-4687},
doi={10.1038/s41586-020-2529-9},
url={https://doi.org/10.1038/s41586-020-2529-9}
}

@Article{Zhong2021,
author={Zhong, Youpeng
and Chang, Hung-Shen
and Bienfait, Audrey
and Dumur, {\'E}tienne
and Chou, Ming-Han
and Conner, Christopher R.
and Grebel, Joel
and Povey, Rhys G.
and Yan, Haoxiong
and Schuster, David I.
and Cleland, Andrew N.},
title={Deterministic multi-qubit entanglement in a quantum network},
journal={Nature},
year={2021},
month={Feb},
day={01},
volume={590},
number={7847},
pages={571-575},
issn={1476-4687},
doi={10.1038/s41586-021-03288-7},
url={https://doi.org/10.1038/s41586-021-03288-7}
}

@Article{Zhong2019,
author={Zhong, Y. P.
and Chang, H.-S.
and Satzinger, K. J.
and Chou, M.-H.
and Bienfait, A.
and Conner, C. R.
and Dumur, {\'E}
and Grebel, J.
and Peairs, G. A.
and Povey, R. G.
and Schuster, D. I.
and Cleland, A. N.},
title={Violating {B}ell's inequality with remotely connected superconducting qubits},
journal={Nature Physics},
year={2019},
month={Aug},
day={01},
volume={15},
number={8},
pages={741-744},
issn={1745-2481},
doi={10.1038/s41567-019-0507-7},
url={https://doi.org/10.1038/s41567-019-0507-7}
}

@article{Magnard2020,
  title = {{Microwave Quantum Link between Superconducting Circuits Housed in Spatially Separated Cryogenic Systems}},
  author = {Magnard, P. and Storz, S. and Kurpiers, P. and Sch\"ar, J. and Marxer, F. and L\"utolf, J. and Walter, T. and Besse, J.-C. and Gabureac, M. and Reuer, K. and Akin, A. and Royer, B. and Blais, A. and Wallraff, A.},
  journal = {Physical Review Letters},
  volume = {125},
  issue = {26},
  pages = {260502},
  numpages = {7},
  year = {2020},
  month = {Dec},
  publisher = {American Physical Society},
  doi = {10.1103/PhysRevLett.125.260502},
  url = {https://link.aps.org/doi/10.1103/PhysRevLett.125.260502}
}

@InProceedings{Kockum2021review,
author="Frisk Kockum, Anton",
editor="Takagi, Tsuyoshi
and Wakayama, Masato
and Tanaka, Keisuke
and Kunihiro, Noboru
and Kimoto, Kazufumi
and Ikematsu, Yasuhiko",
title="Quantum Optics with Giant Atoms---the First Five Years",
booktitle="International Symposium on Mathematics, Quantum Theory, and Cryptography",
year="2021",
publisher="Springer Singapore",
address="Singapore",
pages="125--146",
isbn="978-981-15-5191-8",
doi = {10.1007/978-981-15-5191-8_12}
}

@article{Kockum2018,
  title = {{Decoherence-Free Interaction between Giant Atoms in Waveguide Quantum Electrodynamics}},
  author = {Kockum, Anton Frisk and Johansson, G\"oran and Nori, Franco},
  journal = {Physical Review Letters},
  volume = {120},
  issue = {14},
  pages = {140404},
  numpages = {8},
  year = {2018},
  month = {Apr},
  publisher = {American Physical Society},
  doi = {10.1103/PhysRevLett.120.140404},
  url = {https://link.aps.org/doi/10.1103/PhysRevLett.120.140404}
}

@article{sung2020,
  title = {{Realization of High-Fidelity CZ and $ZZ$-Free iSWAP Gates with a Tunable Coupler}},
  author = {Sung, Youngkyu and Ding, Leon and Braum\"uller, Jochen and Veps\"al\"ainen, Antti and Kannan, Bharath and Kjaergaard, Morten and Greene, Ami and Samach, Gabriel O. and McNally, Chris and Kim, David and Melville, Alexander and Niedzielski, Bethany M. and Schwartz, Mollie E. and Yoder, Jonilyn L. and Orlando, Terry P. and Gustavsson, Simon and Oliver, William D.},
  journal = {Physical Review X},
  volume = {11},
  issue = {2},
  pages = {021058},
  numpages = {32},
  year = {2021},
  month = {Jun},
  publisher = {American Physical Society},
  doi = {10.1103/PhysRevX.11.021058},
  url = {https://link.aps.org/doi/10.1103/PhysRevX.11.021058}
}

@article{Lodahl2017,
    title = {Chiral quantum optics},
    volume = {541},
    copyright = {2017 Macmillan Publishers Limited, part of Springer Nature. All rights reserved.},
    issn = {1476-4687},
    url = {https://www.nature.com/articles/nature21037},
    doi = {10.1038/nature21037},
    
    number = {7638},
    urldate = {2022-01-27},
    journal = {Nature},
    author = {Lodahl, Peter and Mahmoodian, Sahand and Stobbe, Søren and Rauschenbeutel, Arno and Schneeweiss, Philipp and Volz, Jürgen and Pichler, Hannes and Zoller, Peter},
    month = jan,
    year = {2017},
    pages = {473--480},
}

@incollection{Solano2017,
    title = {Chapter {Seven} - {Optical} {Nanofibers}: {A} {New} {Platform} for {Quantum} {Optics}},
    volume = {66},
    copyright = {All rights reserved},
    shorttitle = {Chapter {Seven} - {Optical} {Nanofibers}},
    url = {https://www.sciencedirect.com/science/article/pii/S1049250X1730006X},
    urldate = {2021-03-15},
    booktitle = {Advances {In} {Atomic}, {Molecular}, and {Optical} {Physics}},
    publisher = {Academic Press},
    author = {Solano, Pablo and Grover, Jeffrey A. and Hoffman, Jonathan E. and Ravets, Sylvain and Fatemi, Fredrik K. and Orozco, Luis A. and Rolston, Steven L.},
    month = jan,
    year = {2017},
    doi = {10.1016/bs.aamop.2017.02.003},
    pages = {439--505},
}

@article{Guimond2020,
  title = {A unidirectional on-chip photonic interface for superconducting circuits},
  volume = {6},
  copyright = {2020 The Author(s)},
  issn = {2056-6387},
  url = {https://www.nature.com/articles/s41534-020-0261-9},
  doi = {10.1038/s41534-020-0261-9},
  
  number = {1},
  urldate = {2022-04-25},
  journal = {npj Quantum Information},
  author = {Guimond, P.-O. and Vermersch, B. and Juan, M. L. and Sharafiev, A. and Kirchmair, G. and Zoller, P.},
  month = mar,
  year = {2020},
  pages = {32},
}

@article{Cirac1997,
  title = {{Quantum State Transfer and Entanglement Distribution among Distant Nodes in a Quantum Network}},
  author = {Cirac, J. I. and Zoller, P. and Kimble, H. J. and Mabuchi, H.},
  journal = {Physical Review Letters},
  volume = {78},
  issue = {16},
  pages = {3221--3224},
  numpages = {0},
  year = {1997},
  month = {Apr},
  publisher = {American Physical Society},
  doi = {10.1103/PhysRevLett.78.3221},
  url = {https://link.aps.org/doi/10.1103/PhysRevLett.78.3221}
}

@misc{Dalzell2023,
archivePrefix = {arXiv},
author = {Dalzell, Alexander M. and McArdle, Sam and Berta, Mario and Bienias, Przemyslaw and Chen, Chi-Fang and Gily{\'{e}}n, Andr{\'{a}}s and Hann, Connor T. and Kastoryano, Michael J. and Khabiboulline, Emil T. and Kubica, Aleksander and Salton, Grant and Wang, Samson and Brand{\~{a}}o, Fernando G. S. L.},
eprint = {2310.03011},
title = {{Quantum algorithms: A survey of applications and end-to-end complexities}},
year = {2023}
}

@misc{Eisert2025,
archivePrefix = {arXiv},
author = {Eisert, Jens and Preskill, John},
eprint = {2510.19928},
title = {{Mind the gaps: The fraught road to quantum advantage}},
year = {2025}
}


%% file: supp.bib
@article{swiadek2024enhancing,
  title={Enhancing dispersive readout of superconducting qubits through dynamic control of the dispersive shift: Experiment and theory},
  author={Swiadek, Fran{\c{c}}ois and Shillito, Ross and Magnard, Paul and Remm, Ants and Hellings, Christoph and Lacroix, Nathan and Ficheux, Quentin and Zanuz, Dante Colao and Norris, Graham J and Blais, Alexandre and others},
  journal={PRX Quantum},
  volume={5},
  number={4},
  pages={040326},
  year={2024},
  publisher={APS},
  month = {Nov},
  doi = {https://doi.org/10.1103/PRXQuantum.5.040326},
  url = {https://journals.aps.org/prxquantum/abstract/10.1103/PRXQuantum.5.040326}
}

@article{sete2015quantum,
  title={Quantum theory of a bandpass Purcell filter for qubit readout},
  author={Sete, Eyob A and Martinis, John M and Korotkov, Alexander N},
  journal={Physical Review A},
  volume={92},
  number={1},
  pages={012325},
  year={2015},
  publisher={APS},
  month = {Jul},
  doi = {https://doi.org/10.1103/PhysRevA.92.012325},
  url = {https://journals.aps.org/pra/abstract/10.1103/PhysRevA.92.012325}
}

@article{gardiner1985input,
  title={Input and output in damped quantum systems: Quantum stochastic differential equations and the master equation},
  author={Gardiner, Crispin W and Collett, Matthew J},
  journal={Physical Review A},
  volume={31},
  number={6},
  pages={3761},
  year={1985},
  month = {June},
  publisher={APS},
  doi = {https://doi.org/10.1103/PhysRevA.31.3761},
  url = {https://journals.aps.org/pra/abstract/10.1103/PhysRevA.31.3761}
}

@article{blais2021circuit,
  title={Circuit quantum electrodynamics},
  author={Blais, Alexandre and Grimsmo, Arne L and Girvin, Steven M and Wallraff, Andreas},
  journal={Reviews of Modern Physics},
  volume={93},
  number={2},
  pages={025005},
  year={2021},
  month = {May},
  publisher={APS},
  doi = {https://doi.org/10.1103/RevModPhys.93.025005},
  url = {https://journals.aps.org/rmp/abstract/10.1103/RevModPhys.93.025005}
}

@article{heinsoo2018rapid,
  title={Rapid high-fidelity multiplexed readout of superconducting qubits},
  author={Heinsoo, Johannes and Andersen, Christian Kraglund and Remm, Ants and Krinner, Sebastian and Walter, Theodore and Salath{\'e}, Yves and Gasparinetti, Simone and Besse, Jean-Claude and Poto{\v{c}}nik, Anton and Wallraff, Andreas and others},
  journal={Physical Review Applied},
  volume={10},
  number={3},
  pages={034040},
  year={2018},
  month = {Sep},
  publisher={APS},
  doi = {https://doi.org/10.1103/PhysRevApplied.10.034040},
  url = {https://journals.aps.org/prapplied/abstract/10.1103/PhysRevApplied.10.034040}
}

@article{walter2017rapid,
  title={Rapid high-fidelity single-shot dispersive readout of superconducting qubits},
  author={Walter, Theodore and Kurpiers, Philipp and Gasparinetti, Simone and Magnard, Paul and Poto{\v{c}}nik, Anton and Salath{\'e}, Yves and Pechal, Marek and Mondal, Mintu and Oppliger, Markus and Eichler, Christopher and others},
  journal={Physical Review Applied},
  volume={7},
  number={5},
  pages={054020},
  year={2017},
  month = {May},
  publisher={APS},
  doi = {https://doi.org/10.1103/PhysRevApplied.7.054020},
  url = {https://journals.aps.org/prapplied/abstract/10.1103/PhysRevApplied.7.054020}
}

@book{haroche2006exploring,
  title={Exploring the quantum: atoms, cavities, and photons},
  author={Haroche, Serge and Raimond, J-M},
  year={2006},
  month = {Aug},
  publisher={Oxford university press},
  doi = {https://doi.org/10.1093/acprof:oso/9780198509141.001.0001},
  url = {https://academic.oup.com/book/7346}
}

@article{reed2010fast,
  title={Fast reset and suppressing spontaneous emission of a superconducting qubit},
  author={Reed, Matthew D and Johnson, Blake R and Houck, Andrew A and DiCarlo, Leonardo and Chow, Jerry M and Schuster, David I and Frunzio, Luigi and Schoelkopf, Robert J},
  journal={Applied Physics Letters},
  volume={96},
  pages = {203110},
  number={20},
  year={2010},
  month = {May},
  publisher={AIP Publishing},
  doi = {https://doi.org/10.1063/1.3435463}
}

@article{jeffrey2014fast,
  title={Fast accurate state measurement with superconducting qubits},
  author={Jeffrey, Evan and Sank, Daniel and Mutus, JY and White, TC and Kelly, J and Barends, R and Chen, Y and Chen, Z and Chiaro, B and Dunsworth, A and others},
  journal={Physical Review Letters},
  volume={112},
  number={19},
  pages={190504},
  year={2014},
  month = {May},
  publisher={APS},
  doi = {https://doi.org/10.1103/PhysRevLett.112.190504}
}

@article{bronn2015broadband,
  title={Broadband filters for abatement of spontaneous emission in circuit quantum electrodynamics},
  author={Bronn, Nicholas T and Liu, Yanbing and Hertzberg, Jared B and C{\'o}rcoles, Antonio D and Houck, Andrew A and Gambetta, Jay M and Chow, Jerry M},
  journal={Applied Physics Letters},
  volume={107},
  pages = {172601},
  number={17},
  year={2015},
  month = {Oct},
  publisher={AIP Publishing},
  doi = {https://doi.org/10.1063/1.4934867},
  url = {https://pubs.aip.org/aip/apl/article/107/17/172601/28669/Broadband-filters-for-abatement-of-spontaneous}
}

@book{pozar2021microwave,
  title={Microwave engineering: theory and techniques},
  author={Pozar, David M},
  year={2021},
  month={Nov},
  publisher={John Wiley \& Sons},
  isbn = {978-0-470-63155-3}
}
